\documentclass[a4paper,11pt,onecolumn,accepted=2026-10-16]{quantumarticle}
\pdfoutput=1

\usepackage{graphicx}
\usepackage{dcolumn}% Align table columns on decimal point
\usepackage{bm}% bold math
\usepackage{appendix}
\usepackage{tikzit}
\usepackage{tikz, circuitikz}
\tikzstyle{env}=[copoint,regular polygon rotate=0,minimum width=0.2cm, fill=black]

\tikzstyle{probs}=[shape=semicircle,fill=white,draw=black,shape border rotate=180,minimum width=1.2cm]

\tikzstyle{nudge}=[yshift=0.6mm]

\tikzstyle{every picture}=[baseline=-0.25em,scale=0.5]
\tikzstyle{dotpic}=[] % for backwards-compatibility
\tikzstyle{diredges}=[every to/.style={diredge}]
\tikzstyle{math matrix}=[matrix of math nodes,left delimiter=(,right delimiter=),inner sep=2pt,column sep=1em,row sep=0.5em,nodes={inner sep=0pt},text height=1.5ex, text depth=0.25ex]

\tikzstyle{inline text}=[text height=1.5ex, text depth=0.25ex,yshift=0.5mm]
\tikzstyle{label}=[font=\footnotesize,text height=1.5ex, text depth=0.25ex]
\tikzstyle{left label}=[label,anchor=east,xshift=2mm]
\tikzstyle{right label}=[label,anchor=west,xshift=-2mm]

\tikzstyle{braceedge}=[decorate,decoration={brace,amplitude=2mm,raise=-1mm}]
\tikzstyle{small braceedge}=[decorate,decoration={brace,amplitude=1mm,raise=-1mm}]

\tikzstyle{doubled}=[line width=1.6pt] % set the line width for all doubled (quantum) maps/wires
\tikzstyle{boldedge}=[doubled,shorten <=-0.17mm,shorten >=-0.17mm]
\tikzstyle{boldedgegray}=[doubled,gray,shorten <=-0.17mm,shorten >=-0.17mm]
\tikzstyle{singleedgegray}=[gray]%,shorten <=-0.1mm,shorten >=-0.1mm]

\tikzstyle{semidoubled}=[line width=1.4pt] % set the line width for all doubled (quantum) maps/wires
\tikzstyle{semiboldedgegray}=[semidoubled,gray,shorten <=-0.17mm,shorten >=-0.17mm]

\tikzstyle{boxedge}=[semiboldedgegray]

\tikzstyle{boldedgedashed}=[very thick,dashed,shorten <=-0.17mm,shorten >=-0.17mm]
\tikzstyle{vboldedgedashed}=[doubled,dashed,shorten <=-0.17mm,shorten >=-0.17mm]
\tikzstyle{left hook arrow}=[left hook-latex]
\tikzstyle{right hook arrow}=[right hook-latex]
\tikzstyle{sembracket}=[line width=0.5pt,shorten <=-0.07mm,shorten >=-0.07mm]

\tikzstyle{causal edge}=[->,thick,gray]
\tikzstyle{causal nondir}=[thick,gray]
\tikzstyle{timeline}=[thick,gray, dashed]

\tikzstyle{cedge}=[<->,thick,gray!70!white]

\tikzstyle{empty diagram}=[draw=gray!40!white,dashed,shape=rectangle,minimum width=1cm,minimum height=1cm]
\tikzstyle{empty diagram small}=[draw=gray!50!white,dashed,shape=rectangle,minimum width=0.6cm,minimum height=0.5cm]

\tikzstyle{dot}=[inner sep=0mm,minimum width=2mm,minimum height=2mm,draw,shape=circle]  
\tikzstyle{Wsquare}=[white dot, shape=regular polygon, rounded corners=0.8 mm, minimum size=3.3 mm, regular polygon sides=3, outer sep=-0.2mm]
\tikzstyle{Wsquareadj}=[white dot, shape=regular polygon, rounded corners=0.8 mm, minimum size=3.3 mm, regular polygon sides=3, outer sep=-0.2mm, regular polygon rotate=180]
\tikzstyle{ddot}=[inner sep=0mm, doubled, minimum width=2.5mm,minimum height=2.5mm,draw,shape=circle]

\tikzstyle{black dot}=[dot,fill=black]
\tikzstyle{white dot}=[dot,fill=white,,text depth=-0.2mm]
\tikzstyle{white Wsquare}=[Wsquare,fill=white,,text depth=-0.2mm]
\tikzstyle{white Wsquareadj}=[Wsquareadj,fill=white,,text depth=-0.2mm]
\tikzstyle{green dot}=[white dot] % for backwards-compatibility
\tikzstyle{gray dot}=[dot,fill=gray!40!white,,text depth=-0.2mm]
\tikzstyle{red dot}=[gray dot] % for backwards-compatibility

\tikzstyle{black ddot}=[ddot,fill=black]
\tikzstyle{white ddot}=[ddot,fill=white]
\tikzstyle{gray ddot}=[ddot,fill=gray!40!white]

\tikzstyle{gray edge}=[gray!60!white]

\tikzstyle{small dot}=[inner sep=0.5mm,minimum width=0pt,minimum height=0pt,draw,shape=circle]

\tikzstyle{small black dot}=[small dot,fill=black]
\tikzstyle{small white dot}=[small dot,fill=white]
\tikzstyle{small gray dot}=[small dot,fill=gray!40!white]

\tikzstyle{very small dot}=[inner sep=0.3mm,minimum width=0pt,minimum height=0pt,draw,shape=circle]

\tikzstyle{very small black dot}=[very small dot,fill=black]
\tikzstyle{very small white dot}=[small dot,fill=white]
\tikzstyle{very small gray dot}=[small dot,fill=gray!40!white]

\tikzstyle{causal dot}=[inner sep=0.4mm,minimum width=0pt,minimum height=0pt,draw=white,shape=circle,fill=gray!40!white]

\tikzstyle{phase dimensions}=[minimum size=5mm,font=\footnotesize,rectangle,rounded corners=2.5mm,inner sep=0.2mm,outer sep=-2mm]
\tikzstyle{dphase dimensions}=[minimum size=5mm,font=\footnotesize,rectangle,rounded corners=2.5mm,inner sep=0.2mm,outer sep=-2mm]
\tikzstyle{white phase dot}=[dot,fill=white,phase dimensions]
\tikzstyle{white phase ddot}=[ddot,fill=white,dphase dimensions]

\tikzstyle{white rect ddot}=[draw=black,fill=white,doubled,minimum size=5mm,font=\footnotesize,rectangle,rounded corners=2.5mm,inner sep=0.2mm]
\tikzstyle{gray rect ddot}=[draw=black,fill=gray!40!white,doubled,minimum size=6mm,font=\footnotesize,rectangle,rounded corners=3mm]

\tikzstyle{gray phase dot}=[dot,fill=gray!40!white,phase dimensions]
\tikzstyle{gray phase ddot}=[ddot,fill=gray!40!white,dphase dimensions]
\tikzstyle{grey phase dot}=[gray phase dot]
\tikzstyle{grey phase ddot}=[gray phase ddot]

\tikzstyle{small phase dimensions}=[minimum size=4mm,font=\tiny,rectangle,rounded corners=2mm,inner sep=0.2mm,outer sep=-2mm]
\tikzstyle{small dphase dimensions}=[minimum size=4mm,font=\tiny,rectangle,rounded corners=2mm,inner sep=0.2mm,outer sep=-2mm]

\tikzstyle{small gray phase dot}=[dot,fill=gray!40!white,small phase dimensions]
\tikzstyle{small gray phase ddot}=[ddot,fill=gray!40!white,small dphase dimensions]

\tikzstyle{small map}=[draw,shape=rectangle,minimum height=4mm,minimum width=4mm,fill=white]

\tikzstyle{cnot}=[fill=white,shape=circle,inner sep=-1.4pt]

\tikzstyle{asym hadamard}=[fill=white,draw,shape=NEbox,inner sep=0.6mm,font=\footnotesize,minimum height=4mm]
\tikzstyle{asym hadamard conj}=[fill=white,draw,shape=NWbox,inner sep=0.6mm,font=\footnotesize,minimum height=4mm]
\tikzstyle{asym hadamard dag}=[fill=white,draw,shape=SEbox,inner sep=0.6mm,font=\footnotesize,minimum height=4mm]

\tikzstyle{hadamard}=[fill=white,draw,inner sep=0.6mm,font=\footnotesize,minimum height=4mm,minimum width=4mm]
\tikzstyle{small hadamard}=[fill=white,draw,inner sep=0.6mm,minimum height=1.5mm,minimum width=1.5mm]
\tikzstyle{small hadamard rotate}=[small hadamard,rotate=45]
\tikzstyle{dhadamard}=[hadamard,doubled]
\tikzstyle{small dhadamard}=[small hadamard,doubled]
\tikzstyle{small dhadamard rotate}=[small hadamard rotate,doubled]
\tikzstyle{antipode}=[white dot,inner sep=0.3mm,font=\footnotesize]

\tikzstyle{scalar}=[diamond,draw,inner sep=0.5pt,font=\small]
\tikzstyle{dscalar}=[diamond,doubled, draw,inner sep=0.5pt,font=\small]

\tikzstyle{small box}=[rectangle,inline text,fill=white,draw,minimum height=5mm,yshift=-0.5mm,minimum width=5mm,font=\small]
\tikzstyle{small gray box}=[small box,fill=gray!30]
\tikzstyle{medium box}=[rectangle,inline text,fill=white,draw,minimum height=5mm,yshift=-0.5mm,minimum width=8mm,font=\small]
\tikzstyle{square box}=[small box] % for backwards-compatibility
\tikzstyle{medium gray box}=[small box,fill=gray!30]
\tikzstyle{semilarge box}=[rectangle,inline text,fill=white,draw,minimum height=5mm,yshift=-0.5mm,minimum width=12.5mm,font=\small]
\tikzstyle{large box}=[rectangle,inline text,fill=white,draw,minimum height=5mm,yshift=-0.5mm,minimum width=15mm,font=\small]
\tikzstyle{large gray box}=[small box,fill=gray!30]

\tikzstyle{Bayes box}=[rectangle,fill=black,draw, minimum height=3mm, minimum width=3mm]

\tikzstyle{gray square point}=[small box,fill=gray!50]

\tikzstyle{dphase box white}=[dhadamard]
\tikzstyle{dphase box gray}=[dhadamard,fill=gray!50!white]
\tikzstyle{phase box white}=[hadamard]
\tikzstyle{phase box gray}=[hadamard,fill=gray!50!white]

\tikzstyle{point}=[regular polygon,regular polygon sides=3,draw,scale=0.75,inner sep=-0.5pt,minimum width=9mm,fill=white,regular polygon rotate=180]
\tikzstyle{copoint}=[regular polygon,regular polygon sides=3,draw,scale=0.75,inner sep=-0.5pt,minimum width=9mm,fill=white]
\tikzstyle{dpoint}=[point,doubled]
\tikzstyle{dcopoint}=[copoint,doubled]

\tikzstyle{wide copoint}=[fill=white,draw,shape=isosceles triangle,shape border rotate=90,isosceles triangle stretches=true,inner sep=0pt,minimum width=1.5cm,minimum height=6.12mm]
\tikzstyle{wide point}=[fill=white,draw,shape=isosceles triangle,shape border rotate=-90,isosceles triangle stretches=true,inner sep=0pt,minimum width=1.5cm,minimum height=6.12mm,yshift=-0.0mm]
\tikzstyle{wide point plus}=[fill=white,draw,shape=isosceles triangle,shape border rotate=-90,isosceles triangle stretches=true,inner sep=0pt,minimum width=1.74cm,minimum height=7mm,yshift=-0.0mm]

\tikzstyle{wide dpoint}=[fill=white,doubled,draw,shape=isosceles triangle,shape border rotate=-90,isosceles triangle stretches=true,inner sep=0pt,minimum width=1.5cm,minimum height=6.12mm,yshift=-0.0mm]

\tikzstyle{tinypoint}=[regular polygon,regular polygon sides=3,draw,scale=0.55,inner sep=-0.15pt,minimum width=6mm,fill=white,regular polygon rotate=180] 

\tikzstyle{white point}=[point]
\tikzstyle{white dpoint}=[dpoint]
\tikzstyle{green point}=[white point] % for backwards-compatibility
\tikzstyle{white copoint}=[copoint]
\tikzstyle{gray point}=[point,fill=gray!40!white]
\tikzstyle{gray dpoint}=[gray point,doubled]
\tikzstyle{red point}=[gray point] % for backwards-compatibility
\tikzstyle{gray copoint}=[copoint,fill=gray!40!white]
\tikzstyle{gray dcopoint}=[gray copoint,doubled]

\tikzstyle{white point guide}=[regular polygon,regular polygon sides=3,font=\scriptsize,draw,scale=0.65,inner sep=-0.5pt,minimum width=9mm,fill=white,regular polygon rotate=180]

\tikzstyle{black point}=[point,fill=black,font=\color{white}]
\tikzstyle{black copoint}=[copoint,fill=black,font=\color{white}]

\tikzstyle{tiny gray point}=[tinypoint,fill=gray!40!white]

\tikzstyle{diredge}=[->]
\tikzstyle{ddiredge}=[<->]
\tikzstyle{rdiredge}=[<-]
\tikzstyle{thickdiredge}=[->, very thick]
\tikzstyle{pointer edge}=[->,very thick,gray]
\tikzstyle{pointer edge part}=[very thick,gray]
\tikzstyle{dashed edge}=[dashed]
\tikzstyle{thick dashed edge}=[very thick,dashed]
\tikzstyle{thick gray dashed edge}=[thick dashed edge,gray!40]
\tikzstyle{thick map edge}=[very thick,|->]

\makeatletter
\newcommand{\boxshape}[3]{%
\pgfdeclareshape{#1}{
\inheritsavedanchors[from=rectangle] % this is nearly a rectangle
\inheritanchorborder[from=rectangle]
\inheritanchor[from=rectangle]{center}
\inheritanchor[from=rectangle]{north}
\inheritanchor[from=rectangle]{south}
\inheritanchor[from=rectangle]{west}
\inheritanchor[from=rectangle]{east}
\backgroundpath{% this is new
\southwest \pgf@xa=\pgf@x \pgf@ya=\pgf@y
\northeast \pgf@xb=\pgf@x \pgf@yb=\pgf@y

\@tempdima=#2
\@tempdimb=#3

\pgfpathmoveto{\pgfpoint{\pgf@xa - 5pt + \@tempdima}{\pgf@ya}}
\pgfpathlineto{\pgfpoint{\pgf@xa - 5pt - \@tempdima}{\pgf@yb}}
\pgfpathlineto{\pgfpoint{\pgf@xb + 5pt + \@tempdimb}{\pgf@yb}}
\pgfpathlineto{\pgfpoint{\pgf@xb + 5pt - \@tempdimb}{\pgf@ya}}
\pgfpathlineto{\pgfpoint{\pgf@xa - 5pt + \@tempdima}{\pgf@ya}}
\pgfpathclose
}
}}

\boxshape{NEbox}{0pt}{0pt}
\boxshape{SEbox}{0pt}{-5pt}
\boxshape{NWbox}{5pt}{0pt}
\boxshape{SWbox}{-5pt}{0pt}
\boxshape{EBox}{-3pt}{3pt}
\boxshape{WBox}{3pt}{-3pt}
\makeatother

\tikzstyle{cloud}=[shape=cloud,draw,minimum width=1.5cm,minimum height=1.5cm]

\tikzstyle{map}=[draw,shape=NEbox,inner sep=2pt,minimum height=6mm,fill=white]
\tikzstyle{dashedmap}=[draw,dashed,gray,shape=NEbox,inner sep=2pt,minimum height=6mm,fill=white]
\tikzstyle{medium dashedmap}=[draw,dashed,gray,shape=NEbox,inner sep=2pt,minimum height=6mm,fill=white,minimum width=7mm]
\tikzstyle{semilarge dashedmap}=[draw,dashed,gray,shape=NEbox,inner sep=2pt,minimum height=6mm,fill=white,minimum width=9.5mm]
\tikzstyle{large dashedmap}=[draw,dashed,gray,shape=NEbox,inner sep=2pt,minimum height=6mm,fill=white,minimum width=12mm]
\tikzstyle{very large dashedmap}=[draw,dashed,gray,shape=NEbox,inner sep=2pt,minimum height=6mm,fill=white,minimum width=17mm]

\tikzstyle{dashed map}=[fill=white, draw=gray, shape=rectangle, style=map, dashed]

\tikzstyle{mapdag}=[draw,shape=SEbox,inner sep=2pt,minimum height=6mm,fill=white]
\tikzstyle{mapadj}=[draw,shape=SEbox,inner sep=2pt,minimum height=6mm,fill=white]
\tikzstyle{maptrans}=[draw,shape=SWbox,inner sep=2pt,minimum height=6mm,fill=white]
\tikzstyle{mapconj}=[draw,shape=NWbox,inner sep=2pt,minimum height=6mm,fill=white]

\tikzstyle{medium map}=[draw,shape=NEbox,inner sep=2pt,minimum height=6mm,fill=white,minimum width=7mm]
\tikzstyle{medium map dag}=[draw,shape=SEbox,inner sep=2pt,minimum height=6mm,fill=white,minimum width=7mm]
\tikzstyle{medium map adj}=[draw,shape=SEbox,inner sep=2pt,minimum height=6mm,fill=white,minimum width=7mm]
\tikzstyle{medium map trans}=[draw,shape=SWbox,inner sep=2pt,minimum height=6mm,fill=white,minimum width=7mm]
\tikzstyle{medium map conj}=[draw,shape=NWbox,inner sep=2pt,minimum height=6mm,fill=white,minimum width=7mm]
\tikzstyle{semilarge map}=[draw,shape=NEbox,inner sep=2pt,minimum height=6mm,fill=white,minimum width=9.5mm]
\tikzstyle{semilarge map trans}=[draw,shape=SWbox,inner sep=2pt,minimum height=6mm,fill=white,minimum width=9.5mm]
\tikzstyle{semilarge map adj}=[draw,shape=SEbox,inner sep=2pt,minimum height=6mm,fill=white,minimum width=9.5mm]
\tikzstyle{semilarge map dag}=[draw,shape=SEbox,inner sep=2pt,minimum height=6mm,fill=white,minimum width=9.5mm]
\tikzstyle{semilarge map conj}=[draw,shape=NWbox,inner sep=2pt,minimum height=6mm,fill=white,minimum width=9.5mm]
\tikzstyle{large map}=[draw,shape=NEbox,inner sep=2pt,minimum height=6mm,fill=white,minimum width=12mm]
\tikzstyle{large map conj}=[draw,shape=NWbox,inner sep=2pt,minimum height=6mm,fill=white,minimum width=12mm]
\tikzstyle{very large map}=[draw,shape=NEbox,inner sep=2pt,minimum height=6mm,fill=white,minimum width=17mm]
\tikzstyle{very very large map}=[draw,shape=NEbox,inner sep=2pt,minimum height=6mm,fill=white,minimum width=50mm]
\tikzstyle{large map dag}=[draw,shape=SEbox,inner sep=2pt,minimum height=6mm,fill=white,minimum width=12mm]

\tikzstyle{medium dmap}=[draw,doubled,shape=NEbox,inner sep=2pt,minimum height=6mm,fill=white,minimum width=7mm]
\tikzstyle{medium dmap dag}=[draw,doubled,shape=SEbox,inner sep=2pt,minimum height=6mm,fill=white,minimum width=7mm]
\tikzstyle{medium dmap adj}=[draw,doubled,shape=SEbox,inner sep=2pt,minimum height=6mm,fill=white,minimum width=7mm]
\tikzstyle{medium dmap trans}=[draw,doubled,shape=SWbox,inner sep=2pt,minimum height=6mm,fill=white,minimum width=7mm]
\tikzstyle{medium dmap conj}=[draw,doubled,shape=NWbox,inner sep=2pt,minimum height=6mm,fill=white,minimum width=7mm]
\tikzstyle{semilarge dmap}=[draw,doubled,shape=NEbox,inner sep=2pt,minimum height=6mm,fill=white,minimum width=9.5mm]
\tikzstyle{semilarge dmap trans}=[draw,doubled,shape=SWbox,inner sep=2pt,minimum height=6mm,fill=white,minimum width=9.5mm]
\tikzstyle{semilarge dmap adj}=[draw,doubled,shape=SEbox,inner sep=2pt,minimum height=6mm,fill=white,minimum width=9.5mm]
\tikzstyle{semilarge dmap dag}=[draw,doubled,shape=SEbox,inner sep=2pt,minimum height=6mm,fill=white,minimum width=9.5mm]
\tikzstyle{semilarge dmap conj}=[draw,doubled,shape=NWbox,inner sep=2pt,minimum height=6mm,fill=white,minimum width=9.5mm]
\tikzstyle{large dmap}=[draw,doubled,shape=NEbox,inner sep=2pt,minimum height=6mm,fill=white,minimum width=12mm]
\tikzstyle{large dmap conj}=[draw,doubled,shape=NWbox,inner sep=2pt,minimum height=6mm,fill=white,minimum width=12mm]
\tikzstyle{large dmap trans}=[draw,doubled,shape=SWbox,inner sep=2pt,minimum height=6mm,fill=white,minimum width=12mm]
\tikzstyle{large dmap adj}=[draw,doubled,shape=SEbox,inner sep=2pt,minimum height=6mm,fill=white,minimum width=12mm]
\tikzstyle{large dmap dag}=[draw,doubled,shape=SEbox,inner sep=2pt,minimum height=6mm,fill=white,minimum width=12mm]
\tikzstyle{very large dmap}=[draw,doubled,shape=NEbox,inner sep=2pt,minimum height=6mm,fill=white,minimum width=19.5mm]

\tikzstyle{muxbox}=[draw,shape=rectangle,minimum height=3mm,minimum width=3mm,fill=white]
\tikzstyle{dmuxbox}=[muxbox,doubled]

\tikzstyle{box}=[draw,shape=rectangle,inner sep=2pt,minimum height=6mm,minimum width=6mm,fill=white]
\tikzstyle{dbox}=[draw,doubled,shape=rectangle,inner sep=2pt,minimum height=6mm,minimum width=6mm,fill=white]
\tikzstyle{dmap}=[draw,doubled,shape=NEbox,inner sep=2pt,minimum height=6mm,fill=white]
\tikzstyle{dmapdag}=[draw,doubled,shape=SEbox,inner sep=2pt,minimum height=6mm,fill=white]
\tikzstyle{dmapadj}=[draw,doubled,shape=SEbox,inner sep=2pt,minimum height=6mm,fill=white]
\tikzstyle{dmaptrans}=[draw,doubled,shape=SWbox,inner sep=2pt,minimum height=6mm,fill=white]
\tikzstyle{dmapconj}=[draw,doubled,shape=NWbox,inner sep=2pt,minimum height=6mm,fill=white]

\tikzstyle{ddmap}=[draw,doubled,dashed,shape=NEbox,inner sep=2pt,minimum height=6mm,fill=white]
\tikzstyle{ddmapdag}=[draw,doubled,dashed,shape=SEbox,inner sep=2pt,minimum height=6mm,fill=white]
\tikzstyle{ddmapadj}=[draw,doubled,dashed,shape=SEbox,inner sep=2pt,minimum height=6mm,fill=white]
\tikzstyle{ddmaptrans}=[draw,doubled,dashed,shape=SWbox,inner sep=2pt,minimum height=6mm,fill=white]
\tikzstyle{ddmapconj}=[draw,doubled,dashed,shape=NWbox,inner sep=2pt,minimum height=6mm,fill=white]

\boxshape{sNEbox}{0pt}{3pt}
\boxshape{sSEbox}{0pt}{-3pt}
\boxshape{sNWbox}{3pt}{0pt}
\boxshape{sSWbox}{-3pt}{0pt}
\tikzstyle{smap}=[draw,shape=sNEbox,fill=white]
\tikzstyle{smapdag}=[draw,shape=sSEbox,fill=white]
\tikzstyle{smapadj}=[draw,shape=sSEbox,fill=white]
\tikzstyle{smaptrans}=[draw,shape=sSWbox,fill=white]
\tikzstyle{smapconj}=[draw,shape=sNWbox,fill=white]

\tikzstyle{dsmap}=[draw,dashed,shape=sNEbox,fill=white]
\tikzstyle{dsmapdag}=[draw,dashed,shape=sSEbox,fill=white]
\tikzstyle{dsmaptrans}=[draw,dashed,shape=sSWbox,fill=white]
\tikzstyle{dsmapconj}=[draw,dashed,shape=sNWbox,fill=white]

\boxshape{mNEbox}{0pt}{10pt}
\boxshape{mSEbox}{0pt}{-10pt}
\boxshape{mNWbox}{10pt}{0pt}
\boxshape{mSWbox}{-10pt}{0pt}
\tikzstyle{mmap}=[draw,shape=mNEbox]
\tikzstyle{mmapdag}=[draw,shape=mSEbox]
\tikzstyle{mmaptrans}=[draw,shape=mSWbox]
\tikzstyle{mmapconj}=[draw,shape=mNWbox]

\tikzstyle{mmapgray}=[draw,fill=gray!40!white,shape=mNEbox]
\tikzstyle{smapgray}=[draw,fill=gray!40!white,shape=sNEbox]

\makeatletter

\pgfdeclareshape{cornerpoint}{
\inheritsavedanchors[from=rectangle] % this is nearly a rectangle
\inheritanchorborder[from=rectangle]
\inheritanchor[from=rectangle]{center}
\inheritanchor[from=rectangle]{north}
\inheritanchor[from=rectangle]{south}
\inheritanchor[from=rectangle]{west}
\inheritanchor[from=rectangle]{east}
\backgroundpath{% this is new
\southwest \pgf@xa=\pgf@x \pgf@ya=\pgf@y
\northeast \pgf@xb=\pgf@x \pgf@yb=\pgf@y

\pgfmathsetmacro{\pgf@shorten@left}{\pgfkeysvalueof{/tikz/shorten left}}
\pgfmathsetmacro{\pgf@shorten@right}{\pgfkeysvalueof{/tikz/shorten right}}

\pgfpathmoveto{\pgfpoint{0.5 * (\pgf@xa + \pgf@xb)}{\pgf@ya - 5pt}}
\pgfpathlineto{\pgfpoint{\pgf@xa - 8pt + \pgf@shorten@left}{\pgf@yb - 1.5 * \pgf@shorten@left}}
\pgfpathlineto{\pgfpoint{\pgf@xa - 8pt + \pgf@shorten@left}{\pgf@yb}}
\pgfpathlineto{\pgfpoint{\pgf@xb + 8pt - \pgf@shorten@right}{\pgf@yb}}
\pgfpathlineto{\pgfpoint{\pgf@xb + 8pt - \pgf@shorten@right}{\pgf@yb - 1.5 * \pgf@shorten@right}}
\pgfpathclose
}
}

\pgfdeclareshape{cornercopoint}{
\inheritsavedanchors[from=rectangle] % this is nearly a rectangle
\inheritanchorborder[from=rectangle]
\inheritanchor[from=rectangle]{center}
\inheritanchor[from=rectangle]{north}
\inheritanchor[from=rectangle]{south}
\inheritanchor[from=rectangle]{west}
\inheritanchor[from=rectangle]{east}
\backgroundpath{% this is new
\southwest \pgf@xa=\pgf@x \pgf@ya=\pgf@y
\northeast \pgf@xb=\pgf@x \pgf@yb=\pgf@y

\pgfmathsetmacro{\pgf@shorten@left}{\pgfkeysvalueof{/tikz/shorten left}}
\pgfmathsetmacro{\pgf@shorten@right}{\pgfkeysvalueof{/tikz/shorten right}}

\pgfpathmoveto{\pgfpoint{0.5 * (\pgf@xa + \pgf@xb)}{\pgf@yb + 5pt}}
\pgfpathlineto{\pgfpoint{\pgf@xa - 8pt + \pgf@shorten@left}{\pgf@ya + 1.5 * \pgf@shorten@left}}
\pgfpathlineto{\pgfpoint{\pgf@xa - 8pt + \pgf@shorten@left}{\pgf@ya}}
\pgfpathlineto{\pgfpoint{\pgf@xb + 8pt - \pgf@shorten@right}{\pgf@ya}}
\pgfpathlineto{\pgfpoint{\pgf@xb + 8pt - \pgf@shorten@right}{\pgf@ya + 1.5 * \pgf@shorten@right}}
\pgfpathclose
}
}

\makeatother

\pgfkeyssetvalue{/tikz/shorten left}{0pt}
\pgfkeyssetvalue{/tikz/shorten right}{0pt}

\tikzstyle{kpoint common}=[draw,fill=white,inner sep=1pt,minimum height=4mm]
\tikzstyle{kpoint sc}=[shape=cornerpoint,kpoint common]
\tikzstyle{kpoint adjoint sc}=[shape=cornercopoint,kpoint common]
\tikzstyle{kpoint}=[shape=cornerpoint,shorten left=5pt,kpoint common]
\tikzstyle{kpoint adjoint}=[shape=cornercopoint,shorten left=5pt,kpoint common]
\tikzstyle{kpoint conjugate}=[shape=cornerpoint,shorten right=5pt,kpoint common]
\tikzstyle{kpoint transpose}=[shape=cornercopoint,shorten right=5pt,kpoint common]
\tikzstyle{kpoint symm}=[shape=cornerpoint,shorten left=5pt,shorten right=5pt,kpoint common]

\tikzstyle{black kpoint}=[shape=cornerpoint,shorten left=5pt,kpoint common,fill=black,font=\color{white}]
\tikzstyle{black kpoint adjoint}=[shape=cornercopoint,shorten left=5pt,kpoint common,fill=black,font=\color{white}]
\tikzstyle{black kpointadj}=[shape=cornercopoint,shorten left=5pt,kpoint common,fill=black,font=\color{white}]

\tikzstyle{black dkpoint}=[shape=cornerpoint,shorten left=5pt,kpoint common,fill=black, doubled,font=\color{white}]
\tikzstyle{black dkpoint adjoint}=[shape=cornercopoint,shorten left=5pt,kpoint common,fill=black, doubled,font=\color{white}]
\tikzstyle{black dkpointadj}=[shape=cornercopoint,shorten left=5pt,kpoint common,fill=black, doubled,font=\color{white}] 

\tikzstyle{kpointdag}=[kpoint adjoint]
\tikzstyle{kpointadj}=[kpoint adjoint]
\tikzstyle{kpointconj}=[kpoint conjugate]
\tikzstyle{kpointtrans}=[kpoint transpose]

\tikzstyle{big kpoint}=[kpoint, minimum width=1.2 cm, minimum height=8mm, inner sep=4pt, text depth=3mm]

\tikzstyle{wide kpoint}=[kpoint, minimum width=1 cm, inner sep=2pt]%, text depth=-0.7 mm]
\tikzstyle{wide kpointdag}=[kpointdag, minimum width=1 cm, inner sep=2pt]%, text depth=0.7 mm]
\tikzstyle{wide kpointconj}=[kpointconj, minimum width=1 cm, inner sep=2pt]%, text depth=-0.7 mm]
\tikzstyle{wide kpointtrans}=[kpointtrans, minimum width=1 cm, inner sep=2pt]%, text depth=0.7 mm]

\tikzstyle{gray kpoint}=[kpoint,fill=gray!50!white]
\tikzstyle{gray kpointdag}=[kpointdag,fill=gray!50!white]
\tikzstyle{gray kpointadj}=[kpointadj,fill=gray!50!white]
\tikzstyle{gray kpointconj}=[kpointconj,fill=gray!50!white]
\tikzstyle{gray kpointtrans}=[kpointtrans,fill=gray!50!white]

\tikzstyle{gray dkpoint}=[kpoint,fill=gray!50!white,doubled]
\tikzstyle{gray dkpointdag}=[kpointdag,fill=gray!50!white,doubled]
\tikzstyle{gray dkpointadj}=[kpointadj,fill=gray!50!white,doubled]
\tikzstyle{gray dkpointconj}=[kpointconj,fill=gray!50!white,doubled]
\tikzstyle{gray dkpointtrans}=[kpointtrans,fill=gray!50!white,doubled]

\tikzstyle{white label}=[draw,fill=white,rectangle,inner sep=0.7 mm]
\tikzstyle{gray label}=[draw,fill=gray!50!white,rectangle,inner sep=0.7 mm]
\tikzstyle{black label}=[draw,fill=black,rectangle,inner sep=0.7 mm]

\tikzstyle{dkpoint}=[kpoint,doubled]
\tikzstyle{wide dkpoint}=[wide kpoint,doubled]
\tikzstyle{dkpointdag}=[kpoint adjoint,doubled]
\tikzstyle{wide dkpointdag}=[wide kpointdag,doubled]
\tikzstyle{dkcopoint}=[kpoint adjoint,doubled]
\tikzstyle{dkpointadj}=[kpoint adjoint,doubled]
\tikzstyle{dkpointconj}=[kpoint conjugate,doubled]
\tikzstyle{dkpointtrans}=[kpoint transpose,doubled]

\tikzstyle{kscalar}=[kpoint common, shape=EBox, inner xsep=-1pt, inner ysep=3pt,font=\small]
\tikzstyle{kscalarconj}=[kpoint common, shape=WBox, inner xsep=-1pt, inner ysep=3pt,font=\small]

\tikzstyle{spekpoint}=[kpoint sc,minimum height=5mm,inner sep=3pt]
\tikzstyle{spekcopoint}=[kpoint adjoint sc,minimum height=5mm,inner sep=3pt]

\tikzstyle{dspekpoint}=[spekpoint,doubled]
\tikzstyle{dspekcopoint}=[spekcopoint,doubled]

 \tikzstyle{discard}=[circuit ee IEC, ground,rotate=180,scale=1.5,inner sep=-2mm]
 \tikzstyle{downground}=[circuit ee IEC,thick,ground,rotate=-90,scale=1.5,inner sep=-2mm]

\tikzstyle{maxmix}=[regular polygon,regular polygon sides=3,draw=black,xscale=0.4,yscale=0.3,inner sep=-0.5pt,minimum width=10mm,fill=gray,regular polygon rotate=180]

 \tikzstyle{bigground}=[regular polygon,regular polygon sides=3,draw=gray,scale=0.50,inner sep=-0.5pt,minimum width=10mm,fill=gray]
\tikzstyle{arrs}=[-latex,font=\small,auto]
\tikzstyle{arrow plain}=[arrs]
\tikzstyle{arrow dashed}=[dashed,arrs]
\tikzstyle{arrow bold}=[very thick,arrs]
\tikzstyle{arrow hide}=[draw=white!0,-]
\tikzstyle{arrow reverse}=[latex-]
\tikzstyle{cdnode}=[]

\tikzstyle{green dashed arrow}=[green, arrow dashed]
\tikzstyle{dashed blue}=[blue, dashed]
\tikzstyle{red dashed arrow}=[red, arrow dashed]
\tikzstyle{orange arrow}=[orange, arrs]
\tikzstyle{blue arrow}=[blue, arrs]
\tikzstyle{magenta arrow}=[magenta, arrs]

\tikzstyle{small box}=[shape=rectangle, fill=white, draw=black, minimum height=5mm, yshift=-0mm, minimum width=5mm, font={\small}]
\tikzstyle{medium box}=[shape=rectangle, fill=white, draw=black, minimum height=5mm, yshift=-0mm, minimum width=8mm, font={\small}]
\tikzstyle{semilarge box}=[shape=rectangle, fill=white, draw=black, minimum height=5mm, yshift=-0mm, minimum width=12.5mm, font={\small}]
\tikzstyle{large box}=[shape=rectangle, fill=white, draw=black, minimum height=5mm, yshift=-0mm, minimum width=15mm, font={\small}]
\tikzstyle{very large box}=[shape=rectangle, fill=white, draw=black, minimum height=5mm, yshift=-0mm, minimum width=25mm, font={\small}]
\tikzstyle{very very large box}=[shape=rectangle, fill=white, draw=black, minimum height=5mm, yshift=-0mm, minimum width=35mm, font={\small}]
\tikzstyle{left label}=[font={\footnotesize}, anchor=east, xshift=-0mm]
\tikzstyle{right label}=[font={\footnotesize}, anchor=west, xshift=0mm]
\tikzstyle{bottom label}=[font={\footnotesize}, anchor=north, yshift=0mm]
\tikzstyle{top label}=[font={\footnotesize}, anchor=south, yshift=0mm]

\tikzstyle{bolded}=[-, line width=1.6 pt]

\usepackage{physics}
\usepackage{amssymb}
\usepackage{amsmath}
\usepackage{amsfonts}
\usepackage{braket}
\usepackage{mathtools}
\usepackage{amsthm}
\usepackage{ulem}
\usepackage{bbold}
\usepackage{subcaption}
\usepackage{hyperref}
\usepackage{stmaryrd}
\usepackage{xcolor}
\usepackage{relsize}
\usepackage{epigraph}
\usepackage[disable]{todonotes}

\DeclareMathSymbol{\mlq}{\mathord}{operators}{``}
\DeclareMathSymbol{\mrq}{\mathord}{operators}{`'}

\def\be{\begin{equation}}
\def\ee{\end{equation}}
\def\ba{\begin{align}}
\def\ea{\end{align}}

\newcommand{\id}[1][]{\ensuremath{1_{#1}}}

\DeclareMathOperator{\Lin}{Lin}

\newtheorem{theorem}{Theorem}[section]

\newtheorem{corollary}{Corollary}[section]
\newtheorem{definition}{Definition}[section]
\newtheorem{lemma}{Lemma}[section]
\newtheorem{proposition}{Proposition}[section]

\newtheorem{slogan}{Slogan}
\newtheorem{example}{Example}

\newcommand{\correction}[1]{{#1}}

\DeclareTextFontCommand{\texttt}{\ttfamily\upshape}
\DeclareTextFontCommand{\textrm}{\rmfamily\upshape}

\renewcommand{\id}{\mathbb{1}}
\newcommand{\bbpi}{\mathbb{\Pi}}
\newcommand{\veck}{{\Vec{k}}}
\newcommand{\vecl}{{\Vec{l}}}
\newcommand{\vecq}{{\Vec{q}}}
\newcommand{\hatpi}{\hat{\pi}}
\newcommand{\Qbar}{{\bar{Q}}}
\newcommand{\tK}{{\tilde{K}}}

\newcommand{\ca}{\mathcal A}
\newcommand{\cb}{\mathcal B}

\newcommand{\cf}{\mathcal F}
\newcommand{\cg}{\mathcal G}
\newcommand{\ch}{\mathcal H}

\newcommand{\co}{\mathcal O}
\renewcommand{\cp}{\mathcal P}

\newcommand{\ct}{\mathcal T}

\newcommand{\cz}{\mathcal Z}

\newcommand{\CC}{\mathbb{C}}

\newcommand{\al}{\alpha}
\newcommand{\bet}{\beta}
\newcommand{\ga}{\gamma}

\newcommand{\la}{\lambda}

\newcommand{\Om}{\Omega}
\newcommand{\OM}{\mathlarger{\mathlarger{\mathlarger{\omega}}}}

\newcommand{\Atproj}{\textrm{AtomProj}}

\newcommand{\Sub}{\textrm{Sub}}

\newcommand{\ext}{\textrm{\upshape ext}}
\newcommand{\nul}{\textrm{\upshape null}}

\renewcommand{\int}{\textrm{\upshape int}}

\newcommand{\inv}{^{-1}}

\newcommand{\zerobar}{{X \setminus \{0\}}}

\begin{document}

\title{Partitions in quantum theory}

%\author{Augustin Vanrietvelde, Octave Mestoudjian, Pablo Arrighi}

\author{Augustin Vanrietvelde}
\orcid{0000-0001-9022-8655}
\email{vanrietvelde@telecom-paris.fr}
\affiliation{Télécom Paris, Institut Polytechnique de Paris,  Inria Saclay, Palaiseau, France}
%\affiliation{Quriosity Team, Inria Saclay, France}

\author{Octave Mestoudjian}
\affiliation{Université Paris-Saclay, Inria, CNRS, LMF, 91190 Gif-sur-Yvette, France}
%\affiliation{Quacs Team, Inria Saclay, France}

\author{Pablo Arrighi}
\affiliation{Université Paris-Saclay, Inria, CNRS, LMF, 91190 Gif-sur-Yvette, France}
%\affiliation{Quacs Team, Inria Saclay, France}
% \setlength\epigraphwidth{.50\textwidth}
% \epigraph{The clock hath stricken three -- 'Tis time to part.}{William Shakespeare, \textit{Julius Caesar.}}

\begin{abstract}
Decompositional theories describe the ways in which a global physical system can be split into subsystems, facilitating the study of how different possible partitions of the same system interplay, e.g.\ in terms of inclusions or signalling.
In quantum theory, subsystems are usually framed as sub-C*-algebras of the algebra of operators on the global system.
However, most decompositional approaches have so far restricted their scope to the case of systems cor\-res\-pon\-ding to factor algebras.
We argue that this is a mistake: one should cater for the possibility for non-factor subsystems, arising for instance from symmetry considerations.
Building on simple examples, we motivate and present a definition of partitions into an arbitrary number of parts, each of which is a possibly non-factor sub-C*-algebra.
We discuss its physical interpretation and study its properties, in particular with regard to the structure of algebras' centres.
We prove that partitions, defined at the C*-algebraic level, can be represented in terms of a splitting of Hilbert spaces, using the framework of routed quantum circuits.
For some partitions, however, such a representation necessarily retains a residual pseudo-nonlocality.
We provide an example of this behaviour, given by the partition of a fermionic system into local modes.
\end{abstract}
\maketitle
\tableofcontents

\section{Introduction}

The ability to consider a whole as made out of parts --- a foundational tenet of any physical theory --- can be conceptually framed in two opposite ways. In the \textit{bottom-up} or \textit{compositional} approach, one starts with individual systems, taken to be primitive, and then composes them together to form bigger, joint systems. The \textit{top-down} or \textit{decompositional} approach goes the other way: starting with a large `supersystem' $\Om$, one asks about the ways in which it can be carved up into a bunch of smaller parts. In most of physics --- and especially in quantum information --- the bottom-up approach is the standard, to the point that it usually goes without saying.

There are, however, many situations in which it proves advantageous to think top-down. This is because, in contrast with the compositional one, the decompositional approach does not posit a system's partition as written in stone: it allows one to consider scenarios in which this partition depends on the features of the situation at hand, and/or on the perspective one decides to adopt on it. Consequently, and even more crucially, it is the appropriate setting for \textit{comparing} different ways of partitioning the same system. In other words, not only does it permit us to state that $\Om$ can be partitioned either into $A_1$ and $A_2$ or into $B_1$ and $B_2$, but it then allows us to study their interplay --- e.g., whether $A_1$ is included in $B_1$, whether the information in $A_1$ corresponds to correlations between $B_1$ and $B_2$, whether one can find a common fine-graining of these two bipartitions into a quadripartition, etc.

Research directions that require comparing different partitions are, in fact, ubiquitous; let us just briefly mention a few examples from currently active research subjects in the field of quantum foundations. A first instance is quantum causal modelling \cite{Allen2017, barrett2019, barrett2021cyclic, Ormrod2023}: there, it can be seen that a unitary channel's causal structure comes down to how it relates the partition of its input into subsystems to the partition of its output. Quantum reference frames \cite{Bartlett2007, Giacomini2017, vanrietvelde2018, Carette2023} provide another example from a completely different direction: it has been demonstrated \cite{AliAhmad2021} that they essentially correspond to a relativity of subsystems, i.e.\ to the possibility to carve the world in different ways, with quantum reference frame transformations precisely embodying how these partitions relate.\footnote{See especially Ref.\ \cite{castroruiz2021} for an approach to quantum reference frames that resonates particularly closely with our considerations in this paper. The algebraic approach has also recently led to progress for the obtention of non-diverging entropies in the context of quantum gravity \cite{DeVuyst2024, DeVuyst2024bis} and lattice gauge theories \cite{Araujo-Regado2025}.} Finally, in the controversy \cite{Procopio2014, Rubino2016, oreshkov2019time, paunkovic2020causal, Felce2021, Vilasini2022, Ormrod2023, delaHamette2022, Vilasini2025} on whether an architecture called the Quantum Switch \cite{chiribella2013quantum} has been realised or merely simulated in the lab, it has been widely acknow\-led\-ged that a critical question is that of identifying relevant events, serving as the relata of causal relations. Such events are essentially `subsystems in spacetime' \cite{oreshkov2019time, Wechs2022, wechs2024}; in that perspective, the move from one notion of events to another is yet another case of a repartition.

In the field of (finite-dimensional) quantum information, past works adopting the top-down perspective have put forward the idea that subsystems of a system $\Om$ are best framed as corresponding to \textit{sub-C*-algebras} of the C*-algebra of operators on $\Om$ \cite{viola2001, zanardi2001, zanardi2003} (an idea that first arose, in another context, in algebraic quantum field theory \cite{halvorson2006, fewster2019}). However, these works have in general neglected an important case: the one in which subsystems correspond to \textit{non-factor} C*-algebras (the mathematical and physical meanings of which we will present in detail in a moment). As we shall argue, this is a mistake: non-factor subsystems appear naturally in physics, typically from the imposition of symmetries and superselection rules. More than that, their peculiar features often prove critical at the structural level: in each of the examples we just mentioned of research fields involving partitions, non-factor algebras have been revealed to play a pivotal role --- see e.g.\ Refs.\ \cite{Allen2017, lorenz2020} for quantum causal models, Ref.\ \cite{castroruiz2021} for quantum reference frames, or Ref.\ \cite{Ormrod2023} for the quantum switch controversy. Non-factor C*-algebras obtained from a quantum notion of causal structure also play a pivotal role in a recently proposed new interpretation of quantum theory \cite{Ormrod2024}.

Therefore, it is an important task to pin down what partitions of quantum systems are, and analyse their features, in the presence of non-factor subsystems. This is what this paper achieves. In a sense, our first and most important result might well be the realisation that, as opposed to the standard case of factorisations, even defining partitions in this general setting is a non-trivial task. More specifically, while \textit{bi}partitions (partitions into two parts) of factors admit a natural definition even when the parts are not factors (as in Refs.\ \cite{chiribella2018, zanardi2022, andreadakis2023, zanardi2023}), it is with partitions into three parts or more (which, to our knowledge, we are the first to consider) that most subtleties arise. As we shall explain, these complications are essentially due to a congenital defect of non-factor partitions, which we call `Failure Of Local Tomography' or FOLT: in such partitions, there exists genuinely global information that is not accessible even in the correlations between measurements on the individual parts.

In a nutshell, our contributions here are the following. First, we provide a self-contained presentation of the conceptual and mathematical foundations for using sub-C*-algebras (including the non-factor ones) to model subsystems in quantum theory. Second, we motivate and propose a mathematical definition of partitions of a quantum system into an arbitrary number of parts. Third, we prove some of its properties, showing in particular that \textit{centres} (the parts of C*-algebras that encapsulate their non-factorness) possess a remarkably rigid structure: more specifically, we display a theorem showing how the centre of a conjunction of subsystems can be deduced from the centres of each subsystem. Fourth, we study \textit{representations} of a partition, by which we mean ways to see them as corresponding to splittings of an underlying Hilbert space structure. We show that, using the framework of \textit{routed quantum circuits} \cite{vanrietvelde2021routed, vanrietveldePhD}, representations can be found in which an action on any individual part acts locally on a corresponding Hilbert space; but that, very surprisingly, some partitions are \textit{not fully representable}: in any representation of such partitions, there are actions on the conjunction of several parts that take a non-local form. We show that the partition of a fermionic system into local modes is an example of a non-fully representable partition.

% With care, however, it is possible to incorporate these non-trivial aspects into a consistent framework, which then turns out to display a remarkably rich structure. Drawing lessons from natural physical examples, we motivate and present a sound definition of a partition of a finite-dimensional quantum system into any number of parts. We analyse its physical interpretation and implications, and in particular the consequences of FOLT. We show that in partitions, \textit{centres} (the parts of C*-algebras that encapsulate their non-factorness) possess a remarkably rigid structure: more specifically, we display a theorem showing how the centre of a conjunction of subsystems can be deduced from the centres of each subsystem. Finally, we study \textit{representations} of a partition, by which we mean ways to see them as corresponding to splittings of an underlying Hilbert space structure. We show that, using the framework of \textit{routed quantum circuits} \cite{vanrietvelde2021routed, vanrietveldePhD}, representations can be found in which an action on any individual part acts locally on a corresponding Hilbert space; but that, very surprisingly, some partitions are \textit{not fully representable}, in the sense that some actions on the conjunction of several parts will take a non-local form in any representation of them. We show that the partition of a fermionic system into local modes is an example of a non-fully representable partition.

Given the pervasiveness of non-factor subsystems in physics, and the rich structure of partitions involving those, we expect that the general theory presented here will find much practical use in the framing and analysis of concrete situations involving different partitions of the same system. A first concrete instance, in the field of quantum causal modelling, is provided by the proof of the causal decomposability of 1D quantum cellular automata \cite{causaldecs}, which makes heavy use of the present framework and results.

The structure of this paper is as follows. In Section \ref{sec: preliminaries}, we present the conceptual foundations for framing subsystems as sub-C*-algebras and the mathematical structure of the latter, and we discuss in particular the physical status of subsystems corresponding to non-factor sub-C*-algebras and the significance of FOLT. In Section \ref{sec: 3 trajs}, we present a simple example of a tripartition, in order to build intuition as to its properties and to unveil some requirements for a sound definition of partitions in general. In Section \ref{sec: multipartitions}, we motivate and propose such a definition and study its properties, showing in particular how centres of joint parts can be deduced from centres of individual ones. In Section \ref{sec: representations}, we demonstrate how partitions at the C*-algebraic level can be represented in terms of operators over a split Hilbert space, using the framework of routed circuits. In Section \ref{sec: fermions}, we investigate the partition of a fermionic system into local modes, an example of a non-fully representable partition. We conclude in Section \ref{sec: conclusion}.

\section{Conceptual and mathematical preliminaries} \label{sec: preliminaries}

In this section we introduce, in a pedagogical way, the C*-algebraic perspective on subsystems and some of its properties. We first present the conceptual foundations of the algebraic perspective and discuss its operational meaning. Then, we introduce some of the mathematical toolkit for analysing the structure of C*-algebras. Finally, we discuss the standard case of \textit{bipartitions of factors} in the C*-algebraic framework, and illustrate it on a concrete physical scenario, superposition of trajectories.
Note that this section is written for non-familiar readers; those accustomed to seeing things through the algebraic lens can safely skip part or all of it.
% \footnote{Note that there's much to be said, and some that has been said already, about decompositional approaches in \textit{infinite dimension} --- typically, Tsirelson's problem. We will not venture there at all. In fact, our point is to stress the fact that even in finite dimension, things are not so obvious once we decide to accept subsystems corresponding to non-factor algebras.}

\subsection{(Sub)systems as (sub-)C*-algebras} \label{section: subsystems C*}

The standard framework for finite-dimensional quantum theory is that of Hilbert spaces. However, a decompositional approach is hard to formalise in this picture: typically, given a Hilbert space $\ch_\Om \cong \ch_{A_1} \otimes \ch_{A_2}$, there is no sharp way to characterise $\ch_{A_1}$ as corresponding to a substructure of $\ch_\Om$.\footnote{In particular, it is worth recalling that $\ch_{A_1}$ is not a subspace of $\ch_\Om$. Subspaces of Hilbert spaces appear in sectorisations such as $\ch_\Om = \ch_{A_1} \oplus \ch_{A_2}$, and thus correspond not to parts but to \textit{alternatives}.}

The decompositional approach fits more nicely within the algebraic picture \cite{emch1972algebraic}. The shift to the algebraic picture is given by moving from the Hilbert space $\ch_\Om$ to $\OM := \Lin(\ch_\Om)$, the space of linear operators on it. In other words, a quantum system is then to be seen as a \textit{C*-algebra}, which in finite dimension is essentially a complex vector space equipped with an inner multiplication and an adjunction. (We leave the mathematical definition and presentation of C*-algebras to the next section, in order to first stress conceptual ideas.) From this perspective, subsystems arise naturally as corresponding to \textit{sub-}C*-algebras, i.e.\ sub-vector spaces closed under multiplication and under the adjoint: typically, in the example of $\ch_\Om = \ch_{A_1} \otimes \ch_{A_2}$, one can define system $A_1$ as corresponding to $\ca_1 := \left\{ f_{A_1} \otimes \id_{A_2} \,|\, f \in \Lin(\ch_{A_1}) \right\} \subseteq \OM$, which can easily be verified to be a sub-C*-algebra. As for $A_2$, it corresponds to the symmetrically defined $\ca_2$ --- which, as we shall see in Section \ref{sec: bipartitions of factors}, arises naturally as the complement of $\ca_1$ through the use of commutants.

The idea of framing subsystems as corresponding to sub-C*-algebras of operators is of course hardly new. It is in particular one of the core ingredients of algebraic quantum field theory (AQFT) \cite{halvorson2006, fewster2019}, where algebras are associated to regions of spacetime. The same idea has also been investigated in the context of finite-dimensional quantum information, in particular in order to pin down `virtual subsystems' immune to noise \cite{viola2001, zanardi2001, zanardi2003}.\footnote{This has laid the foundation for a rich literature discussing how a \textit{preferred} partition into subsystems --- and therefore a preferred notion of locality --- can arise, e.g.\ from the spectrum of the global system's Hamiltonian \cite{cotler2017, carroll2020, adil2024, loizeau2024} or from correlation measures \cite{zanardi2022, andreadakis2023, zanardi2023} (see also Ref.\ \cite{Franzman2024}, which argues that this preferred partition should be evolving in time). Our considerations here are agnostic with respect to that question: we focus on defining what a partition is and what its features are, not on what might make a certain choice of a partition more natural than others.} We can also mention other approaches to subsystems, in which C*-algebras are replaced with Lie algebras \cite{barnum2002, barnum2003, viola2004}, convex cones of states \cite{barnum2005}, groupoids \cite{ciaglia2020}, sets of transformations in general physical theories \cite{Kraemer2017, chiribella2018, gogioso2019}, or restrictions with respect to a canonical basis \cite{arrighi2021, arrighi2022}.

In this paper, the claim that subsystems correspond to sub-C*-algebras is accepted as a starting point. Justifying that claim, e.g.\ from operational considerations, is of course a task of great importance (which has, at least partially, been undertaken in past work on the subject).\footnote{In fact, soon after this paper appeared on arXiv, a compelling derivation of the algebraic approach from causal considerations was put forward \cite{Ormrod2025}.} However, a proper study of that question would require an article of its own. We therefore leave it out of the scope of our present work, so we can focus on examining the consequences of upholding it, especially in the case of multipartitions.

The operational interpretation of associating a sub-C*-algebra $\ca \subseteq \OM$ to a subsystem $A$ of $\Om$ is that $\ca$ contains the \correction{`actions'} on $\Om$ that only affect $A$. Depending on the kind of physical data that one takes as \correction{primary}, \correction{one might mean different things by the word `actions' in the previous interpretation.} If one emphasises the role of observables (as is typical of AQFT, for instance), the interpretation is that observables on $A$ are the Hermitian elements of $\ca$. If one emphasises unitary evolutions, it is that evolutions of $A$ are the unitary elements of $\ca$. If one emphasises quantum transformations described as completely positive maps (as is typical in quantum-informational approaches), it is that the transformations of $A$ are maps that admit a Kraus decomposition whose Kraus operators are in $\ca$. \correction{Finally, if one emphasises predicates encoded as POVM elements, it is that predicates on $A$ are POVM elements belonging to $\ca$.} Here, we shall \correction{remain agnostic with respect to which interpretation should be favoured.}\footnote{\correction{Studying the relationship between these interpretations, and their potential equivalence,} is a subtle issue, which relates to the previously mentioned problem of operationally justifying the use of sub-C*-algebras to describe subsystems. We also leave this important question for future work.}

This interpretation can be illustrated by resorting to agents. To $A$, we can associate an agent Alice that `only has access to $A$'; $\ca$ then contains the observables (resp.\ unitary evolutions or Kraus operators of quantum transformations) that Alice can measure (resp.\ apply) on $\Om$ through that access. We will make much use of this picture since it greatly helps intuition. Note, however, that it does not commit us to upholding agents as a fundamental ingredient of physics; it can very well be taken as nothing more than an illustrative tool.

\correction{Finally, an important consequence of the adoption of the algebraic perspective is that it opens the door to a new case of subsystems: those corresponding to \textit{non-factor algebras}. This is what we will discuss in the next subsection; a physical discussion of these subsystems, as arising from conservation rules or partial classicality, will be provided in Section \ref{sec: non-factor subsystems}.}

\subsection{A classification of C*-algebras}

\subsubsection{Wedderburn-Artin}

Let us recall the definition of a C*-algebra in finite dimension. Note that we could equivalently talk about von Neumann algebras, which in finite \correction{dimensions} are the same notion.

\begin{definition}
    \correction{A (finite-dimensional) \emph{complex algebra} $\OM$ is a finite-dimensional vector space equipped with a binary multiplication $\cdot: \OM \times \OM \to \OM$ satisfying
    \begin{itemize}
        \item $ \forall x, y, z \in \OM, x \cdot (y \cdot z) = (x \cdot y) \cdot z \, ;$
        \item $ \forall x, y, z \in \OM, x \cdot (y + z) = (x \cdot y) + (x \cdot z) \, ;$
        \item $ \forall x, y, z \in \OM, (x + y) \cdot z = (x \cdot z) + (y \cdot z) \, ;$
        \item $ \forall \la \in \mathbb{C}, \forall x, y \in \OM, \la (x \cdot y) = (\la x) \cdot y = x \cdot (\la y) \, ;$
        \item $ \exists \id, \forall x \in \OM, x \cdot \id = \id \cdot x = x \, .$
    \end{itemize}}
\end{definition}

\begin{definition}[*-algebras]
    A (finite-dimensional) \emph{*-algebra} $\OM$ is a finite-dimensional complex algebra, equipped with an involution $\dag$ satisfying:
    \begin{itemize}
        \item $\forall x, y \in \OM, \, (x + y)^\dag = x^\dag + y^\dag$;
        \item $\forall x, y \in \OM, \, (x y)^\dag = y^\dag x^\dag$;
        \item $\forall x \in \OM, \forall \la \in \mathbb{C}, (\la x)^\dag = \la^* x^\dag$.
    \end{itemize}

    A \correction{(unital)}\footnote{\correction{Throughout this paper, we will always take our definition of a subalgebra to impose unitality: the ambient algebra's identity should be an element of the subalgebra. For conciseness, we will not repeat this every time.}} \textit{sub-*-algebra} of $\OM$ \correction{is a  sub-vector space $\ca \subseteq \OM$ containing the identity, and closed under the multiplication and dagger: $\forall x, y \in \ca, xy \in \ca$ and $x^\dag \in \ca$.}
\end{definition}

\begin{definition}[*-algebra homomorphisms]
    A linear map $h$ between two *-algebras is a \emph{homomorphism of *-algebras} if it preserves multiplication ($h(xy) = h(x) h(y)$) and the dagger ($h(x^\dag) = h(x)^\dag$). If $h$ is bijective, then it is an \emph{isomorphism of *-algebras}.
\end{definition}

    \begin{example}[$\Lin(\ch)$ as a *-algebra]
    \correction{Given a finite-dimensional Hilbert space $\ch$, take $\Lin(\ch)$ to be the complex vector space of linear maps from $\ch$ to $\ch$. Taking $\cdot$ to be the composition of endomorphisms and defining $\dagger$ to be the standard dagger, i.e.
    \be \forall f \in \Lin(\ch), f^\dag \textrm{ is the unique element satisfying } \forall u, v \in \ch, \langle u, f(v) \rangle = \langle f^\dag(u), v \rangle \, , \ee 
    makes it into a *-algebra.}
\end{example}

\begin{definition}[C*-algebras]
    In finite dimension, a \emph{C*-algebra} $\OM$ is a *-algebra that is isomorphic (for some $\ch$) to a sub-*-algebra of the *-algebra $\Lin(\ch)$ \correction{defined above}.

    Any sub-*-algebra of $\OM$ is then a C*-algebra, and can thus be called a sub-C*-algebra of $\OM$.\footnote{The reader more used to infinite dimension will notice that we picked a definition of C*-algebras that makes no reference to the requirement of featuring a norm that plays well with the dagger and yields a Banach space structure. We did this deliberately, in order to stress the point that in finite dimension, and as far as one is interested in \textit{sub}-C*-algebras, it is not necessary to care about norms and the topologies they induce. This is essentially due to the fact that in finite dimension, all norms over any complex vector space induce equivalent topologies, and all complex vector spaces are complete.} Similarly, *-algebra homomorphisms and isomorphisms between C*-algebras are called C* homomorphisms and isomorphisms.
\end{definition}

A natural example of a finite-dimensional C*-algebra is that of an algebra $\OM = \Lin(\ch_\Om)$ of operators over a finite-dimensional Hilbert space $\ch_\Om$. C*-algebras that are isomorphic to such algebras of operators are called \textit{factors}. Importantly, not all finite-dimensional C*-algebras are factors; in general, they are direct sums of factors. The idea that such \textit{non-factor} C*-algebras should be taken seriously is a central tenet of this work.

\begin{theorem}[Wedderburn-Artin \cite{farenick}] \label{th: WA}
    Given a finite-dimensional C*-algebra $\OM$, there exists a list of finite-dimensional Hilbert spaces $(\ch_\Om^k)_{k \in K}$, indexed by a finite set $K$, such that 
    \be \label{eq: WA} \OM \cong \bigoplus_{k \in K} \Lin \left(\ch_\Om^k \right) \, .\ee
\end{theorem}

This can be restated in an even more concrete way, which allows us to see $\OM$ as an algebra of \textit{block-diagonal} operators over a single Hilbert space with a preferred sectorisation.

\begin{corollary}
    Writing $\ch_\Om := \bigoplus_{k \in K} \ch_\Om^k$ in the previous theorem,
    \be \label{eq: blockdiag rep}\OM \cong \left\{ f \in \Lin(\ch_\Om) \,\,\middle|\,\, \forall k, f \left(\ch_\Om^k\right) \subseteq \ch_\Om^k \right\} \, . \ee
\end{corollary}

In this representation, $\OM$ becomes the set of operators on $\ch_\Om := \bigoplus_{k \in K} \ch_\Om^k$ that map each of the $\ch_\Om^k$'s within itself, i.e.\ the ones whose matrices in an adapted basis are block-diagonal.

\correction{Let us end with a concrete example to build intuition.

\begin{example}[Sub-C*-algebras of $\Lin (\CC^3)$] \label{example: C3}
    Up to changes of basis, $\Lin(\CC^3)$ admits five sub-C*-algebras:
    \begin{enumerate}
        \item The trivial algebra $\ca_1 = \CC \cdot \id = \left\{ \begin{pmatrix}
            \al & 0 & 0 \\
            0 & \al & 0 \\
            0 & 0 & \al
        \end{pmatrix} \,\, \middle| \,\, \al \in \CC \right\}$;
        \item $\ca_2 = \left\{ \begin{pmatrix}
            \al & 0 & 0 \\
            0 & \bet & 0 \\
            0 & 0 & \ga
        \end{pmatrix} \,\, \middle| \,\, \al, \bet, \ga \in \CC \right\}$;
        \item $\ca_3 = \left\{ \begin{pmatrix}
            \al & 0 & 0 \\
            0 & \bet & 0 \\
            0 & 0 & \bet
        \end{pmatrix} \,\, \middle| \,\, \al, \bet \in \CC \right\}$;
        \item $\ca_4 = \left\{ \begin{pmatrix}
            \al & 0 & 0 \\
            0 & \bet & \delta \\
            0 & \epsilon & \ga
        \end{pmatrix} \,\, \middle| \,\, \al, \bet, \ga, \delta, \epsilon \in \CC \right\}$;
        \item the full algebra, $\ca_5 = \Lin(\CC^3)$.
    \end{enumerate}
$\ca_1$ and $\ca_5$ are factors, while $\ca_2$, $\ca_3$, and $\ca_4$ are not, being respectively isomorphic to $\Lin(\CC) \oplus \Lin(\CC) \oplus \Lin(\CC)$, $\Lin(\CC) \oplus \Lin(\CC)$, and $\Lin(\CC) \oplus \Lin(\CC^2)$.
\end{example}
}

\subsubsection{A representational tool: routes} \label{sec: routes}

This block-diagonalness finds a natural formalisation using \textit{routes} \cite{vanrietvelde2021routed}, which will be an important tool \correction{throughout} this paper. Let us provide a basic introduction. The idea is that if we have two sectorised Hilbert spaces $\ch_A = \bigoplus_{k \in K} \ch_A^k$ and $\ch_B = \bigoplus_{l \in L} \ch_B^l$ and a relation\footnote{Relations are generalisations of functions. A relation maps each element of its domain to a (possibly empty) set of elements of its output domain.} $\eta: K \to L$ --- which we can equivalently represent as a Boolean matrix $(\eta_k^l)_{k \in K, l \in L}$ ---, then we can interpret $\eta$ as encoding sectorial constraints on how linear maps $\ch_A \to \ch_B$ are allowed to connect input sectors to output sectors, with its zeroes stipulating the disallowed connections. The set of linear maps satisfying these constraints is
\be \label{eq: following a route} \Lin_\eta(\ch_{A}, \ch_B) := \left\{ f \in \Lin(\ch_{A}, \ch_B) \,\, \middle| \,\, \forall k \in K, \,\, f \left( \ch_{A^{k}}\right) \subseteq \bigoplus_{l \in \eta(k)} \ch_{B^l} \right\} \, . \ee

A relation $\eta$ seen in this way is called a route, and we say that $f$ follows the route $\eta$ if it is in $\Lin_\eta(\ch_{A}, \ch_B)$. If $\ch_A = \ch_B$\footnote{By this we mean that their sectorisations are also the same, and are labelled by the same set.}, we simply write $\Lin_\eta(\ch_A) = \Lin_\eta(\ch_{A}, \ch_A)$.  Note that $\Lin_\eta(\ch_A)$ forms a C*-algebra if and only if $\eta$ is a partial equivalence relation.

Using the identity relation $\delta$ (corresponding to the Kronecker-delta Boolean matrix $(\delta_k^l)_{k,l}$), (\ref{eq: blockdiag rep}) can be rewritten as
\be \label{eq: routed rep}\OM \cong  \Lin_\delta(\ch_\Om) \, . \ee
This means that one can make use of the diagrammatic representation of \textit{routed circuits} \cite{vanrietvelde2021routed} to deal with non-factor C*-algebras. In this representation, sectorised Hilbert spaces are represented by wires whose names feature superscripts; for instance, $\ch_A = \bigoplus_{k \in K} \ch_A^k$ will be denoted $A^k$.\footnote{The $k$ here should be thought of as a free variable, denoting the existence of different possible values for $k$, and not a single specific value of it. This is similar to the Einstein notation convention for linear tensors, in which $g_{\mu \nu}$ refers to a whole tensor and not to one of its coefficients.} A map $f: A^k \to B^l$ that we ask to follow a given route $\eta_k^l$ will then be represented as
\be \begin{tikzpicture}
	\begin{pgfonlayer}{nodelayer}
		\node [style=small box] (0) at (0, 0) {$f$};
		\node [style=none] (1) at (0, -2) {};
		\node [style=none] (2) at (0, -0.5) {};
		\node [style=none] (3) at (0, 0.5) {};
		\node [style=left label] (4) at (0, 2) {$B^l$};
		\node [style=left label] (5) at (0, -2) {$A^k$};
		\node [style=none] (6) at (-2, 0) {$\eta_k^l$};
		\node [style=none] (7) at (0, 2) {};
	\end{pgfonlayer}
	\begin{pgfonlayer}{edgelayer}
		\draw (2.center) to (1.center);
		\draw (3.center) to (7.center);
	\end{pgfonlayer}
\end{tikzpicture} \,\, . \ee

Furthermore, in the typical case in which the routes feature Kronecker deltas between indices, these can be represented nicely through the graphical shortcut of `index-matching': a delta-route between $k$ and $k'$ gets replaced with a repetition of the $k$ index, implicitly stipulating that the $k$-values have to be equal:
\be \begin{tikzpicture}
	\begin{pgfonlayer}{nodelayer}
		\node [style=none] (0) at (0, 2) {};
		\node [style=small box] (1) at (0, 0) {$f$};
		\node [style=none] (2) at (0, -2) {};
		\node [style=none] (3) at (0, -0.5) {};
		\node [style=none] (4) at (0, 0.5) {};
		\node [style=left label] (5) at (0, 2) {$B^{k}$};
		\node [style=left label] (6) at (0, -2) {$A^k$};
		% \node [style=none] (7) at (-2.75, 1.5) {$\mathlarger{\mathlarger{\mathlarger{\mathlarger{\mathlarger{`}}}}}$};
		% \node [style=none] (8) at (2.25, 1.5) {$\mathlarger{\mathlarger{\mathlarger{\mathlarger{\mathlarger{\textrm{'}}}}}}$};
	\end{pgfonlayer}
	\begin{pgfonlayer}{edgelayer}
		\draw (3.center) to (2.center);
		\draw (4.center) to (0.center);
	\end{pgfonlayer}
\end{tikzpicture}  \quad := \quad  \begin{tikzpicture}
	\begin{pgfonlayer}{nodelayer}
		\node [style=none] (7) at (0, 2) {};
		\node [style=small box] (0) at (0, 0) {$f$};
		\node [style=none] (1) at (0, -2) {};
		\node [style=none] (2) at (0, -0.5) {};
		\node [style=none] (3) at (0, 0.5) {};
		\node [style=left label] (4) at (0, 2) {$B^{k'}$};
		\node [style=left label] (5) at (0, -2) {$A^k$};
		\node [style=none] (6) at (-2, 0) {$\delta_k^{k'}$};
	\end{pgfonlayer}
	\begin{pgfonlayer}{edgelayer}
		\draw (2.center) to (1.center);
		\draw (3.center) to (7.center);
	\end{pgfonlayer}
\end{tikzpicture}\, . \ee
This allows us to rewrite (\ref{eq: routed rep}) in a nice diagrammatic way,
\be \label{eq: diag routed rep}\OM \quad \cong \quad   \left\{ \quad  \begin{tikzpicture}
	\begin{pgfonlayer}{nodelayer}
		\node [style=none] (0) at (0.5, 2) {};
		\node [style=small box] (1) at (0.5, 0) {$f$};
		\node [style=none] (2) at (0.5, -2) {};
		\node [style=none] (3) at (0.5, -0.5) {};
		\node [style=none] (4) at (0.5, 0.5) {};
		\node [style=left label] (5) at (0.5, 2) {$\Om^{k}$};
		\node [style=left label] (6) at (0.5, -2) {$\Om^k$};
	\end{pgfonlayer}
	\begin{pgfonlayer}{edgelayer}
		\draw (3.center) to (2.center);
		\draw (4.center) to (0.center);
	\end{pgfonlayer}
\end{tikzpicture}  \quad \right\} \, , \ee
\correction{where the repetition of the $k$'s indicates that the $f$'s in this set should be taken within $\Lin_\delta(\ch_\Om)$, and not $\Lin(\ch_\Om)$.} This diagrammatic notation, which might look like overkill at this point, will prove valuable in more involved cases.

\subsubsection{Centres}
It is important for our purposes to emphasise the critical aspects of C*-algebras, which yield crucial structural handles on them. First among these is the notion of \textit{centre}. The centre of $\OM$, $\cz(\OM)$, is the set of elements of $\OM$ that commute with everything in it:
\be \cz(\OM) := \left\{ f \in \OM \,\,\middle|\,\, \forall g \in \OM, fg = gf \right\} \, .\ee

The centre plays a key role in characterising an algebra. It is easy to see that it itself forms a commutative C*-algebra, i.e.\ one whose elements all commute pairwise. This allows us to characterise it very nicely using the following slogan.

\begin{slogan}
    A commutative C*-algebra is just a bunch of orthogonal projectors.
\end{slogan}

Let us make this more formal. We say that $\pi \in \OM$ is an orthogonal projector if $\pi \pi = \pi$ and $\pi^\dag = \pi$, and that the projectors $\pi^1$ and $\pi^2$ are orthogonal if $\pi^1 \pi^2 = 0$.\footnote{\correction{Unfortunately, this standard terminology is therefore slightly confusing: the term `orthogonality' refers both to orthogonality of one projector (self-adjointness), and orthogonality between two projectors (multiplying to zero). To us, `a family of orthogonal projectors' means both at once: each projector is self-adjoint, and any two different ones multiply to zero.}} Our slogan then takes its roots in the following result, proven in Appendix \ref{app: proof commutative algs}.

\begin{theorem}[Commutative C*-algebras] \label{th: commutative algs}
    Let $\cz$ be a commutative (finite-dimensional) C*-algebra. Then there exists a unique (finite) set of orthogonal projectors $\{\pi^k\}_{k \in K}$, pairwise orthogonal and non-null, forming a basis of $\cz$. 
\end{theorem}

We call this unique set $\Atproj(\cz)$, the set of $\cz$'s \textit{atomic projectors}. They are the second important structural handle that we will be using.

\correction{\begin{example}[Centres and atomic projectors of sub-C*-algebras of $\Lin(\CC^3)$]
    Going back to our Example \ref{example: C3} of sub-C*-algebras of $\Lin(\CC^3)$, we have 
    \begin{itemize}
        \item     $\cz(\ca_1) = \cz (\ca_5) = \ca_1$, with atomic projectors $\{\id\}$; 
    \item $\cz(\ca_3) = \cz (\ca_4) = \ca_3$, with atomic projectors $\{\ketbra{0}{0}, \ketbra{1}{1} + \ketbra{2}{2}\}$;
    \item $\cz(\ca_2) = \ca_2$, with atomic projectors $\{\ketbra{0}{0}, \ketbra{1}{1},  \ketbra{2}{2}\}$.
    \end{itemize}
\end{example}
}
Atomic projectors have a direct connection to the direct sum structure displayed in Theorem \ref{th: WA}. Indeed, if, for a C*-algebra $\OM$, we take $\{\pi^k\}_{k \in K}$ to be the atomic projectors of its centre $\cz(\OM)$, and consider the algebras
\be \forall k \in K, \quad  \pi^k \OM := \{ \pi^k f \,\,|\,\, f \in \OM \} \, , \ee
then it is easy to see that these form (non-unital) factor sub-C*-algebras of $\OM$, and that
\be \label{eq:centraldec} \OM \cong \bigoplus_{k \in K} \pi^k \OM \, . \ee
As is evident from the way the $\pi^k$'s are defined in Appendix \ref{app: proof commutative algs}, the $\pi^k \OM$'s are then precisely $\OM$'s blocks, i.e., the direct sum in (\ref{eq:centraldec}) corresponds to that in (\ref{eq: WA}).

Thus, the atomic projectors of an algebra's centre precisely designate its blocks. In particular, a C*-algebra is a factor if and only if its centre is trivial, i.e.\ only contains multiples of the identity.

\subsection{Bipartitions of factors} \label{sec: bipartitions of factors}

\subsubsection{Definition}
We can now apply our newfound tools to a first discussion of bipartitions. \correction{We will start by assuming that the system to be partitioned is a factor, as this is the standard case, studied elsewhere in the literature. A major outcome of this paper will be the eventual extension of this definition to the case in which the overall system is not a factor.}

Taking a system $\Om$ corresponding to a factor C*-algebra $\OM$, our question is: what would it mean to partition it into two complementary parts $A_1$ and $A_2$? As we already argued, these should first correspond to two sub-C*-algebras $\ca_1$ and $\ca_2$ of $\OM$. In addition, \correction{it is commonly assumed \cite{viola2001, zanardi2001, zanardi2003} that complementarity should be pinned down by employing} \textit{commutants}.

Given $\ca \subseteq \OM$, $\ca$'s commutant within $\OM$ is the set $\ca'$ of elements of $\OM$ that commute with all elements of $\ca$,
\be \ca' := \{f \in \OM \,\,|\,\, \forall a \in \ca, fa = af \} \, . \ee
It is natural, from any of the different conceptual perspectives on subsystems enunciated earlier, to argue that the algebras corresponding to the two subsystems in a bipartition should be each other's commutants. For instance, in the perspective where subsystems are defined by the observables on them, it is natural to say that the subsystem complementary to $A_1$ corresponds to the observables that commute with the observables on $A_1$; and a similar reasoning can be made in a perspective that emphasises the role of unitary evolutions. This leads us to the following definition.

\begin{definition}[Bipartitions of factors] \label{def: bipart of factors}
    Let $\OM$ be a factor. We say that a pair of two sub-C*-algebras $(\ca_1, \ca_2)$ of it forms a \emph{bipartition},\footnote{\correction{Of course, partitions as defined here should not be confused with set-theoretic partitions, i.e.\ decompositions of a set into a disjoint union of subsets. Set-theoretic partitions would make little sense for our goals.}} denoted as $(\ca_1, \ca_2) \vdash \OM$, if
    \be \ca_2 = \ca_1' \, . \ee
\end{definition}

Note that the symmetry of this definition relies on the important fact that any sub-C*-algebra $\ca$ of a factor $\OM$ is \textit{closed under the bicommutant}, i.e.
\be \ca'' = \ca \, . \ee
This ensures that `the complementary of the complementary of $A$ is indeed $A$.' This will not be the case for $\OM$ non-factor. Note that this is also a trademark of finite dimension.\footnote{This is a direct consequence of von Neumann's bicommutant theorem \cite{averson}, which states that in  a factor, any subalgebra's bicommutant is equal to its closure in the weak and strong topologies; in finite dimension, every sub-C*-algebra is closed with respect to any norm, and thus is equal to its bicommutant.}

A basic and important fact about bipartitions of factors is that they share a common centre.

\begin{proposition} \label{prop: common centre}
    Let $(\ca_1, \ca_2) \vdash \OM$ be a bipartition of a factor. Then $\cz(\ca_1) = \cz(\ca_2)$.
\end{proposition}
\begin{proof}
    $\cz(\ca_1) := \ca_1 \cap \ca_1' = \ca_1 \cap \ca_2 = \ca_2' \cap \ca_2 =: \cz(\ca_2)$.
\end{proof}
In particular, $\ca_1$ is a factor if and only if $\ca_2$ is a factor; in that case, we call the partition a \textit{factorisation}. 

\correction{\begin{example}[Bipartitions of $\Lin(\CC^3)$] Going back to Example \ref{example: C3}, in $\Lin(\CC^3)$ we have $\ca_1' = \ca_5$, $\ca_3' = \ca_4$, $\ca_2' = \ca_2$. Therefore, up to change of basis and swapping the parts, $\Lin(\CC^3)$ admits three possible bipartitions: $(\ca_1, \ca_5)$ (the trivial factorisation), $(\ca_3, \ca_4)$, and $(\ca_2, \ca_2)$. (We will come back later to this seemingly strange case of `the two parts being the same algebra' and why it makes sense.)
    
\end{example}}

\subsubsection{Representation} \label{sec: reps of bipartitions}

We now turn to finding a representation theorem for such bipartitions. By this, we mean finding a way of seeing $\OM$ as a concrete algebra of operators on a Hilbert space, such that this Hilbert space itself is `made of parts' corresponding to $A_1$ and $A_2$.\footnote{This use of the word `representation' should of course not be confused with the study of representation theory.}

\begin{theorem}[Representations of bipartitions of a factor\footnote{Theorem \ref{th: standard rep bipartitions} can be obtained as a special case of the more general Theorem \ref{th: representation} which we will prove later on.}] \label{th: standard rep bipartitions}
     Let $(\ca_1, \ca_2) \vdash \OM$ be a bipartition of a factor. Then there exist a finite set $K$, two lists of Hilbert spaces $(\ch_{A_1^k})_{k \in K}$ and $(\ch_{A_2^k})_{k \in K}$, and a C*-isomorphism
   \be  \iota \colon \OM \to \Lin \left( \bigoplus_{k \in K} \ch_{A_1^k} \otimes \ch_{A_2^k} \right) \ee
   such that

   \begin{subequations}
       \be \iota \left(\ca_1 \right) = \left\{ \bigoplus_{k \in K} f^k \otimes \id_{{A_2^k}} \,\, \middle| \,\, \forall k, f^k \in \Lin(\ch_{A_1^k}) \right\} \, ; \ee
       \be \iota \left(\ca_2 \right) = \left\{ \bigoplus_{k \in K} \id_{{A_1^k}} \otimes g^k \,\, \middle| \,\, \forall k, g^k \in \Lin(\ch_{A_2^k}) \right\} \, . \ee 
   \end{subequations}
\end{theorem}

We see that we find a representation of $\OM$ as a space of operators on 
\be \ch_\Om = \bigoplus_{k \in K} \ch_{A_1^k} \otimes \ch_{A_2^k} \, . \ee
This structure of `sectorisation then factorisation' is typical of partitions. We can see that the general recipe for getting two complementary systems is to first slice up a Hilbert space into sectors, then to factorise each sector in two. Operators on the first (second) part are then those that are acting on the first (second) factor of each sector.

Note that the sectors here are precisely those designated by the atomic projectors of the common centre (by Proposition \ref{prop: common centre}) $\cz(\ca_1) = \cz(\ca_2)$. In particular, in the case of a factorisation, that common centre is trivial and thus so is the direct sum, and we recover a standard representation as a tensor product.
\correction{
\begin{example}[Representation of a bipartition of $\Lin(\CC^3)$]
    Going back to our Example \ref{example: C3}, take the bipartition $(\ca_3, \ca_4) \vdash \Lin(\CC^3)$. Then the representation introduced in Theorem \ref{th: standard rep bipartitions} is given by $K= \{0,1\}$, $\ch_{A_3^0} = \ch_{A_3^1} = \ch_{A_4^0} = \CC$, $\ch_{A_4^1} = \CC^2$. Writing an orthonormal basis of $\ch_{A_3}$ as $\{\ket{0} \in \ch_{A_3^0}, \ket{1} \in \ch_{A_3^1}\}$, and one of $\ch_{A_4}$ as $\{\ket{0} \in \ch_{A_4^0}, \ket{1} \in \ch_{A_4^1}, \ket{2} \in \ch_{A_4^1}\}$, $\iota$ is given by $\iota(f) = U f U^\dag$ where $U$ is the isometry defined by
    \begin{subequations}
        \be U \ket{0} = \ket{0} \ket{0} \, ,\ee
        \be U \ket{1} = \ket{1} \ket{1} \, ,\ee
        \be U \ket{2} = \ket{1} \ket{2} \, .\ee
    \end{subequations}
\end{example}}

In the general case where the direct sum is non-trivial, one practical issue with this representation is that it blends together tensor products and direct sums, which is impractical to work with. A trick around this is to rather see the operators as acting on an extended space
\be \ch_\Om^\ext := \underbrace{\left( \bigoplus_{k \in K} \ch_{A_1^k} \right)}_{=: \ch_{A_1}} \otimes \underbrace{\left( \bigoplus_{l \in K} \ch_{A_2^l} \right)}_{=: \ch_{A_2}} \, , \ee
but to require them to be null on the $\ch_{A_1^k} \otimes \ch_{A_2^l}$'s such that $k \neq l$. We denote
\be \label{eq: sec corrs} \tilde{\bbpi} := \sum_k \bbpi_{A_1}^{k} \otimes \bbpi_{A_2}^{k} \in \Lin(\ch_\Om^\ext) \, , \ee
where $\bbpi_{A_1}^k$ (resp.\ $\bbpi_{A_2}^k$) is the projector of $\ch_{A_1}$ onto $\ch_{A_1^k}$ (resp.\ of $\ch_{A_2}$ onto $\ch_{A_2^k}$),\footnote{We use the `bold pi' notation $\bbpi$ in order not to confuse these projectors with the atomic projectors of algebras, to which they are closely related but different. For instance, an atomic projector $\pi_1^{k_1}$ of $\cz(\ca_1)$ and the corresponding $\bbpi_{A_1}^{k_1}$ are related in Corollary \ref{cor: routed rep bipartitions} by $\iota(\pi_1^{k_1}) = (\bbpi_{A_1}^{k_1} \otimes \id_{A_2}) \tilde{\bbpi}$.} and we write $\Delta : K \times K \to K \times K$ as the route given by the Boolean matrix 
\be \Delta_{k_1 k_2}^{k_1' k_2'} := \delta_{k_1 k_2} \delta^{k_1' k_2'} \, ,\ee
which forces its two input values to be equal, and its two output values to be equal, but otherwise imposes no relationship between the former and the latter. Using the routed notation of (\ref{eq: following a route}), our trick takes the following form.

\begin{corollary} \label{cor: routed rep bipartitions}
    With the assumptions of the previous theorem and the notations above, there exists a C*-isomorphism
    \be  \label{eq: routed rep bipartitions Omega} \iota \colon \quad \OM \to \Lin_\Delta(\ch_\Om^\ext) \ee   
   such that 
   \begin{subequations}
       \be \label{eq: routed rep bipartitions A_1}\iota \left(\ca_1 \right) = \left\{ (f \otimes \id_{A_2}) \, \tilde{\bbpi} \,\, \middle| \,\,  f \in \Lin_\delta(\ch_{A_1}) \right\} \, ; \ee
       \be \label{eq: routed rep bipartitions A_2}\iota \left(\ca_2 \right) = \left\{ (\id_{A_1} \otimes g) \,  \tilde{\bbpi} \,\, \middle| \,\,  g \in \Lin_\delta(\ch_{A_2}) \right\} \, . \ee 
   \end{subequations}
\end{corollary}

This allows us to depict the Hilbert space splittings in terms of tensor products only, and the $\ca_n$'s as acting on either factor of this tensor product, but at the cost of introducing \textit{sectorial constraints} encoded by the $\delta$ routes, and \textit{sectorial correlations} encoded by the projector (\ref{eq: sec corrs}), or equivalently by the route $\Delta$.

Using the notations introduced in Section \ref{sec: routes}, (\ref{eq: routed rep bipartitions Omega}) takes the form
\be \iota \colon \quad \OM \quad \to \quad \left\{ \quad \begin{tikzpicture}
	\begin{pgfonlayer}{nodelayer}
		\node [style=none] (0) at (-1, 2) {};
		\node [style=large box] (1) at (0, 0) {$f$};
		\node [style=none] (2) at (-1, -2) {};
		\node [style=none] (3) at (-1, -0.5) {};
		\node [style=none] (4) at (-1, 0.5) {};
		\node [style=left label] (5) at (-1, 2) {$A_1^{k'}$};
		\node [style=none] (7) at (1, 2) {};
		\node [style=none] (8) at (1, -2) {};
		\node [style=none] (9) at (1, -0.5) {};
		\node [style=none] (10) at (1, 0.5) {};
		\node [style=right label] (11) at (1, 2) {$A_2^{k'}$};
		\node [style=left label] (12) at (-1, -2) {$A_1^{k}$};
		\node [style=right label] (13) at (1, -2) {$A_2^{k}$};
	\end{pgfonlayer}
	\begin{pgfonlayer}{edgelayer}
		\draw (3.center) to (2.center);
		\draw (4.center) to (0.center);
		\draw (9.center) to (8.center);
		\draw (10.center) to (7.center);
	\end{pgfonlayer}
\end{tikzpicture} \quad \right\} \, ,\ee
and (\ref{eq: routed rep bipartitions A_1}) becomes\footnote{A close look at this equation reveals that the index-matching notation led to some ambiguity about which maps follow which routes; see Section 6 in Ref.\ \cite{vanrietvelde2021routed} for more details. We will overlook this ambiguity here and trust pernickety readers to parse what the routes should be exactly.}
\be \label{eq: routed rep bipartitions diag A_1} \iota \left(\ca_1 \right) = \left\{ \quad \begin{tikzpicture}
	\begin{pgfonlayer}{nodelayer}
		\node [style=none] (0) at (-1, 3.5) {};
		\node [style=none] (2) at (-1, -0.75) {};
		\node [style=none] (3) at (-1, 1) {};
		\node [style=none] (4) at (-1, 2) {};
		\node [style=left label] (5) at (-1, 3.5) {$A_1^k$};
		\node [style=none] (7) at (1, -3) {};
		\node [style=none] (8) at (1, 3.5) {};
		\node [style=right label] (10) at (1, 3.5) {$A_2^k$};
		\node [style=left label] (11) at (-1, 0) {$A_1^{k}$};
		\node [style=small box] (13) at (-1, 1.5) {$f$};
		\node [style=large box] (14) at (0, -1.25) {$\tilde{\bbpi}$};
		\node [style=none] (15) at (-1, -3) {};
		\node [style=none] (16) at (-1, -1.75) {};
		\node [style=left label] (17) at (-1, -3) {$A_1^{k}$};
		\node [style=none] (18) at (1, -3) {};
		\node [style=right label] (20) at (1, -3) {$A_2^{k}$};
	\end{pgfonlayer}
	\begin{pgfonlayer}{edgelayer}
		\draw (3.center) to (2.center);
		\draw (4.center) to (0.center);
		\draw (8.center) to (7.center);
		\draw (16.center) to (15.center);
	\end{pgfonlayer}
\end{tikzpicture} \quad \right\} \, , \ee
with (\ref{eq: routed rep bipartitions A_2}) of a symmetric form.

One can note that $\tilde{\bbpi}$'s only role in (\ref{eq: routed rep bipartitions diag A_1}) is to enforce sectorial correlations --- or in broad words, to force $A_1$'s and $A_2$'s indices to be equal; in that sense, it is redundant with the route stipulated by repetition of the $k$ indices. In order to alleviate notations, we will thus drop it and take the convention that when routes on a diagram enforce sectorial correlations between otherwise unconnected wires, this diagram should be understood as featuring an implicit pre-composition with the projector enforcing these correlations (here, $\tilde{\bbpi}$). This allows us to rewrite (\ref{eq: routed rep bipartitions diag A_1}) in the form
\be \label{eq: routed rep bipartitions diag A_1 no pi} \iota \left(\ca_1 \right) = \left\{ \quad \begin{tikzpicture}
	\begin{pgfonlayer}{nodelayer}
		\node [style=none] (0) at (-1, 2) {};
		\node [style=none] (2) at (-1, -0.5) {};
		\node [style=none] (3) at (-1, 0.5) {};
		\node [style=left label] (4) at (-1, 2) {$A_1^k$};
		\node [style=none] (5) at (1, -2) {};
		\node [style=none] (6) at (1, 2) {};
		\node [style=right label] (7) at (1, 2) {$A_2^k$};
		\node [style=left label] (8) at (-1, -2) {$A_1^{k}$};
		\node [style=small box] (9) at (-1, 0) {$f$};
		\node [style=none] (14) at (1, -2) {};
		\node [style=right label] (15) at (1, -2) {$A_2^{k}$};
		\node [style=none] (16) at (-1, -2) {};
	\end{pgfonlayer}
	\begin{pgfonlayer}{edgelayer}
		\draw (3.center) to (0.center);
		\draw (6.center) to (5.center);
		\draw (2.center) to (16.center);
	\end{pgfonlayer}
\end{tikzpicture} \quad \right\} \, . \ee

\subsubsection{FOLT} \label{sec: FOLT}
It is important to discuss a seemingly disturbing feature, which will be central to what follows: if $\ca_1$ (and therefore also $\ca_2$) is not a factor, then $\ca_1$ and $\ca_2$ taken together do \textit{not} span $\OM$: denoting, for any $\ca, \cb \subseteq \OM$, $\ca \vee \cb$ to be the closure of $\ca \cup \cb$ under C*-algebraic operations (dagger, multiplication and addition), we have\footnote{\correction{Here we use von Neumann's bicommutant theorem \cite{averson}: in a finite-dimensional factor, C*-algebraic closure is the same thing as closure under the bicommutant. We then use the identity $(F \cup G)' = F' \cap G'$, which is direct to prove.}}
\be \label{eq: FOLT in bipartitions} \ca_1 \vee \ca_2 \correction{ = (\ca_1 \cup \ca_2)'' = (\ca_1' \cap \ca_2')' = (\ca_1' \cap \ca_1)'} = \cz(\ca_1)' = \cz(\ca_2)' \subsetneq \OM \, . \ee

In other words, $\ca_1$ and $\ca_2$ taken together are unable to recover the non-block diagonal operators in the sectorisation given by their common centre. This disturbing fact that `$\OM$ is more than the sum of its parts'\footnote{Note that this goes beyond the well-known idea that in quantum theory, entanglement means that `the whole is more than the sum of its parts.' Indeed, in the standard case of factorisations, the possibility for entanglement does not prevent the fact that one has $\ca_1 \vee \ca_2 = \OM$. Or, to put it more technically, quantum theory with tensor products, while featuring entanglement, still satisfies the principle of local tomography.} when the parts are allowed to be non-factors will be important to keep in mind; we will call it `Failure Of Local Tomography', or FOLT.\footnote{The operational meaning of FOLT, its fundamental or merely circumstantial status, and its ontological implications are a subject of active discussion (see e.g.\ Ref.\ \cite{Centeno2024} and references therein); entering this discussion in detail would lead us too far astray. In particular, we are being deliberately loose as to how the algebraic property we call FOLT relates to FOLT as defined at the strictly operational level, in order to stress the algebraic structure at hand, in the spirit of our resolution about operational considerations described in Section \ref{section: subsystems C*}.}

One might very well find FOLT unpalatable, and want to forbid the cases in which it appears. This is an understandable stance --- FOLT will indeed, down the line, lead to numerous complications and counter-intuitive effects ---, but we need to stress that \textit{adopting this view is exactly the same as requiring that subsystems only correspond to factor algebras}. Indeed, asking for equality to hold in (\ref{eq: FOLT in bipartitions}) amounts precisely to asking for $\cz(\ca_1)$ to be trivial (i.e.\ only containing multiples of the identity), that is, for $\ca_1$ (and consequently $\ca_2$) to be a factor. Thus, the dilemma is between sticking with factors and tolerating FOLT. Note that the same thing could be said about \textit{redundancy} in our bipartitions, i.e.\ the fact that algebras corresponding to complementary subsystems have a non-trivial intersection (given by $\cz(\ca_1)$).

This is an apt moment to discuss how our approach to subsystems compares to previous C*-algebra-based proposals. Indeed, a significant fraction of those precisely impose local tomography as an axiom in order to forbid non-factor algebras. Refs.\ \cite{zanardi2001, zanardi2003}, for instance, explicitly require $\ca_1 \vee \ca_2 = \OM$ (an assumption called `completeness' in Ref.\ \cite{zanardi2003}), thereby restricting themselves to considering factorisations (which they call `tensor product structures'). This assumption remains prevalent in subsequent literature, e.g.\ Refs.\ \cite{cotler2017, carroll2020, adil2024, loizeau2024, Franzman2024}, but some works \cite{zanardi2022, andreadakis2023, zanardi2023} incorporate non-factor bipartitions (which they call `generalised tensor product structures'). Subsystems corresponding to non-factor sub-C*-algebras are also presented as an important example in Ref.\ \cite{chiribella2018}, within an approach that defines subsystems through subsets of quantum transformations.

% The case of AQFT is more subtle. There, non-factor algebras are allowed, but an extra requirement, the \textit{split property} \cite{fewster2019}, is introduced in order to render them innocuous. Without entering the details of the framework, the broad idea is that the non-factorness of the algebra associated to a spacetime region should always be `screened' at an arbitrarily short distance, using a factor algebra. This has the effect of essentially banning redundancies, i.e.\ non-trivial intersections, between algebras associated to independent regions of spacetime. Indeed, these redundancies describe correlations between the observables of these regions, and the usual justification for upholding the split property \cite{fewster2019} is that such correlations are not acceptable in a fundamental theory of physics, in which it should be possible to reprepare the states of these regions and thus decorrelate them.

\subsection{The interpretation of non-factor (sub)systems} \label{sec: non-factor subsystems}

While factors match our usual notion of a quantum system, it might not be clear to the reader how one should physically interpret the non-factorness of the algebra associated to a physical system. There are two, complementary, possible interpretations: one in terms of conservation rules, the other in terms of partial classicality.

\subsubsection{Conservation rules, and the superposition of trajectories example}

Let us start with conservation rules. The idea is that a system whose dynamics is subject (either in principle or in practice) to the constraint of preserving a certain quantity will then be best described with a non-factor C*-algebra. Indeed, let us take a system $\Om$, normally described with $\ch_\Om$, but of which we stipulate that its dynamics has to preserve the value of a certain observable $\co \in \textrm{Herm}(\ch_\Om)$. This means that any unitary operator encoding this dynamics should commute with $\co$. Writing $\ch_\Om$'s sectorisation into $\co$'s eigenspaces as $\ch_\Om := \bigoplus_{k \in K} \ch_\Om^k$, it is easy to see that this commutation relation is equivalent to the requirement $U \in \OM$, with $\OM$ the C*-algebra of block-diagonal operators with respect to this sectorisation, defined as in (\ref{eq: blockdiag rep}).

The non-factorness of the algebra, in this interpretation, is ascribable to a certain degree of degeneracy induced by the preservation condition. One is still allowed to perform any quantum operation within a given eigenspace of $\co$ --- i.e.\ `the whole block is allowed' ---, but not to map between two eigenspaces --- i.e.\ `cross-block terms are forbidden'.

How can non-factor algebras as \textit{sub}systems of a certain factor supersystem be understood in this perspective? The \textit{superposition of trajectories} scenario \cite{chiribella2019shannon} gives a good example. Suppose our supersystem $\Om$ starts as the combination of a localised qudit $P$ --- which we call a particle --- and a control qubit $C$, and that we engineer a dynamics in which the qubit's state in the computational basis coherently controls which of two paths $A_1$ and $A_2$ the particle follows. Agents at $A_1$ and $A_2$ can act on the particle if it is there, but cannot destroy it or create a new one. It is natural to think of the $A_n$'s as having access to subsystems of $\Om$. Going down this path leads to finding that these subsystems correspond to non-factor subalgebras of $\OM = \Lin(\ch_\Om)$.

Take $A_1$, for instance: if she gets the particle, she can act on its inner degrees of freedom --- i.e.\ implement a unitary on the qudit ---; if the particle goes the other way, she only sees a vacuum and can do nothing; and importantly, she can also measure whether the particle is there, and dephase between the two alternatives in case of a superposition. In the representation corresponding to the `particle/control' partition, $\ch_\Om := \ch_C \otimes \ch_P$, her set of possible unitary operations is of the form
\be \textrm{Unit}_{A_1} = \left\{ \ketbra{1}{1}_C \otimes U_P + e^{i \theta} \ketbra{2}{2}_C \otimes \id_P \,\,\middle|\,\, U \in \textrm{Unit}(\ch_P), \theta \in [0, 2 \pi[ \right\} \, ,\ee
which is precisely the unitary group of the sub-C*-algebra 
\be \correction{\ca_1 = \Big\{ \ketbra{1}{1}_C \otimes f_P + \al \ketbra{2}{2}_C \otimes \id_P \,\,\Big|\,\, f \in \Lin(\ch_P), \alpha \in \CC \Big\} \subseteq \OM \, ,} \ee
while the possible actions of $A_2$ correspond to the commutant algebra,
\be \correction{\ca_2 = \ca_1' = \Big\{ \al \ketbra{1}{1}_C \otimes \id_P + \ketbra{2}{2}_C \otimes f_P \,\,\Big|\,\, f \in \Lin(\ch_P), \alpha \in \CC \Big\} \, .} \ee
Their intersection and common centre is generated by the atomic projectors $\ketbra{1}{1}_C \otimes \id_P$ and $\ketbra{2}{2}_C \otimes \id_P$.

Thus, thinking of $A_1$ and $A_2$ as subsystems of $\Om$ naturally leads to linking them to non-factor sub-C*-algebras of $\OM$. Non-factorness is ascribable to the particle-number preservation condition imposed on the agents' operations. Redundancy --- the fact that both share a non-trivial set of common operations, corresponding to the common centre of their algebras --- is due to the fact that both can locally measure the `which path did the particle go' observable and dephase with respect to it. As for FOLT, it denotes the fact that in this scenario, the phase between the two paths is not only inaccessible to either agent, but cannot even be seen in correlations between results of any local measurements; one can only access it by acting globally on the two trajectories.

Preservation of the number of particles strongly suggests that there is a sense in which this number becomes a classical variable, at least as far as local agents are concerned. This ties in with the other way of interpreting non-factorness of subsystems.

\subsubsection{Partial classicality} \label{sec: partial classicality}

There is a sense in which the non-factorness of a C*-algebra can be seen as corresponding to partial classicality of a system, although one has to be careful about what is meant by classicality here. The idea is that, if one reasons in terms of the observables accessible to an agent acting on the system, only the block-diagonal ones are allowed, meaning that the system cannot be measured to be in a superposition of the blocks --- although it can be measured to be in a superposition of states within a given block. Thus, there is partial classicality in the sense that one cannot measure a superposition of answers to the `which block is the system in?' question. For instance, in our paradigmatic example of the superposition of trajectories, an individual agent cannot measure a superposition of answers to the `do I hold the particle?' question.

However, one should be aware of the specific features of this notion of classicality. First, the non-factorness also means that one cannot map between the blocks at all, even in a non-coherent way. This is at odds with the standard notion of a classical variable, which can of course be mapped between its different classical states. Second, it should be emphasised that even though an agent in possession of a non-factor system cannot \textit{measure} relative phases between the blocks, she can \textit{modify} them, \correction{i.e.\ implement (local) dephasings (for instance apply a phase controlled on whether she holds the particle or not);} this is not a classical operation. Third, as we saw in the example of the superposition of trajectories, FOLT means that even though the phase between the blocks is not accessible \textit{locally} in each system in a bipartition --- and not even in the correlations between local measurements ---, it still is accessible through operations on the global system $\Om$. This ties in directly with the previous point: the local dephasings cannot be measured locally, but are still meaningful as they affect the global system.

This can all best be seen in the extreme example of a bipartition into subsystems corresponding to \textit{commutative} sub-C*-algebras. This is the other extreme compared to the case of a factor subsystem: here, each algebra is a direct sum of trivial, one-dimensional factors, with the algebra's elements being diagonal,
\be f = \sum_k \alpha_k \pi^k \, , \ee
as already stated in Theorem \ref{th: commutative algs}. Furthermore, the algebras are then equal:
\be \ca_1 = \ca_2 = \cz(\ca_1) = \cz(\ca_2) \, . \ee

Although this sounds like an odd state of affairs, it in fact has a natural interpretation. The idea is that a specific orthonormal basis of $\ch_\Om$ has been chosen, and that each agent has access to a copy of the system's state in that basis. Thus, both agents have access to the same set of actions, which amount only to measuring in that basis or to dephasing with respect to it. Because all these possible actions commute, this is not in tension with the claim that the subsystems are separate. Keeping in mind our cautionary remarks about what is meant by `classical' here, one can think of $A_1$ and $A_2$ as completely classical systems.\footnote{\correction{This extremal case illustrates the tension between different ways of seeing `subsystems as subalgebras'. In our framework, a `classical subsystem', corresponding to a commutative subalgebra, is one on which non-trivial permutations are not allowed as actions, but dephasings are. In contrast, in a more traditional view, a classical system corresponding to the same algebra is one in which permutations are allowed, while dephasings are not (or are taken as irrelevant).}}

\section{An example of a multipartition: the superposition of three trajectories} \label{sec: 3 trajs}

Although we hope that there was some added value in the self-contained nature of the previous section, it was merely rehashing conceptions and results which can be found elsewhere in the literature. Our point will now be to show how these ideas can be further pushed to provide a consistent and satisfactory theory of \textit{multi}partitions, consisting in more than two parts.

As we will shortly see, the definition of bipartitions does not generalise straightforwardly to multipartitions; subtleties will arise and force us to make decisions about what we deem a good multipartition. To motivate these decisions and start building some intuition, we start with a concrete case of a situation one would naturally call a multipartition.

\subsection{The scenario}

This concrete case is simply given by the natural extension of our previous `superposition of trajectories' to the case of three parties. It consists in considering a system $\Om$ initially partitioned into a qudit $P$ and a qutrit $C$, and subjected to a dynamics in which $P$ is sent in one (or a superposition) of three trajectories $A_1$, $A_2$ and $A_3$, depending on $C$'s state. Similarly to before, it is natural to think of agents at each of the $A_n$'s as having access to subsystems of $\Om$.

What does, say, the C*-algebra $\ca_1$ corresponding to $A_1$'s actions look like? As before, $A_1$ can only act on the inner degrees of freedom on $P$ if she holds it, measure whether she holds it, and dephase with respect to that alternative. In the representation corresponding to the `particle/control' partition, $\ch_\Om = \ch_C \otimes \ch_P$, this corresponds to the non-factor C*-algebra
\be \begin{split}
    \ca_1 &= \left\{ \ketbra{1}{1}_C \otimes f_P + \alpha (\ketbra{2}{2}_C + \ketbra{3}{3}_C) \otimes \id_P \,\,\middle|\,\, f \in \Lin(\ch_P), \alpha \in \mathbb{C} \right\} \\
    &=\correction{ \left\{ \begin{pmatrix}
        f_P & 0 & 0 \\
        0 & \al \id_P & 0 \\
        0 & 0 & \al \id_P 
    \end{pmatrix} \right\} \, ,}
\end{split} \ee
\correction{where in the second line we used a block-matrix notation, with coordinates of the overall matrix corresponding to the computational basis of $\ch_C$, and the action on $P$ written within each block.}

We can already note a feature that will prove typical of multipartitions. $A_1$ can only distinguish, as well as dephase between, the `$C=1$' and the `$C \in \{2,3\}$' alternatives. This translates the fact that, in the case where the particle is not sent to her, she has no way to figure whether it has been sent to $A_2$ or to $A_3$ (or to a superposition of both). Thus the centre $\cz(\ca_1)$ of her algebra (which we can think of as displaying the `classical' --- in the sense of Section \ref{sec: partial classicality} --- information she has access to) is given by the atomic projectors $\ketbra{1}{1}_C \otimes \id_P$ and $(\ketbra{2}{2}_C + \ketbra{3}{3}_C) \otimes \id_P$. Because the situation for $A_2$ and $A_3$ is symmetric, we can see that
\be \label{eq: neq of centres in tripartition} \forall m \neq n, \, \cz(\ca_m) \neq \cz(\ca_n) \,, \ee
which stands in contrast to the case of bipartitions, in which one has Proposition \ref{prop: common centre}.

\subsection{Conjunctions of systems and FOLT} \label{sec: tripartite conjunction}

There is another question we should mention: that of `conjunctions of systems'. Our theory should be able to specify a notion of the system `$A_1$ and $A_2$ taken together', denoted $A_{\{1,2\}}$. This would correspond to an agent that can act globally on trajectories 1 and 2, but not on trajectory 3. As usual, such an agent can measure, and dephase with respect to, whether she holds the particle, and perform any action that preserves its presence when she holds it. Because the agent is acting globally on both trajectories, the latter set of actions now includes moving the particle from path 1 to path 2, or measuring, as well as creating, superpositions between these two paths. This corresponds to the C*-algebra
\be \begin{split}
    \ca_{\{1,2\}} &= \left\{ \sum_{i,j \in \{1,2\}} \ketbra{j}{i}_C \otimes f^{ji}_P + \alpha  \ketbra{3}{3}_C \otimes \id_P \,\,\middle|\,\, \forall i,j,  \, f^{ji} \in \Lin(\ch_P), \alpha \in \mathbb{C} \right\} \\
    &= \correction{ \left\{ \begin{pmatrix}
        f^{11}_P & f^{12}_P & 0 \\
        f^{21}_P & f^{22}_P & 0 \\
        0 & 0 & \al \id_P 
    \end{pmatrix} \right\} \, .}
\end{split}   \ee
$\cz(\ca_{\{1,2\}})$ is given by the atomic projectors $\pi = (\ketbra{1}{1}_C + \ketbra{2}{2}_C) \otimes \id_P$ and $\Bar{\pi} = \ketbra{3}{3}_C \otimes \id_P$, which makes it equal to $\cz(\ca_3)$; in fact, one can check that $(\ca_{\{1,2\}}, \ca_3) \vdash \OM$ according to our previous Definition \ref{def: bipart of factors}.

\correction{Another noteworthy feature, which will be important in our later discussion of definitions of partitions, is that $\cz(\ca_1)$ and $\cz(\ca_2)$ do not contain $\cz(\ca_{\{1,2\}})$:
\be \label{eq: overall centre not in parts centres} \cz(\ca_{\{1,2\}}) \not\subseteq \cz(\ca_1) \quad \textrm{and} \quad \cz(\ca_{\{1,2\}}) \not\subseteq \cz(\ca_2) \, .\ee
A consequence of this is that $\ca_1' \cap \ca_{\{1,2\}}$, the commutant of $\ca_1$ within $\ca_{\{1,2\}}$, is strictly larger than $\ca_{2}$, as it necessarily includes $\cz(\ca_{\{1,2\}})$. However --- and this fact will be important later --- equality can be restored if the commutant is specialised to each block of $\ca_{\{1,2\}}$, i.e.\ we have
\be \pi \ca_1' \cap \ca_{\{1,2\}} = \pi \ca_2 \quad \textrm{and} \quad  \bar{\pi} \ca_1' \cap \bar{\pi} \ca_{\{1,2\}} = \bar{\pi} \ca_2 \, .  \ee

}

Finally, we emphasise that here again, `$\ca_{\{1,2\}}$ is more than the sum of $\ca_1$ and $\ca_2$', in the sense that
\be \label{eq: FOLT in tripartitions} \begin{split}
    \ca_1 \vee \ca_2 &= \left\{ \sum_{i \in \{1,2\}} \ketbra{i}{i}_C \otimes f^{i}_P + \alpha  \ketbra{3}{3}_C \otimes \id_P \,\,\middle|\,\, \forall i,  \, f^{i} \in \Lin(\ch_P), \alpha \in \mathbb{C} \right\} \\
  &=\correction{ \left\{ \begin{pmatrix}
        f^1_P & 0 & 0 \\
        0 & f^2_P & 0 \\
        0 & 0 & \al \id_P 
    \end{pmatrix} \right\} } \subsetneq \ca_{\{1,2\}} \, . \end{split} \ee
This corresponds to the fact that $A_{\{1,2\}}$ has access to the coherence between trajectories 1 and 2, something that is not available even in correlations between potential measurements of $A_1$ and $A_2$; and that, correspondingly, $A_{\{1,2\}}$ can also map, as well as create superpositions, between these trajectories, something that no coordinated action of $A_1$ and $A_2$ can do.

This feature should not make us any more spooked than we (potentially) already were: it is just another avatar of FOLT, which we showed in Section \ref{sec: FOLT} to be a necessary consequence of admitting subsystems corresponding to non-factor algebras. In that sense, it bears no novelty compared to the case of bipartitions, with (\ref{eq: FOLT in tripartitions}) an analogue of (\ref{eq: FOLT in bipartitions}).

\subsection{Representation} \label{sec: rep for tripartition}

The final question is whether we can find an adapted representation for our tripartition, i.e.\ a way to see $\OM$ as a space of operators on a Hilbert space made of three parts corresponding to the $A_n$'s, in analogy to what we obtained for bipartitions in Section \ref{sec: reps of bipartitions}. As it turns out, we can: denoting $\ch_{A_n^0} \cong \mathbb{C}$ and $\ch_{A_n^1} \cong \ch_P$, denoting the Boolean matrix
\be \label{eq: seccorrs tripartition} \forall k_1, k_2, k_3 \in \{0,1\}, \quad \sigma_{\Vec{k}}:= \begin{cases}
    1 \textrm{ if exactly one of the } k_n \textrm{'s is equal to } 1 , \\
    0 \textrm{ otherwise,}
\end{cases} \ee
and
\be \ch_\Om  := \bigoplus_{k_1, k_2, k_3 \textrm{ such that } \sigma_{\Vec{k}} = 1} \ch_{A_1^{k_1}} \otimes \ch_{A_2^{k_2}} \otimes \ch_{A_3^{k_3}}  , \ee
we can find a representation $\OM \cong \Lin(\ch_\Om)$ in which
\be \ca_1 \cong \left\{ \bigoplus_{k_1, k_2, k_3 \textrm{ such that } \sigma_{\Vec{k}} = 1} f^{k_1}_{A_1^{k_1}} \otimes \id_{A_2^{k_2}} \otimes \id_{A_3^{k_3}} \,\,\middle|\,\, \forall k_1, f^{k_1} \in \Lin(\ch_{A_1}^{k_1}) \right\} \, ,\ee
and the other $\ca_n$'s have a symmetric form (we leave the proof out, as it will derive from the general representation theorem we shall prove later). This gives us an analogue of Theorem \ref{th: standard rep bipartitions}.

We can also, like we did in Corollary \ref{cor: routed rep bipartitions}, turn this into a representation on an extended space that would not feature the awkward blend of direct sums and tensor products. Indeed, defining
\be \ch_\Om^\ext := \underbrace{\left( \bigoplus_{k_1 \in \{0,1\}} \ch_{A_1^{k_1}}\right)}_{=: \ch_{A_1}}   \otimes \underbrace{\left( \bigoplus_{k_2 \in \{0,1\}} \ch_{A_2^{k_2}}\right)}_{=: \ch_{A_2}}   \otimes \underbrace{\left( \bigoplus_{k_3 \in \{0,1\}} \ch_{A_3^{k_3}} \right)}_{=: \ch_{A_3}} \, , \ee
\be \label{eq: sec corrs tripartite} \tilde{\bbpi} := \sum_{k_1, k_2, k_3 \textrm{ such that } \sigma_{\Vec{k}} = 1} \bbpi_{A_1}^{k_1} \otimes \bbpi_{A_2}^{k_2} \otimes \bbpi_{A_3}^{k_3} \in \Lin(\ch_\Om^\ext) \, ,\ee
and the relation $\eta^{\rm tot} : \{0,1\}^3 \to \{0,1\}^3$ defined by 
\be \left( \eta^{\rm tot} \right)_{\Vec{k}}^{\Vec{l}} := \sigma_{\Vec{k}} \, \sigma_{\Vec{l}} \, ,\ee
we have a representation given by the C*-algebra isomorphism
\be \iota: \quad \OM \quad \to \quad \Lin_{\eta^{\rm tot}}(\ch_\Om^\ext) \, , \ee
in which
       \be \label{eq: iota individual} \iota \left( \ca_1 \right) = \left\{ (f_{A_1} \otimes \id_{A_2} \otimes \id_{A_3}) \, \tilde{\bbpi} \,\, \middle| \,\,  f \in \Lin_\delta(\ch_{A_1}) \right\}  \ee
and the other $\ca_n$'s have a symmetric form.

Furthermore, in this representation, we also obtain a clean form for the algebras corresponding to joint systems, like $\ca_{\{1,2\}}$; taking the relation $\eta$ on $\{0, 1\}^2$ to be the partial trace of $\eta^{\rm tot}$, i.e.
\be \label{eq: def eta tripartite}  \eta_{k_1 k_2}^{l_1 l_2}  := \sum_{k_3} \left( \eta^{\rm tot} \right)_{k_1 k_2 k_3}^{l_1 l_2 k_3} \, ,
    \ee
which can be checked to be a partial equivalence relation on $\{0,1\}^2$, we can pick $\iota$ \correction{(among the possible choices of $\iota$'s satisfying (\ref{eq: iota individual})) such that it additionally satisfies}
\be \iota \left( \ca_{\{1,2\}} \right)  = \left\{ (f_{A_1 A_2} \otimes \id_{A_3}) \, \tilde{\bbpi} \,\, \middle| \,\,  f \in \Lin_\eta(\ch_{A_1} \otimes \ch_{A_2}) \right\} \, , \ee
so that we can also see $\ca_{\{1,2\}}$ as acting on the $A_1$ and $A_2$ Hilbert spaces. \correction{(This result is a special case of the upcoming Theorem \ref{th: representation}.)}

How should we understand the route $\eta_{k_1 k_2}^{l_1 l_2}$? One can compute it to be equal to $1$ if and only if either all its indices are zero, or exactly one of its inputs and exactly one of its outputs are equal to 1. In other words, it encodes two constraints on a linear map $f$: first, that neither in the inputs nor in the outputs of $f$ can there be two particles (which would correspond to having two indices equal to 1); second, that the presence or absence of the particle should be preserved by $f$. We see that this corresponds exactly to the operational constraints on agents acting globally on trajectories 1 and 2 that we described earlier.

This is from a physical perspective; from a mathematical perspective, on the other hand, we defined $\eta$ in (\ref{eq: def eta tripartite}) solely from the sectorial correlations matrix $\sigma$, by considering the partial trace of $\eta^{\rm tot}$, itself obtained out of $\sigma$. We shall see that this procedure works in every multipartition to find the route followed by a certain joint-system algebra.

Finally, let us look at how these algebras can be represented graphically in routed quantum circuits, using the notations introduced in Section \ref{sec: routes}. Here, we additionally use `global index constraints', which are ways to graphically sum up the routes in a diagram by just writing equations relating possible index values \cite{vanrietvelde2022}. Additionally, as in (\ref{eq: routed rep bipartitions diag A_1 no pi}), we consider that pre-processing with $\tilde{\bbpi}$ is implied by the global index constraints. Our representation then takes the form
\begin{subequations}
    \be \iota \left( \ca_1 \right)    =  \left\{ \quad \begin{tikzpicture}
	\begin{pgfonlayer}{nodelayer}
		\node [style=none] (0) at (0, 2.5) {};
		\node [style=none] (4) at (3, -2.5) {};
		\node [style=none] (5) at (3, 2.5) {};
		\node [style=right label] (7) at (0, -2.5) {${k_2}$};
		\node [style=small box] (8) at (-3, 0) {$f$};
		\node [style=none] (9) at (3, -2.5) {};
		\node [style=right label] (10) at (3, -2.5) {${k_3}$};
		\node [style=none] (11) at (0, -2.5) {};
		\node [style=none] (12) at (-3, -2.5) {};
		\node [style=none] (13) at (-3, 2.5) {};
		\node [style=none] (15) at (-3, -2.5) {};
		\node [style=right label] (16) at (-3, -2.5) {${k_1}$};
		\node [style=right label] (17) at (0, 2.5) {${k_2}$};
		\node [style=right label] (18) at (3, 2.5) {${k_3}$};
		\node [style=right label] (19) at (-3, 2.5) {${k_1}$};
		\node [style=left label] (20) at (11.5, 0) {$k_1 + k_2 + k_3 = 1$};
		\node [style=none] (21) at (-3, 0.5) {};
		\node [style=none] (22) at (-3, -0.5) {};
	\end{pgfonlayer}
	\begin{pgfonlayer}{edgelayer}
		\draw (5.center) to (4.center);
		\draw (0.center) to (11.center);
		\draw (13.center) to (21.center);
		\draw (15.center) to (22.center);
	\end{pgfonlayer}
\end{tikzpicture} \quad \right\} \, ; \ee
    \be \iota \left( \ca_{\{1,2\}} \right)   =  \left\{ \quad \begin{tikzpicture}
	\begin{pgfonlayer}{nodelayer}
		\node [style=none] (0) at (0, 2.5) {};
		\node [style=none] (1) at (3, -2.5) {};
		\node [style=none] (2) at (3, 2.5) {};
		\node [style=right label] (3) at (0, -2.5) {${k_2}$};
		\node [style=very large box] (4) at (-1.5, 0) {$f$};
		\node [style=none] (5) at (3, -2.5) {};
		\node [style=right label] (6) at (3, -2.5) {${k_3}$};
		\node [style=none] (7) at (0, -2.5) {};
		\node [style=none] (8) at (-3, -2.5) {};
		\node [style=none] (9) at (-3, 2.5) {};
		\node [style=none] (10) at (-3, -2.5) {};
		\node [style=right label] (11) at (-3, -2.5) {${k_1}$};
		\node [style=right label] (12) at (0, 2.5) {${k_2'}$};
		\node [style=right label] (13) at (3, 2.5) {${k_3}$};
		\node [style=right label] (14) at (-3, 2.5) {${k_1'}$};
		\node [style=left label] (15) at (11.5, -2.5) {$k_1 + k_2 + k_3 = 1$};
		\node [style=none] (16) at (-3, 0.5) {};
		\node [style=none] (17) at (-3, -0.5) {};
		\node [style=left label] (18) at (11.5, 2.5) {$k_1' + k_2' + k_3 = 1$};
	\end{pgfonlayer}
	\begin{pgfonlayer}{edgelayer}
		\draw (2.center) to (1.center);
		\draw (0.center) to (7.center);
		\draw (9.center) to (16.center);
		\draw (10.center) to (17.center);
	\end{pgfonlayer}
\end{tikzpicture}\quad \right\} \, ; \ee
    \be \iota \left( \OM \right)   =  \left\{ \quad\begin{tikzpicture}
	\begin{pgfonlayer}{nodelayer}
		\node [style=none] (0) at (0, 2.5) {};
		\node [style=none] (1) at (3, -2.5) {};
		\node [style=none] (2) at (3, 2.5) {};
		\node [style=right label] (3) at (0, -2.5) {$k_2$};
		\node [style=very very large box] (4) at (0, 0) {$f$};
		\node [style=none] (5) at (3, -2.5) {};
		\node [style=right label] (6) at (3, -2.5) {$k_3$};
		\node [style=none] (7) at (0, -2.5) {};
		\node [style=none] (8) at (-3, -2.5) {};
		\node [style=none] (9) at (-3, 2.5) {};
		\node [style=none] (10) at (-3, -2.5) {};
		\node [style=right label] (11) at (-3, -2.5) {$k_1$};
		\node [style=right label] (12) at (0, 2.5) {$k_2'$};
		\node [style=right label] (13) at (3, 2.5) {$k_3'$};
		\node [style=right label] (14) at (-3, 2.5) {$k_1'$};
		\node [style=left label] (15) at (11.5, -2.5) {$k_1 + k_2 + k_3 = 1$};
		\node [style=none] (16) at (-3, 0.5) {};
		\node [style=none] (17) at (-3, -0.5) {};
		\node [style=left label] (18) at (11.5, 2.5) {$k_1' + k_2' + k_3' = 1$};
	\end{pgfonlayer}
	\begin{pgfonlayer}{edgelayer}
		\draw (2.center) to (1.center);
		\draw (0.center) to (7.center);
		\draw (9.center) to (16.center);
		\draw (10.center) to (17.center);
	\end{pgfonlayer}
\end{tikzpicture}\quad \right\} \, . \ee 
\end{subequations}

\section{A theory of multipartitions} \label{sec: multipartitions}

In this section, we propose a theory of what it means in general to take a partition of a quantum system into more than two parts, with some of these possibly corresponding to non-factor algebras.

\subsection{Motivating our choices}

Suppose we have a quantum system $\Om$, described by an algebra $\OM$, and that we want to partition it into subsystems $A_n$, with the labels $n$ belonging to a finite set $X$. The first important point is that \textit{specifying the algebras corresponding to the individual $A_n$'s is not sufficient}; indeed, we also need to specify what the conjunctions of systems are, i.e.\ what it means to act globally on a collection $S \subseteq X$ of systems, like we did with $\ca_{\{1,2\}}$ in our concrete example above. In the standard case of factorisations, this is superfluous, as one then always has $\ca_S = \bigvee_{n \in S} \ca_n$; however, the generic presence of FOLT (cf.\ Sections \ref{sec: FOLT} and \ref{sec: tripartite conjunction}) means we cannot rely on this. Thus, a multipartition should take its labels not in X, but in its powerset $\cp(X)$, i.e.\ be a mapping
\be
\begin{split}
        \ca : \,\,\cp(X) &\to \Sub\left( \OM \right) \\
    S &\mapsto \ca_S \, ,
\end{split}
\ee
where we write $\Sub(\OM)$ for the set of sub-C*-algebras of $\OM$.

How can we make sure that such a mapping yields a good multipartition? The safest way would be to check that, for any two disjoint sets of labels $S, T \subseteq X$, $(\ca_S, \ca_T)$ forms a bipartition of $\ca_{S \sqcup T}$. However, so far we only defined bipartitions of a \textit{factor}, and $\ca_{S \sqcup T}$ might in general not be one --- for instance, in our concrete example of a tripartition in Section \ref{sec: 3 trajs}, $\ca_{\{1,2\}}$ is not. Thus, our first goal should be to get a good definition of bipartitions $(\ca_1, \ca_2)$ of $\OM$ when $\OM$ is not a factor; our definition of multipartitions will directly derive from it.

A first idea would be to simply reuse our definition for the factor case, i.e.\ to ask for $\ca_1 = \ca_2'$ and $\ca_2 = \ca_1'$. However, this turns out to be too strong. Indeed, \correction{any commutant taken within $\OM$ necessarily includes its centre $\cz(\OM)$, as the latter commutes with all of $\OM$ by definition. Thus} our first idea would directly entail that
\be \label{eq: parts sharing centre} \cz(\OM) \subseteq \cz(\ca_1) \quad \textrm{and} \quad \cz(\OM) \subseteq \cz(\ca_2)  \, , \ee
which is, for instance, \textit{not} satisfied in our paradigmatic example, as can be seen in  (\ref{eq: overall centre not in parts centres}).

How can we understand this discrepancy? \correction{Remember that, following Section \ref{sec: partial classicality}, we can think of $\cz(\OM)$ as encoding the classical information available in $\Om$. For instance, in our paradigmatic example of Section \ref{sec: 3 trajs}, $\cz(\ca_{\{1,2\}})$ is the information that is classical in the eyes of the agent that can act on trajectories 1 and 2; namely, the information about whether the particle is in trajectories 1 and 2, or in the trajectory 3. Now, (\ref{eq: parts sharing centre}) would have the physical meaning that both $A_1$ and $A_2$ have full access to this classical information --- for instance, both can measure it. We see however that this is not the case: $A_1$ only knows whether the particle is in her trajectory; if it isn't, she has no way to tell whether it is in $A_2$ or in $A_3$.}

\correction{The moral here is that one should allow for the `overall classical' information --- encapsulated by $\cz(\OM)$ ---} to be split in whatever way between the \correction{two subsystems}, including but not restricted to the possibility that both subsystems fully share it. \correction{A naive use of commutants cannot express such a condition, since, as we saw, any commutant will directly include the whole centre of the ambient algebra.}

\correction{The way to bar commutants from including the centre of the ambient algebra is to only take them individually within each block of the algebra, i.e.\ within each $\pi^k \OM$, where the $\pi^k$'s are $\cz(\OM)$'s atomic projectors. Each of these blocks can be thought of as an alternative in which $\Om$'s classical information takes a specific value. For instance, looking at $\ca_{\{1,2\}}$ in Section \ref{sec: 3 trajs}, it consists of two blocks: one identified by the atomic projector $(\ketbra{1}{1}_C + \ketbra{2}{2}_C) \otimes \id_P$ (corresponding to the alternative `$A_{\{1,2\}}$ holds the particle'), the other by $\ketbra{3}{3}_C \otimes \id_P$ (`$A_{\{1,2\}}$ does not hold the particle'). Since each block is a factor by construction, we can safely use the commutant condition within it, with the meaning being: once the overall classical information is fixed to a certain value, we are back to a fully quantum system (consisting of the other degrees of freedom) that can be partitioned using the standard Definition \ref{def: bipart of factors}. This is captured by the condition
\be \label{eq: def bipart comm prelim} \forall k \in K, \, \,\pi^k \ca_1' = \pi^k \ca_2 \, . \ee

This first condition ensures that the quantum information within $\Om$ is suitably split between $A_1$ and $A_2$. However, it now leaves out another requirement: that $\Om$'s classical information --- encapsulated by $\cz(\OM)$ --- is also suitably split. While we have seen that (\ref{eq: parts sharing centre}) is too strong, asking for nothing is too weak, since it allows for the overall classical information to be `nowhere at all' in the individual parts.}\footnote{\correction{To see this, take a basic example of an overall algebra that is already fully classical and consists of a single bit: $\OM = \cz(\OM)  = \left\{\begin{pmatrix}
    \al & 0 \\ 0 & \beta 
\end{pmatrix} \, \middle| \, \alpha, \beta \in \mathbb{C}\right\}$. Taking $\ca_1 = \ca_2 = \{\lambda \id | \lambda \in \mathbb{C}\}$ to be fully trivial systems, containing no information at all, it can be seen that these still satisfy (\ref{eq: def bipart comm prelim}). Thus, only imposing (\ref{eq: def bipart comm prelim}) in our definition would allow for `bipartitions of a classical system into two trivial systems', which is clearly unacceptable.}} \correction{In our example of Section \ref{sec: 3 trajs}, the overall classical information of $\ca_{\{1,2\}}$ is for instance accessible from local information: from the data of whether $A_1$ holds the particle, and that of whether $A_2$ holds it, one can infer whether $A_{\{1,2\}}$ does. This conceptual idea that `the overall classical information should be accessible from the conjunction of local information' finds its formal counterpart in the fact that the overall centre is within the span of local ones, namely that\footnote{\correction{A seemingly more general condition would be $\cz(\OM) \subseteq \ca_1 \vee \ca_2$, but under the assumption of (\ref{eq: def bipart comm prelim}), the latter is in fact equivalent to (\ref{eq: def bipart centre prelim}).}}
\be \label{eq: def bipart centre prelim} \cz(\OM) \subseteq \cz(\ca_1) \vee \cz(\ca_2) \, . \ee

In a nutshell, (\ref{eq: def bipart comm prelim}) and (\ref{eq: def bipart centre prelim}) can thus be thought of as ensuring the proper splitting of, respectively, $\Om$'s quantum and classical information.}

% In order to avoid the inconveniences caused by the naive use of the commutant definition, we should thus only apply it within individual blocks of $\OM$, which are factors by definition; i.e., writing $\Atproj(\cz(\OM)) = \{\pi^k\}_{k \in K}$, we should ask for $\pi^k \ca_1' = \pi^k \ca_2 \, \forall k$. This condition alone, however, is not sufficient; for instance, it would entail that when $\OM$ is a commutative algebra, it can be partitioned into two trivial subalgebras $\{\lambda \id | \lambda \in \mathbb{C}\}$. What we are missing is the condition that $\cz(\OM)$ itself is spanned by $\cz(\ca_1)$ and $\cz(\ca_2)$ --- in other words, that the classical information stored in $A_1$ and $A_2$ includes the classical information in $\OM$. 

\subsection{Definition}

Thus, we define bipartitions in general in the following way. 

\correction{\begin{definition}[Projections of algebras]
    Let $\cf\subseteq \OM$ be two C* algebras and let $\pi \in \cf'$ be an orthogonal projector. $\cf$'s projection by $\pi$ is then
    \be \pi \cf := \{ \pi f | f \in \cf\} = \{ f \pi | f \in \cf\} \, .\ee
\end{definition}}

\begin{definition}[Bipartitions] \label{def: bipartitions}
    Let $\OM$ be a C*-algebra, and $\ca_1, \ca_2 \subseteq \OM$ sub-C*-algebras of it. We say that $(\ca_1, \ca_2)$ forms a \emph{bipartition} of $\OM$ (denoted $(\ca_1, \ca_2) \vdash \OM$) if the following two conditions are satisfied:
\begin{subequations}
    \be \label{eq: def bipart comm} \forall k \in K, \, \,\pi^k \ca_1' = \pi^k \ca_2 \, ; \ee
    \be \label{eq: def bipart centre} \cz(\OM) \subseteq \cz(\ca_1) \vee \cz(\ca_2) \, ; \ee
\end{subequations}
where the atomic projectors of $\OM$'s centre are denoted as $\Atproj(\cz(\OM)) = \{\pi^k\}_{k \in K}$, \correction{and $\cz(\ca_1) \vee \cz(\ca_2)$ is the C*-algebraic closure of $\cz(\ca_1) \cup \cz(\ca_2)$}.
\end{definition}

Note that this matches Definition \ref{def: bipart of factors} when $\OM$ is a factor. We can now directly leverage this to get a definition of multipartitions.

\begin{definition}[Partitions]\label{def: partitions}
    Let $\OM$ be a C*-algebra. A \emph{partition} of it, labelled by the finite set $X$, is a mapping
    \be
\begin{split}
        \ca : \,\,\cp(X) &\to \Sub\left( \OM \right) \\
    S &\mapsto \ca_S \, ,
\end{split}
\ee
satisfying the following conditions:
\be \ca_X = \OM \, ;\ee
\be \label{eq: def partition empty} \ca_\emptyset = \{\lambda \id \,|\, \lambda \in \mathbb{C}\} \, ;\ee
\be \forall S, T \subseteq X \textrm{, disjoint, } (\ca_S, \ca_T) \vdash \ca_{S \sqcup T} \, .\ee
We then denote $(\ca_S)_{S \subseteq X} \vdash \OM$.
\end{definition}

\subsection{Properties} \label{sec: properties}

Let us now investigate the properties of partitions under our definition (all proofs for this Section are given in Appendix \ref{app: properties}). First, we can ask how bipartitions relate to commutants, since we know that we don't necessarily have $\ca_2 = \ca_1'$. The idea is that the leeway $\ca_2$ has in not being equal to $\ca_1'$ corresponds to how much of $\cz(\OM)$ it includes.

\begin{proposition} \label{prop: partitions and commutants}
    If $(\ca_1, \ca_2) \vdash \OM$, then
    \be \ca_1' = \cz(\OM) \vee \ca_2 \, .  \ee
\end{proposition}

\correction{Thus $\ca_2 \subseteq \ca_1'$, but this is an equality if and only if $\cz(\OM) \subseteq \ca_2$.

Another important fact to point out is that} in a bipartition of a non-factor $\OM$, \textit{the $\ca_n$'s are not necessarily closed}, in the sense that 
\be \ca_n'' = \cz(\OM) \vee \ca_n \, ,\ee
so that \correction{whenever $\OM$ is not a factor and $\cz(\OM) \not\subseteq \ca_n$, we have} $\ca_n \subsetneq \ca_n''$. This non-closure under the bicommutant can be seen as the structural root for a certain number of troublesome features, but it is inescapable; ruling it out would mean requiring $\cz(\OM) \subseteq \ca_n \, \, \forall n$, which is too strong for our needs, \correction{as it is not satisfied in our three-trajectories example when we take $\OM = \ca_{\{1,2\}}$, cf.\ (\ref{eq: overall centre not in parts centres}).}

Next, we want to investigate how centres of the different algebras are related. This can be summed up in a useful slogan.

\begin{slogan}
    In a partition of a factor $\OM$, the centre of an algebra $\ca_S$ corresponds to \textit{redundant elements}, i.e.\ to elements of $\ca_S$ that are also in the complementary algebra $\ca_{\Bar{S}}$.

    For $\OM$ non-factor, the centre of $\ca_S$ might, in addition, include bits of $\cz(\OM)$.
\end{slogan}

Let us make this rigorous, starting with the case of bipartitions, where we have the following Proposition generalising Proposition \ref{prop: common centre}. To alleviate notations, we take the habit of writing $\cz_S := \cz(\ca_S)$.

\begin{proposition} \label{prop: centre Z_1}
    If $(\ca_1, \ca_2) \vdash \OM$, then
    \be  \cz_1 = \ca_1 \cap \left( \cz(\OM) \vee \ca_2 \right) = \ca_1 \cap \left( \cz(\OM) \vee \cz_2 \right) \, . \ee
\end{proposition}

Given Definition \ref{def: partitions}, this directly entails that in a partition, writing $\Bar{S} := X \setminus S$, we have
\be \forall S \subseteq X,  \quad \cz_S = \ca_S \cap \left( \cz(\OM) \vee \ca_{\Bar{S}} \right) = \ca_S \cap \left( \cz(\OM) \vee \cz_{\Bar{S}} \right) \, .\ee
In addition, we can also use this to see how the centres of joint systems are determined by those of individual ones together with $\cz(\OM)$. This is a remarkable property, showing that the structure of centres is actually quite rigidly constrained by the definition of a partition.

\begin{proposition} \label{prop: Z from individual Zs}
    If $(\ca_S)_{S \subseteq X} \vdash \OM$, then
    \be \label{eq: Z from individual Zs} \forall S \subseteq X,\quad \cz_S = \left( \bigvee_{n \in S} \cz_n \right) \cap \left( \bigvee_{m \not\in S} \cz_m \vee \cz(\OM) \right) \, .\ee
\end{proposition}

In other words, \textit{the centre of $\ca_S$ is made up exactly of these elements that can be found both in the centres of the individual algebras within it, and in the combination of the individual algebras outside of it with the centre of $\OM$ itself.} In particular, information that is present only within the $\cz_n$'s with $n \in S$ but not elsewhere is \textit{not} in $\cz_S$ --- i.e.\ $\ca_S$ includes elements that do not commute with the projectors encoding it. 

It is important to understand what this `disappearance of centres' means physically. The idea is that if there is some classical information shared only between systems with labels $x \in S$, and not present in the centre of the whole system $\Om$ (i.e.\ not considered to be classical overall), then any agent acting globally on the whole of $A_S$ (or on a supersystem of it) does not see it as classical and can map between its sectors. One can see this in action in our example of $\ca_{\{1,2\}}$ in Section \ref{sec: tripartite conjunction}, in which an agent acting globally on trajectories 1 and 2 can map between these trajectories: in that case, the classical information that becomes quantum in the eyes of the superagent is that about whether the particle is in trajectory 1 or in trajectory 2.

\section{A representation theorem for partitioned algebras}
\label{sec: representations}

We defined partitions in an algebraic manner; we now investigate how they can be concretely represented as sets of operators over Hilbert spaces. In Sections \ref{sec: reps of bipartitions} and \ref{sec: rep for tripartition}, we showed how bipartitions and our paradigmatic example of a tripartition admitted representations over \correction{tensor products of} Hilbert spaces, in which each algebra could be seen as the one acting on the corresponding \correction{factor} of the Hilbert space. Here, we discuss the generalisation of this result to all partitions.

As it turns out, some features of what we did in this earlier example can always be reproduced, but not all. More precisely, we can always find a representation in which each of the algebras for individual parts is well-localised, i.e.\ is acting over the corresponding Hilbert space \correction{factor.} Furthermore, in such a representation, we can also compute the routes for the algebras corresponding to joint systems in a systematic way. However, it is not always possible to find a representation in which each of the algebras corresponding to joint systems is the one acting \correction{exactly on the corresponding tensor factors.}

For the remainder of this section, we fix a C*-algebra $\OM$ and a partition $(\ca_S)_{S \subseteq X}$ of it. 

\subsection{Representations of centres}

In Section \ref{sec: rep for tripartition}, we saw that the representation of our example of a tripartition involved the use of non-trivial Boolean matrices, such as the Boolean matrix $\sigma$ encoding sectorial correlations defined in (\ref{eq: seccorrs tripartition}), and the route $\eta$ used in the representation of the joint algebra $\ca_{\{1,2\}}$, defined in (\ref{eq: def eta tripartite}). Here, we show how this can be generalised.

As we showed in Proposition \ref{prop: Z from individual Zs}, for any set $S \subseteq X$, the centre $\cz_S$ of the joint systems $\ca_S$ is determined by the individual centres $\cz_n, n \in X$ and the centre of the global algebra $\cz_X = \cz(\OM)$. There is an analogous result for the route used to describe $\cz_S$: as we will now show, this route is determined by the sole data of $\cz(\OM)$ and of redundancies between the $\cz_n$'s.

We start by indexing each of the $\Atproj(\cz_n)$'s,
\be \Atproj(\cz_n) = \left\{ \pi_n^{k_n} \right\}_{k_n \in K_n} \, . \ee
We also define
\begin{subequations}
    \be K := \bigtimes_{n \in X} K_n \, , \ee
    \correction{\be \forall \vec{k} \in K, \quad \pi^\veck := \prod_{n \in X} \pi_n^{k_n} \, , \ee
    \be \Tilde{K} := \left\{ \Vec{k} \in K \,\, \middle| \,\,  \pi^\veck \neq 0 \right\} \, \subseteq K \, , \ee}
\end{subequations}
and we write the Boolean indicator of $\Tilde{K}$ in $K$ as $\sigma$, defined by
\be \forall (k_n)_{n \in X} \in K, \quad \sigma_{\vec{k}} = \begin{cases}
    1 \textrm{ if } \vec{k} \in \Tilde{K} \\
    0 \textrm{ otherwise.}
\end{cases} \ee 
This encodes sectorial correlations --- or in more structural terms, redundancies in the centres ---, and generalises (\ref{eq: seccorrs tripartition}). We can then generalise (\ref{eq: def eta tripartite}) in the following way.  From $\cz_X = \cz(\OM)$, we can define an equivalence relation $\sim$ on $\tilde{K}$ through
\be \label{eq: recipe etaX} \begin{split}
    &\forall \vec{k}, \vec{l} \in \tilde{K}, \quad \vec{k} \sim \vec{l} \\\
    &\iff \exists \mu \in \Atproj(\cz_X) \textrm{ such that both } \pi^\veck \preccurlyeq \mu \textrm{ and } \pi^{\vec{l}} \preccurlyeq \mu \textrm{ hold} \, ,
\end{split}  \ee
\correction{where $\pi \preccurlyeq \mu$ means $\pi \mu = \pi$.} In other words, $\veck$ and $\vec{l}$ are equivalent if the corresponding projectors are within the same block of $\OM$. We extend this equivalence relation on $\tilde{K}$ to a partial equivalence relation $\sim_X$ on $K$ \correction{by 
\be \forall \veck, \vec{l} \in K, \quad \veck \sim_X \vecl \quad \iff \quad \veck, \vecl \in \tilde{K} \textrm{ and } \veck \sim \vecl \, .\ee}
From $\sim_X$, for any $S \subseteq X$, we define a relation \correction{(which, as we shall state in Theorem \ref{th: computing centres}, is itself a partial equivalence relation)} $\sim_S$ on $\bigtimes_{n \in S} K_n$:
\be \label{eq: recipe etaS} \begin{split} \forall S \subseteq X, \quad &\forall \veck, \vec{k'} \in \bigtimes_{n \in S} K_n, \\
&\vec{k} \sim_S \vec{k'} \quad \iff \quad \exists (l_n)_{n \in X \setminus S}, \,\, \vec{k} \times \vec{l} \sim_X \vec{k'} \times \vec{l} \, . \end{split} \ee
More concretely, $\sim_S$ is simply the partial trace (in the theory of Boolean matrices) of $\sim_X$ over its indices in $X \setminus S$. 

For a given $S$, we choose a set of indices \correction{$Q_S$} for the set of $\sim_S$'s equivalence classes, denoting these equivalence classes as $\{ E_{q_S} \}_{q_S \in Q_S}$. With these notations, we have the following theorem.

\begin{theorem} \label{th: computing centres}
For any $S \subseteq X$, $\sim_S$ is a partial equivalence relation. In addition, \correction{with the above notations,} we have $\Atproj(\cz_S) = \left\{\pi_S^{q_S} \right\}_{q_S \in Q_S}$ where
    \be \label{eq: computing centres} \forall q_S \in Q_S, \quad \pi_S^{q_S} := \sum_{\vec{k} \in E_{q_S}} \left( \prod_{n \in S} \pi_n^{k_n} \right) \, . \ee
\end{theorem}

The proof \correction{of Theorem \ref{th: computing centres}} is given in Appendix \ref{app: computing centres}.

In other words, the $\cz_S$'s are determined solely from the specification of $\sim_X$, which itself is the combination of the data contained in $\sigma$ (encoded in $\sim_X$'s partialness) and in $\cz(\OM)$ (encoded in $\sim_X$'s equivalence classes). $\sim_S$ is inferred from $\sim_X$ as its partial trace (as a Boolean matrix) on the $n \not\in S$ coefficients.

\subsection{The main theorem}

\correction{With the $\sim_S$'s defined, we can state our general representation theorem. We recall that, following Theorem \ref{th: WA}, we have
\be \forall n \in X, \quad \ca_n \cong \bigoplus_{k_n \in K_n} \Lin\left(\ch_{A_n}^{k_n}\right) \, .\ee
We then denote 
\be \ch_{A_n} := \bigoplus_{k_n \in K_n}  \ch_{A_n}^{k_n} \, ,\ee 
define $\bbpi_{A_n}^{k_n}$ as the orthogonal projector of $\ch_{A_n}$ onto $\ch_{A_n}^{k_n}$, and}
    \be \label{eq: sec corrs multipartite} \tilde{\bbpi} := \sum_{\veck \textrm{ such that } \sigma_{\vec{k}} = 1} \left( \bigotimes_{n \in X} \bbpi^{k_n}_{A_n} \right) \in \Lin\left(\bigotimes_{n \in X} \ch_{A_n} \right) \, .\ee

\begin{theorem}[Representations of partitions] \label{th: representation}

    Let $(\ca_S)_{S \subseteq X} \vdash \OM$. With the notations above, there exists an isomorphism of C*-algebras
    \be \iota: \,\, \OM \to \Lin_{\sim_X}\left( \bigotimes_{n \in X} \ch_{A_n} \right) \ee
    mapping the individual algebras to a well-localised form:
    \be \label{eq: rep individual} \forall n \in X, \quad \iota(\ca_{n}) = \left\{ \Big(f \otimes \id_{A_m, \, m \neq n}\Big) \, \tilde{\bbpi} \,\,\, \middle| \,\,\,  f \in \Lin_{\delta}\left(\ch_{A_n}\right) \right\} \, . \ee

         \correction{Furthermore, let us define $\tilde{\ca}_S$'s as the analogously shaped `well-localised forms' of the $\ca_S$'s:
    \be \label{eq: well-localised form}
        \forall S \subseteq X, \quad \tilde{\ca}_S := \left\{ \,\Big(f \otimes \id_{A_m, \, m \not\in S}\Big) \,  \, \tilde{\bbpi} \,\,\, \middle| \,\,\,  f \in \Lin_{\sim_S}\left(\bigotimes_{n \in S} \ch_{A_n}\right) \right\} \, .  \ee
Given a $\iota$ satisfying (\ref{eq: rep individual}),  there exists for every $S \subseteq X$ a unitary $\phi_S \in \bigvee_{n \in X} \iota \left(\cz_n\right)$, such that }
    \be \label{eq: rep composite} \begin{split}
        \forall S \subseteq X, \quad \iota(\ca_S) &= \left\{ \phi_S  \,\Big(f \otimes \id_{A_m, \, m \not\in S}\Big) \,  \phi_S^\dagger \, \tilde{\bbpi} \,\,\, \middle| \,\,\,  f \in \Lin_{\sim_S}\left(\bigotimes_{n \in S} \ch_{A_n}\right) \right\} \\
        &= \hat{\phi}_S \left( \tilde{\ca_S} \right)\, ,
    \end{split} \ee
    where $\hat{\phi}_S : f \mapsto \phi_S f \phi_S^\dagger$ is the conjugation by $\phi_S$.
\end{theorem}

This Theorem is proven in Appendix \ref{app: proof representation}. Note how the indicator of sectorial correlations $\sigma$ and the routes $\sim_S$, as introduced in the previous section, play a crucial role here as well.

The unitaries $\phi_S$ appearing in (\ref{eq: rep composite}) are a surprising and important element. They should essentially be understood as dephasings;\footnote{\correction{These should not be confused with the `local dephasings' that agents in possession of a non-factor subsystem may implement, as discussed in Section \ref{sec: partial classicality}. The $\phi_S$'s are not actions that may or may not be applied on a subsystem: they are only involved in the mathematical representation of subalgebras, and are fixed for a fixed representation $\iota$ of a partition.}} indeed, as unitary elements of the commutative algebra $\bigvee_{n \in X} \iota(\cz_n)$, they \correction{must} be of the form
\be \phi_S = \sum_{\vec{k} \in \tilde{K}} e^{i \alpha_S(\vec{k})} \,\, \bigotimes_{n \in X} \bbpi_{A_n}^{k_n} \ee
for a certain phase function $\alpha_S: \tilde{K} \to [0, 2\pi[$. These dephasings can sometimes be eliminated for an appropriate choice of $\iota$, in which case we say that the partition is \textit{fully representable}.

\begin{definition}[Fully representable partitions] \label{def: representable}

    A partition $(\ca_S)_{S \subseteq X} \vdash \OM$ is \emph{fully repre\-sentable} if the isomorphism $\iota$ in Theorem \ref{th: representation} can furthermore be picked such that
    \be \forall S \subseteq X, \quad \iota(\ca_S) = \tilde{\ca}_S = \left\{ \Big(f \otimes \id_{A_n, \, n \not\in S}\Big) \, \tilde{\bbpi} \,\,\, \middle| \,\,\,  f \in \Lin_{\sim_S}\left(\bigotimes_{n \in S} \ch_{A_n}\right) \right\} \, . \ee
\end{definition}

Surprisingly, not all partitions are fully representable; there exist some such that, for any $\iota$, there exists at least one $S$ such that $\iota(\ca_S) \neq \tilde{\ca}_S$. A specific counterexample is given in the next section.

\section{\correction{A counterexample to full representability:} the partition of a fermionic system into local modes} \label{sec: fermions}

In this section, we discuss the multipartition of a fermionic system into local modes. Its main interest is that it is an instance of a multipartition that is \textit{not} fully representable in the sense of Definition \ref{def: representable}. That the failure of full representability arises in such a basic situation demonstrates, in particular, that it should not be seen as an issue affecting the formalism of multipartitions, but rather as a standard property of it.

All the proofs for \correction{this section} are presented in Appendix \ref{app: fermions}.

\subsection{The algebra of physical operators on a fermionic system}

Let $N$ be a non-zero natural number and $X = \{1,...,N\}$. The Fock space corresponding to an $N$-mode fermionic system is the Hilbert space $\mathcal{H}$ generated by the $2^N$ vectors $\ket{n_1 n_2 ... n_N}$ where $n_i \in \{0,1\}$. The annihilation and creation operators $a_i, a^{\dagger}_i$, associated to each mode $1 \leq i \leq N$, satisfy the canonical anti-commutation relations
\begin{subequations}
\begin{align}    
    a_ia_j + a_ja_i = 0 \,, \\
    a^{\dagger}_ia^{\dagger}_j + a^{\dagger}_ja^{\dagger}_i = 0 \,,\\
    a_ia^{\dagger}_j + a^{\dagger}_ja_i = \delta_{ij} \, ,
\end{align}
\end{subequations}
and span the algebra $\Lin(\ch)$. Note that $\Lin(\ch)$ includes operators that do not satisfy the super-selection rule of a fermionic system (i.e.\ do not preserve the parity of the number of fermions) and is thus not the physical algebra. To determine the physical operators on a certain subset of modes $S \subseteq X$, we define 
\begin{equation}
    \forall S \subseteq X, \quad \mathcal{M}_S := \{a_ia_j, a_ia^{\dagger}_j, a^{\dagger}_ia_j, a^{\dagger}_ia^{\dagger}_j | i,j \in S\} \,,
\end{equation}
and $\OM := \textrm{Span}(\mathcal{M}_X)$ (where Span denotes the algebraic span), the algebra of physical operators (the ones that preserve the parity of the number of fermions) on our $N$-mode fermionic system; while for $S \subset X$, we define $\mathcal{A}_{S} = \textrm{Span}(\mathcal{M}_S)$, the subalgebra of $\OM$ of the physical operators acting on the modes of $S$. For instance, we have
\begin{subequations}
\begin{align}
    \mathcal{A_{\emptyset}} &= \mathbb{C}\id \,,\\
    \mathcal{A}_{\{1\}} &= \textrm{Span}(\{\id , a^{\dagger}_1a_1, a_1a^{\dagger}_1\}) \,, \\
    \mathcal{A}_{\{1,2\}} &= \textrm{Span}(\{\id , a^{\dagger}_1a_1, a_1a^{\dagger}_1, a^{\dagger}_2a_2, a_2a^{\dagger}_2, a_1a_2, a^{\dagger}_1a^{\dagger}_2, a_1^{\dagger}a_2, a_1a^{\dagger}_2\}) \, ,
\end{align}
\end{subequations}
etc.

Before presenting our main result, we characterise the centres of the $\mathcal{A}_S$ algebras and their atomic projectors.

\begin{proposition} \label{prop: atomic projectors fermions}
    Let $\emptyset \neq S \subseteq X$. The atomic projectors of $\cz(\ca_S)$ are $\{\pi_S, \id - \pi_S\}$ where, denoting $n_i := a_i^\dag a_i$,
    \begin{equation}
    \pi_S = \sum_{\emptyset \neq Y \subseteq S} (-2)^{|Y|-1} \prod_{i \in Y} n_i \,.
\end{equation}
\end{proposition}

As shown in Appendix \ref{app: proof atomic projectors fermions}, \correction{these projectors} have an intuitive meaning: $\pi_S$ is the projector on the `odd number of fermions in the modes of S' sector, and $\id - \pi_S$ is the projector on the `even number of fermions in the modes of S' sector. We see that the reason $\mathcal{Z}(\ca_S)$ is non-trivial is that operators of $\mathcal{A}_S$ are required to satisfy the super-selection rule on the modes of $S$, which precisely means commuting with $\pi_S$.

The $\pi_S$'s can also be computed with the following method.

\begin{proposition} \label{prop: conjunctions of centres fermions}
    Let $S$ and $T$ be two disjoint subsets of $X$; the atomic projectors of $\cz(\ca_{S \sqcup T})$ can be computed from those of $\cz(\ca_S)$ and $\cz(\ca_T)$ by
    \begin{equation}
       \pi_{S \sqcup T} = \pi_S + \pi_T - 2 \pi_S \pi_T \, .
    \end{equation}
\end{proposition}

Following the same intuition, this proposition translates the fact that if $S$ and $T$ are two disjoint sets of modes, the parity of the number of fermions on the modes of $S \sqcup T$ is the sum (modulo 2) of the parity of the number of fermions in the modes of $S$ and the modes of $T$, respectively. In particular the superselection on the system of modes $S \sqcup T$ is weaker than the combination of the two superselections on systems $S$ and $T$ respectively, \correction{as it also allows for the operators that change the parity of the number of fermions in both the modes of $S$ and the modes of $T$}: this is yet another instance of FOLT.\footnote{See e.g.\ Ref.\ \cite{D'Ariano_2014} for a discussion of FOLT in fermionic systems.}

Reasonably enough, the partition of a fermionic system into local modes defines a sound partition.

\begin{proposition} \label{prop: partition fermions}
    \correction{Let $N \in \mathbb{N}^*$ and $X = \{1,...,N\}$, then $(\mathcal{A}_S)_{S \subseteq X} \vdash \OM = \ca_X$.}
\end{proposition}

\subsection{Representations of the algebra of physical operators}
The partition of a fermionic system into local modes is not fully representable.
\begin{proposition}\label{prop: non-representable fermions}
    For $N \geq 3$, the partition $(\mathcal{A}_S)_{S \subseteq X} \vdash \OM$ is not fully representable.
\end{proposition}

We want to emphasise that this result is \textit{not} about the (rather trivial) fact that, in order to ensure anticommutation between an $a_i$ and an $a_j$, at least one of them has to be represented as acting on the Hilbert space of the other. Indeed, an $a_i$ by itself is not a physical operator, and thus not an element of $\ca_i$ (or even of $\OM$). What is shown here is that \textit{even when restricting to physical operators}, at least some of them have to be represented as acting on Hilbert spaces outside of the modes they are affecting.

% \correction{This fact connects with results in the existing literature. For instance, Ref.~\cite{friis2013fermionic} showed an explicit failure (for three modes) of consistent density-matrix mappings from fermionic space to qubits, established via a sign-consistency argument, and Ref.~\cite{guaita2024locality} proved that an exact, fully local fermion-to-qubit encoding exists if and only if the associated locality graph is a tree, which forbids it for any three mutually-coupled modes.\footnote{Our framework involves no notion of a locality graph, so that any pair of fermionic modes can \textit{a priori} be coupled. See the notion of connectedness in Ref.~\cite{causaldecs} for an incorporation of some locality notions within our framework.} Relatedly, Ref.~\cite{mlodinow2020quantum}, focusing on dynamics rather than states, proved a no-go theorem for strictly local realizations of parity-preserving fermionic dynamics on lattices of dimension greater than one, evaded in only by embedding in a much larger Hilbert space. None of these results is stated in our C*-algebraic language. Although a formal connection is lacking, it is reasonable to expect that this failure of state- or dynamics-encodings is tightly linked with the incapacity to locally represent the C* subalgebras that these density matrices or  belong to.}

\correction{The obstruction exhibited by Proposition \ref{prop: non-representable fermions} is not new, although it appears here in a form adapted to our notion of full representability. Several closely related results have appeared previously. Refs.~\cite{friis2013fermionic, friis2016reasonable} show, at the level of fermionic states and reduced states, that no fermion-to-qubit isomorphism can in general preserve the subsystem structure for all mode bipartitions simultaneously; and more generally, Ref.~\cite{guaita2024locality} considers locality-preserving encodings of fermionic observables on a graph and proves that such an encoding exists if and only if the graph is a tree.

Our Proposition \ref{prop: non-representable fermions} is the corresponding statement for the multipartition of all $N$ fermionic modes: full representability would in particular provide a locality-preserving encoding of the two-mode physical operators on the complete locality graph,\footnote{Our framework involves no notion of a locality graph, or of any spatial structure, so that we have \textit{a priori} all-to-all connectivity. See the notion of connectedness in Ref.~\cite{causaldecs} for an incorporation of some considerations of spatial structure within our framework.}  which is impossible for $N \geq 3$ by the result of Ref.~\cite{guaita2024locality}. We nevertheless give a proof in our language because it makes explicit the precise obstruction to full representability and connects directly with the general representation theorem of Section \ref{sec: representations}.}

% This fact is in itself not new: it has recently appeared in the literature in different guises --- see e.g.\ Refs.\ \cite{friis2013fermionic, friis2016reasonable, mlodinow2020quantum, guaita2024locality}. These references proved obviously related but technically disjoint statements; \correction{Proposition \ref{prop: non-representable fermions}}, which we proved using our own language for clarity, states the form of the result that matches our framework.

The failure of full representability, therefore, does not mean that our framework captures non-physical examples: on the contrary, such a failure is a common feature of quantum theory, arising in natural physical theories, such as that of fermionic systems. This should not be surprising since full representability concerns not the morphological and relational features of a partition --- or in other words, its algebraic structure --- but only our ability to frame it in terms of Hilbert spaces. A non-fully representable partition is perfectly well-behaved at the algebraic level. The conclusion is simply that the formalism of Hilbert spaces sometimes fails to capture this good behaviour, and should therefore be taken with a grain of salt.\footnote{That Hilbert spaces can betray us is a common theme in infinite dimension, where they are routinely traded for an algebraic picture, essentially due to convergence issues. The case at hand shows that a point \correction{leading to the same conclusion} can be made even in finite dimension, where the use of Hilbert spaces can lead to pseudo-nonlocality.}

\section{Conclusion} \label{sec: conclusion}

Our contributions can be grouped into four classes. First, we made a case for the inclusion of non-factor subsystems in a decompositional approach to quantum theory, and discussed their operational meaning. Second, we motivated and put forward a theory of partitions of a quantum system into any number of parts. Third, we investigated the structural properties of this theory, in particular the well-structured behaviour of centres and its physical meaning. Fourth, we demonstrated how, using the framework of routed quantum circuits, algebraic partitions can be represented in terms of Hilbert spaces; but also how, for some partitions, this representation necessarily involves a residual amount of pseudo-nonlocality in the actions on joint parts, with the display of a concrete example of such a partition.

A remarkable payoff of this work is how it shows that, when non-factorness is involved, partitions display a strikingly rich mathematical structure. It would be surprising if this rich mathematical structure did not have a counterpart in a rich physical structure. Thus, we can reasonably hope that our framework holds potential for discovering as well as elucidating important physical facts in more concrete cases, especially in the presence of symmetries, and in particular in the fields we mentioned in the introduction. The fact that our framework is a central component of a proof of causal decomposability, elucidating the structural features of certain quantum causal structures \cite{causaldecs}, is a first demonstration of its practical use.

We only provided a theory of partitions into `eternal subsystems', in the sense that there was no attempt here to consider `$A$ at $t_1$' as a different subsystem than `$A$ at $t_2$'. Yet, there is reasonable conceptual ground to hold that these constitute different loci of intervention. Following this cue would lead to a theory of `time-thin subsystems' --- which should more aptly be called `events' ---, a prospect which we think holds both great potential for discoveries, and major technical challenges. In particular, this would be the proper way to frame the `spacetime repartitions' that we hinted at in our brief introductory discussion of the Quantum Switch controversy, and to bridge with the framework of time-delocalised subsystems \cite{oreshkov2019time, Wechs2022, wechs2024}. This might also connect with the argument made in Ref.\ \cite{Franzman2024} that partitions should be evolving in time.

Another interesting direction for future research would be to explore whether there exists an algebraic characterisation of fully representable partitions: are there some algebraic conditions on a partition (requesting e.g.\ commutation between the algebras corresponding to various parts) that are satisfied if and only if the partition is fully representable? This would provide an enlightening interpretation of the existence of an underlying Hilbert space structure as equivalent to specific morphological properties. Despite our best efforts, however, we have so far not been able to pin down such a characterisation. It is possible that none exists, which would entail that `good behaviour' at the Hilbert space level has no meaning at the morphological level.

Finally, it would be important to lift our self-imposed restraint from entering too much the realm of operational justifications, for instance through seeing how the ideas of Ref.~\cite{Ormrod2025} could be extended to justify our definition of multipartitions. More generally, investigating how this paper's ideas can or cannot extend to operational frameworks, in which subsystems are defined through subsets of quantum \textit{transformations} and the commutation thereof \cite{Kraemer2017, chiribella2018}, would be important. An idea to bring in further operational considerations while keeping the structural power of algebraic approaches would be to investigate whether the requirement of commutation between subalgebras could be replaced with \textit{commutation up to a global phase}, since the latter is operationally meaningless.

\section*{Acknowledgements}
It is a pleasure to thank Alexei Grinbaum, Robin Lorenz, Nick Ormrod, and Tein van der Lugt for helpful discussions and comments. Special thanks go to Ved Kunte for mentioning in passing Ref.\ \cite{friis2016reasonable} and putting an end to months of head-scratching about the failure of full representability.

AV is supported by the STeP2 grant (ANR-22-EXES-0013) of Agence Nationale de la Recherche (ANR), the PEPR integrated project EPiQ (ANR-22-PETQ-0007) as part of Plan France 2030, the ANR grant TaQC (ANR-22-CE47-0012), and the ID \#62312 grant from the John Templeton Foundation, as part of the \href{https://www.templeton.org/grant/the-quantum-information-structure-of-spacetime-qiss-second-phase}{‘The Quantum Information Structure of Spacetime’ Project (QISS)}. OM and PA are partially funded by the European Union through the MSCA SE project QCOMICAL, by the French National Research Agency (ANR): projects TaQC ANR-22-CE47-0012 and within the framework of `Plan France 2030', under the research projects EPIQ ANR-22-PETQ-0007, OQULUS ANR-23-PETQ-0013, HQI-Acquisition ANR-22-PNCQ-0001 and HQI-R\&D
ANR-22-PNCQ-0002, and by the ID \#62312 grant from the John Templeton Foundation, as part of the \href{https://www.templeton.org/grant/the-quantum-information-structure-of-spacetime-qiss-second-phase}{‘The Quantum Information Structure of Spacetime’ Project (QISS)}. The opinions expressed in this publication are those of the authors and do not necessarily reflect the views of the John Templeton Foundation.

\section*{Author contribution statement}
All authors contributed to the research and preparation of the manuscript. LLMs were used solely in the final revision prior to publication, to assist the authors in copyediting the manuscript and bibliography.

\bibliographystyle{utphys}
\bibliography{refs}

\appendix
\section{Proofs}

\subsection{Proof of Theorem \ref{th: commutative algs}} \label{app: proof commutative algs}

\begin{proof}
    Let $\cz$ be a commutative algebra. By Theorem \ref{th: WA}, there exists an isomorphism $\iota : \cz \to \bigoplus_{k \in K} \Lin(\ch_Z^k)$, and each of the $\ch_Z^k$'s is necessarily one-dimensional, as matrices over dimension at least 2 are not commutative. Defining $\pi^k:= \iota\inv(1_{Z^k} \oplus (\bigoplus_{k'\neq k} 0_{Z^{k'}}))$, it is then straightforward to check that the $\pi^k$'s are pairwise orthogonal and non-null and form a basis of $\cz$.

    For uniqueness, suppose there exists another family $(\mu^l)_{l \in L}$ satisfying the required properties. Since the $\pi^k$'s form a basis of $\cz$, we have for every $l$, $\mu^l = \sum \al_{kl} \pi^k$: orthogonality of the $\mu^l$'s then leads to $\alpha_{kl} \in \{0, 1\}$, and pairwise orthogonality to $l \neq l' \implies \alpha_{kl} \alpha_{k l'} = 0$. Since the $\mu^l$'s are, as a basis, as many as the $\pi^k$ and non-null, the only possibility is that there exists a bijection $f: K \to L$ such that $\mu^{f(k)} = \pi^{k}$, so that the families are the same up to relabelling.
\end{proof}

\subsection{Proofs for Section \ref{sec: properties}} \label{app: properties}

We start with the proof of Proposition \ref{prop: partitions and commutants}.

\begin{proof}
    Let us first note that for any $\cf \subseteq \OM$, $\cz(\OM) \vee \cf = \biguplus_k \pi^k \cf$, where the $\pi^k$'s are $\cz(\OM)$'s atomic projectors and $\uplus$ denotes the linear span of the union. Indeed, since the $\pi^k$'s linearly generate $\cz(\OM)$, any element of $\cz(\OM) \vee \cf$ can be written as $\sum \pi^k f^k$ with the $f^k$'s in $\cf$.

    Using this, we have $\ca_2 \subseteq \cz(\OM) \vee \ca_2 = \biguplus_k \pi^k \ca_2 = \biguplus_k \pi^k \ca_1' = \cz(\OM) \vee \ca_1' = \ca_1'$, with the last equality stemming from the fact that $\cz(\OM) \subseteq \ca_1'$.
\end{proof}

The proof of Proposition \ref{prop: centre Z_1} is then the following.

\begin{proof}
Using Proposition \ref{prop: partitions and commutants}, the first equality is direct from the fact that $\cz_1 = \ca_1 \cap \ca_1'$. For the equality $\cz_1 = \ca_1 \cap (\cz(\OM) \vee \cz_2)$, direct inclusion comes from the fact that $\cz_1 \subseteq \ca_1$ by definition and $\cz_1 \subseteq \cz(\OM) \vee \cz_2$ (the latter is proven in Ref.\ \cite{causaldecs}'s Proposition B.5), and the reverse inclusion from the fact that $\ca_1 \cap (\cz(\OM) \vee \cz_2) \subseteq \ca_1 \cap (\cz(\OM) \vee \ca_2) =\cz_1$.
\end{proof}

Finally, this allows us to prove Proposition \ref{prop: Z from individual Zs}.

\begin{proof}
   \correction{We start with the reverse inclusion. Fixing $n \in S$, we have $\cz_n \subseteq \ca_n \subseteq \ca_S$. On the other hand, for $n \not\in S$, $\cz_n \subseteq \ca_n \subseteq \ca_{\bar{S}}$, and by Proposition \ref{prop: partitions and commutants} applied to $(\ca_S, \ca_{\bar{S}}) \vdash \OM$, $\ca_{\bar{S}} \subseteq \cz(\OM) \vee \ca_{\bar{S}} = \ca_S'$ so that $\cz_n \subseteq \ca_S'$. We thus find that the RHS of (\ref{eq: Z from individual Zs}) is included in $\ca_S \cap \ca_S' = \cz_S$.} For the direct inclusion, a straightforward induction on (\ref{eq: def bipart centre}) yields that $\cz_S \subseteq \bigvee_{n \in S} \cz_n$; while from Proposition \ref{prop: centre Z_1}, we have $\cz_S \subseteq \cz_{\bar{S}} \vee \cz(\OM) \subseteq (\bigvee_{n \in \bar{S}} \cz_n) \vee \cz(\OM)$.
\end{proof}

\subsection{A Lemma for Theorem \ref{th: computing centres}}

We fix a partition $(\ca_S)_{S \subseteq X} \vdash \OM$ and an $S$. We denote $\tilde{K}_S := \{(k_n)_{n \in S} | \prod_n \pi_n^{k_n} \neq 0\} \subseteq K_S := \bigtimes_{n \in S} K_n$.

\correction{Let us denote $\bar{Q}_S$ as a set of labels for $\Atproj(\cz(\ca_S))$. Using (\ref{eq: def bipart centre}) and a straightforward induction, we have $\cz_S \subseteq \bigvee_{n \in S} \cz_n$. Thus, for a given $q_S \in \Qbar_S$, we can write the corresponding $\pi_S^{q_S} \in \Atproj(\cz_S)$ as a
\be \label{eq: computing centres proof} \pi_S^{q_S} = \sum_{\veck \in F_{q_S}} \prod_{n \in S} \pi_n^{k_n} \ee
for a certain $F_{q_S} \subseteq \tK_S$. The conditions $\pi_S^{q_S} \pi_S^{q_S'} = 0$ for $q_S \neq q_S'$ and $\sum_{q_S} \pi_S^{q_S} = \id$ imply that the $F_{q_S}$'s form a partition of $\tK_S$; they thus correspond to an equivalence relation on it, which we extend to a partial equivalence relation $\approx_S$ on $K_S$. }

Let us start with a Lemma characterising $\approx_S$.

\begin{lemma} \label{lem: eta as cross blocks}
    Given $\Vec{k}, \Vec{k}' \in K_S$, 
    \be \label{eq: eta as cross blocks} \Vec{k} \approx_S \Vec{k}' \quad \iff \quad \exists f \in \ca_S, \, \left(\prod_n \pi_n^{k_n'} \right) f \left(\prod_n \pi_n^{k_n} \right) \neq 0 \, . \ee
\end{lemma}

\begin{proof}
    Suppose $\Vec{k} \approx_S \Vec{k}'$; then they are in the same equivalence class corresponding to a certain $\pi_S^{q_S} \in \Atproj(\cz_S)$ for some $q_S \in \Qbar_S$, and $\pi_S^{q_S} \ca_S$ is a factor. Up to an isomorphism, it is thus a $\Lin(\ch)$ for some Hilbert space $\ch$, on which $\prod_n \pi_n^{k_n}$ and $\prod_n \pi_n^{k_n'}$ define two non-null subspaces, and there exists a non-null $f \in \pi_S^{q_S} \ca_S$ mapping from one to the other and thus satisfying the RHS of (\ref{eq: eta as cross blocks}).

    On the other hand, suppose $\Vec{k} \not\approx_S \Vec{k}'$; then either at least one of them is not in $\tK_S$, in which case the RHS of (\ref{eq: eta as cross blocks}) cannot hold; or they are in two different equivalence classes, corresponding respectively to $\pi_S^{q_S}, \pi_S^{q_S'} \in \Atproj(\cz_S)$. In this case we have, for any $f \in \ca_S$,
    \be
    \begin{split}
        \left(\prod_n \pi_n^{k_n'} \right) f \left(\prod_n \pi_n^{k_n} \right) &= \left(\prod_n \pi_n^{k_n'} \right) \pi_S^{q_S'} f \pi_S^{q_S} \left(\prod_n \pi_n^{k_n} \right) \\
        &= \left(\prod_n \pi_n^{k_n'} \right) f \pi_S^{q_S'} \pi_S^{q_S} \left(\prod_n \pi_n^{k_n} \right) \\
        &= 0 \, . 
    \end{split}
    \ee  \qedhere
\end{proof}

\subsection{Proof of Theorem \ref{th: computing centres}} \label{app: computing centres}

\correction{Using the notations of the previous subsection, our goal is to prove that $\sim_S = \approx_S $.} We take $\Vec{k}, \Vec{k}' \in K_S$ and first suppose $\Vec{k} \not\sim_S \Vec{k}'$. Then for any $f \in \ca_S$, we have
\be \label{eq: comp proof form of centres}
\begin{split}
    \left(\prod_{n \in S} \pi_n^{k_n'} \right) f \left(\prod_{n \in S} \pi_n^{k_n} \right) &= \sum_{\Vec{l} \in \bigtimes_{n \not\in S} K_n} \left(\prod_{n \not\in S} \pi_n^{l_n} \right) \left(\prod_{n \in S} \pi_n^{k_n'} \right) f \left(\prod_{n \in S} \pi_n^{k_n} \right) \\
    &= \sum_{\Vec{l} \in \bigtimes_{n \not\in S} K_n} \left(\prod_{n \not\in S} \pi_n^{l_n} \right) \left(\prod_{n \in S} \pi_n^{k_n'} \right) f \left(\prod_{n \not\in S} \pi_n^{l_n} \right) \left(\prod_{n \in S} \pi_n^{k_n} \right) \, ,
\end{split}
\ee 
where we used the fact that the $\prod_{n \not\in S} \pi_n^{l_n}$'s form a partition of the identity, then the fact that, as elements of $\ca_{\Bar{S}}$, they commute with elements of $\ca_S$. Using Lemma \ref{lem: eta as cross blocks} applied to $S = X$ (in that case we have $\approx_X = \sim_X$ by definition) yields that (\ref{eq: comp proof form of centres}) is null and thus, by Lemma \ref{lem: eta as cross blocks}, that $\Vec{k} \not\approx_S \Vec{k}'$.

Conversely, suppose $\Vec{k} \sim_S \Vec{k}'$. We can then fix $\Vec{l} \in \bigtimes_{n \not\in S} K_n$ such that $(\Vec{k}, \Vec{l}) \sim_X (\Vec{k}', \Vec{l})$. This means that $\left(\prod_{n \in S} \pi_n^{k_n} \right) \left(\prod_{n \not\in S} \pi_n^{l_n} \right)$ and $\left(\prod_{n \in S} \pi_n^{k_n'} \right) \left(\prod_{n \not\in S} \pi_n^{l_n} \right)$ are both non-null and lesser or equal, as projectors, to a certain $\pi \in \Atproj(\cz(\OM))$. Since, by Proposition B.5 of Ref.\ \cite{causaldecs}, $\cz(\ca_S) \subseteq \cz(\OM) \vee \cz_{X \setminus S} \subseteq \cz(\OM) \vee \left( \bigvee_{n \not\in S} \cz_n \right)$, $\pi \left(\prod_{n \not\in S} \pi_n^{l_n} \right)$ is lesser or equal than a certain $\mu \in \Atproj(\cz(\ca_S))$.

$\mu \left(\prod_{n \in S} \pi_n^{k_n} \right)$ and $\mu \left(\prod_{n \in S} \pi_n^{k'_n} \right)$ are then non-null (as they are greater or equal, respectively, to the non-null $\pi \left(\prod_{n \not\in S} \pi_n^{l_n} \right) \left(\prod_{n \in S} \pi_n^{k_n} \right)$ and $\pi \left(\prod_{n \not\in S} \pi_n^{l_n} \right) \left(\prod_{n \in S} \pi_n^{k_n'} \right)$), and $\mu \ca_S$ is a factor; therefore there exists a $f \in \mu \ca_S$ such that $\mu \left(\prod_{n \in S} \pi_n^{k_n} \right) f \mu \left(\prod_{n \in S} \pi_n^{k_n'} \right) \neq 0$. Since $f \in \mu \ca_S$, $ \mu f = f$; and in addition $\mu$ commutes with all elements of $\ca_S$, so we can rewrite this as $\left(\prod_{n \in S} \pi_n^{k_n} \right) f \left(\prod_{n \in S} \pi_n^{k_n'} \right) \neq 0$, with $f \in \mu \ca_S \subseteq \ca_S$. Therefore, Lemma \ref{lem: eta as cross blocks} yields that $\Vec{k} \approx_S \Vec{k}'$.

\correction{Thus $\Vec{k} \sim_S \Vec{k}' \iff \Vec{k} \approx_S \Vec{k}'$, so $\sim_S = \approx_S$. Therefore $\sim_S$ is a partial equivalence relation, and (up to relabelling) the set of labels $Q_S$ for its equivalence classes is equal to ${\bar{Q}}_S$, and each of its equivalence classes $E_{q_S}$ is equal to $F_{q_S}$, so that (\ref{eq: computing centres proof}) yields (\ref{eq: computing centres}).}

\subsection{Proof of Theorem \ref{th: representation}} \label{app: proof representation}

We will use the following lemmas.

\begin{lemma}\label{lem: homos on factors}
    Let $\ca$ be a factor. A homomorphism of C*-algebras with domain $\ca$ is either null or injective.
\end{lemma}

\begin{proof}
    This is a direct consequence of Ref.\ \cite{causaldecs}'s Proposition B.1 when applied to a homomorphism with a factor domain.
\end{proof}

\correction{\begin{lemma}\label{lem: commuting with all parts}
    Let $(\ca_S)_{S \subseteq X} \vdash \OM$. Then
    \be \label{eq: commuting with all parts} \left(\bigvee_{n \in X} \ca_n \right)' = \bigvee_{n \in X} \cz(\ca_n) \, .\ee
\end{lemma}

\begin{proof}
    The proof is by induction on the size of $X$. For the base case $X = \emptyset$, we have by (\ref{eq: def partition empty}) that $\OM$ is the trivial algebra so that (\ref{eq: commuting with all parts}) trivially holds. Let us now suppose the lemma holds up to a size $N$ and take $X$ of size $N+1$.
    
    For the reverse inclusion, any $\cz(\ca_n)$ commutes with $\ca_n$ by assumption, and with the other $\ca_m$'s by $\cz(\ca_n) \subseteq \ca_n \subseteq \ca_m'$. We now focus on the forward one. We denote (\ref{eq: commuting with all parts})'s LHS as $\cf$; it is direct to see that $\cf = \bigcap_n \ca_n'$.  Taking an arbitrary element $0 \in X$, we therefore have $\cf \subseteq \ca_0' = \cz(\OM) \vee \ca_{X \setminus \{0\}} \subseteq \cz(\ca_0) \vee \ca_{X \setminus \{0\}}$ by Proposition \ref{prop: partitions and commutants} then (\ref{eq: def bipart centre}), so that
    \be \cf \subseteq \left(\cz(\ca_0) \vee \ca_{X \setminus \{0\}}\right) \cap \left( \bigcap_{n \neq 0} \ca_n' \right) \, . \ee

    Furthermore, $\cz(\ca_0)$ is included in the RHS here, since it is included in $\ca_0 \subseteq \ca_n' \forall n \neq 0$; and it also commutes with that RHS since it commutes with $\cz(\ca_0)$ (i.e.\ itself) and with $\ca_{X \setminus \{0\}}$. $\cz(\ca_0)$ is therefore contained in the RHS' centre. Denoting the atomic projectors of $\cz(\ca_0)$ as $\pi^k$'s, we have by Ref.~\cite{causaldecs}'s Lemma B.3,
    \be \label{eq: proof commuting with all parts} \begin{split}
        \cf \subseteq \left(\cz(\ca_0) \vee \ca_{X \setminus \{0\}}\right) \cap \left( \bigcap_{n \neq 0} \ca_n' \right)  &= \biguplus_k \pi^k \left( \left(\cz(\ca_0) \vee \ca_{X \setminus \{0\}}\right) \cap \left( \bigcap_{n \neq 0} \ca_n' \right) \right) \\
        &= \biguplus_k  \left(\pi^k \ca_{X \setminus \{0\}}\right) \cap \left( \bigcap_{n \neq 0} \ca_n' \right) \, .
    \end{split}  \ee

    We fix a $k$ and an element $1 \in X$ and apply Ref.~\cite{causaldecs}'s Lemma D.5 with $\cf_1 = \ca_1$, $\cf_2 = \ca_{X \setminus \{0\}}$, and $\cg = \ca_0'$. $\pi^k \in \cz(\ca_0')$, so that the $\mu$ in this lemma, defined as the minimal projector in $\cz(\ca_0')$ such that $\pi^k \mu = \pi^k$, is equal to $\pi^k$. Thus this lemma yields
    \be \left( \pi^k \ca_\zerobar \right) \cap \left( \bigcap_{n \neq 0} \ca_n' \right) = \pi^k \left( \ca_\zerobar \cap \ca_1' \right) \cap \left( \bigcap_{n \neq 0, 1} \ca_n' \right) \, . \ee
    We can then apply Ref.~\cite{causaldecs}'s Lemma D.5  with $\cf_1 = \ca_2$, $\cf_2 = \ca_\zerobar \cap \ca_1'$, and repeat this operation to reach
    \be \begin{split}
        \left( \pi^k \ca_\zerobar \right) \cap \left( \bigcap_{n \neq 0} \ca_n' \right) &= \pi^k \left( \ca_\zerobar \cap \left( \bigcap_{n \neq 0} \ca_n' \right) \right) \\
        &= \pi^k \left( \ca_\zerobar \cap \left( \bigvee_{n \neq 0} \cz(\ca_n) \right) \right)  \\
        &\subseteq \bigvee_n \cz(\ca_n) \, ,
    \end{split}  \ee
    where the middle equality comes from our induction hypothesis. Combined with (\ref{eq: proof commuting with all parts}), this yields the inclusion we had set out to prove and thus the induction step.
\end{proof}}

We also denote, for every $\ca \subseteq \OM$ and orthogonal projector $\pi$, $\hat{\pi} \ca := \{\pi f \pi | f \in \ca\}$.

We take a partition $(\ca_S)_{S \subseteq X} \vdash \OM$, and denote $\Atproj(\cz(\ca_n)) = \{\pi_n^{k_n}\}_{k_n \in K_n}$ and $\Atproj(\cz(\OM)) = \{\mu^l\}_{l \in L}$.  For every $\veck := (k_n)_{n \in X} \in K$, we denote $\pi^{\Vec{k}} := \prod_{n\in X} \pi_n^{k_n}$. We recall that $\Tilde{K} = \{\Vec{k} | \sigma_{\veck} = 1\}= \{\Vec{k} | \pi^{\Vec{k}} \neq 0\}$. Without loss of generality, we can, by the Artin-Wedderburn theorem, take $\OM = \Lin_\delta(\bigoplus_{l \in L} \ch^l)$. The family $(\pi^\veck)_{\veck \in \Tilde{K}}$ then provides a sectorisation of $\ch$, which therefore can \correction{without loss of generality be assumed to decompose as} the corresponding direct sum $\ch = \bigoplus_{\veck \in \Tilde{K}} \ch^\veck$ so that
\be \label{eq: form of piveck OM} \forall \veck \in \Tilde{K}, \quad \hatpi^\veck (\OM) = \Lin(\ch^\veck) \oplus \left( \bigoplus_{\vecq \neq \veck} 0_{\ch^\vecq} \right) \, . \ee

Again by the Wedderburn-Artin theorem, for every $n \in X$, there exists a family of Hilbert spaces $(\ch_{A_n}^{k_n})_{k_n \in K_n}$ such that for all $k_n \in K_n$, $\pi_n^{k_n} \ca_n \cong \Lin (\ch_{A_n}^{k_n})$, and $\ca_n \cong \bigoplus_{k_n \in K_n} \Lin (\ch_{A_n}^{k_n})$. We fix a $\Vec{k} \in \Tilde{K}$. For every $n$, $\pi_n^{k_n} \ca_n$ is a factor, and, since $\pi^{\Vec{k}} \correction{= \pi_n^{k_n} \cdot \prod_{m\neq n} \pi_m^{k_m}}$ commutes with its elements \correction{(as $\pi_n^{k_n}$ does because it is in $\cz(\ca_n)$, and the other $\pi_m^{k_m}$'s do because each is in an $\ca_m \subseteq \ca_n'$)}, it is straightforward to prove that $f \mapsto \pi^{\Vec{k}} f$ defines a homomorphism \correction{from it to $\hatpi^\veck(\OM)$}, non-null as it maps $\pi_n^{k_n}$ to $\pi^{\Vec{k}} \neq 0$. By Lemma \ref{lem: homos on factors}, its image $\pi^{\Vec{k}} \ca_n$ is thus a factor, isomorphic as well to $\Lin(\ch_{A_n}^{k_n})$.

\correction{Let us define $\cf := (\bigvee_n \pi^\veck \ca_n)' \cap \hatpi^\veck(\OM)$. Fixing an $n$ and a $k_n'$, $\cf \subseteq (\pi^{k_n'} \ca_n)'$; indeed either $k_n \neq k_n'$, in which case the product between elements of $\cf$ and elements of $\pi^{k_n'} \ca_n$ is always null; or $k_n = k_n'$, in which case we have, for $f \in \cf, a \in \ca_n$, $f \pi^{k_n} a = f \pi^\veck \pi^{k_n} a = f \pi^\veck a = \pi^\veck a f= a \pi^\veck f = a \pi^{k_n} \pi^\veck f = a \pi^{k_n} f = \pi^{k_n} a f$. Since $\ca_n = \bigvee_{k_n} \pi^{k_n} \ca_n$ by Wedderburn-Artin, we find $\cf \subseteq \bigcap_n \ca_n' = \bigvee_n \cz_n$ by Lemma \ref{lem: commuting with all parts}, so we find that $\cf$ is commutative. }

We then consider the product map from $\bigotimes_{n \in X} \hatpi^{\Vec{k}}  \ca_n$ to $\hatpi^{\Vec{k}} (\OM)$ defined by $\bigotimes_{n \in X} f_n \mapsto \prod_n f_n$. Using the fact that the $\pi^{\Vec{k}}  \ca_n$'s commute \correction{(since the $\ca_n$'s commute pairwise and $\pi^\veck$ commutes with all of them)}, it is straightforward to prove that it is a non-null homomorphism of C*-algebras. Its domain, being a tensor product of factors, is a factor; it is thus injective by Lemma \ref{lem: homos on factors}. \correction{In addition, its image $\bigvee_n \pi^{\Vec{k}} \ca_n$ is a subfactor of the factor $\hat{\pi}^{\Vec{k}} (\OM)$, whose commutant within the latter is $\cf$, so $\cf$ must be a factor too; since we also know $\cf$ is commutative, it has to be trivial. Therefore, by the bicommutant theorem applied within the factor $\hat{\pi}^{\Vec{k}} (\OM)$, this image is the whole of $\hat{\pi}^{\Vec{k}} (\OM)$.}

The product map is therefore an isomorphism of C*-algebras. We thus find that 
\be \hatpi^{\Vec{k}} (\OM) \cong \bigotimes_{n} \pi^{\Vec{k}}  \ca_n \cong \bigotimes_{n} \Lin \left(\ch_{A_n}^{k_n} \right) \cong \Lin\left(\bigotimes_{n} \ch_{A_n}^{k_n}\right) \, , \ee 
with the corresponding isomorphism $\iota^\veck: \hatpi^{\Vec{k}} (\OM )\to \Lin\left(\bigotimes_{n} \ch_{A_n}^{k_n}\right)$ satisfying by construction
\be \iota^\veck \left(\pi^{\Vec{k}}  \ca_n \right) = \Lin(\ch_{A_n}^{k_n}) \otimes \left( \bigotimes_{m \neq n} \id_{A_m^{k_m}} \right) \, . \ee

Furthermore, (\ref{eq: form of piveck OM}) makes it clear that there also exists an isomorphism $\tilde{\iota}^\veck :  \Lin(\ch^\veck) \to \hatpi^{\Vec{k}} (\OM)$, given by $\tilde{\iota}^\veck(f) = f \oplus \bigoplus_{\vecq \neq \veck} 0_{\ch^\vecq}$. We thus have an isomorphism between factors $\iota^\veck \circ \tilde{\iota}^\veck : \Lin(\ch^\veck) \to \Lin\left(\bigotimes_{n} \ch_{A_n}^{k_n}\right)$, which, by Skolem-Noether, can be written as a conjugation $f \mapsto u^\veck f (u^\veck)\inv$ for a certain unitary map $u^\veck : \ch^\veck \to \bigotimes_n \ch_{A_n}^{k_n}$.  Defining the unitary map $u := \bigoplus_{\veck \in \tilde{K}} u^\veck$ from $\ch$ to $\bigoplus_{\veck \in \tilde{K}} \bigotimes_n \ch_{A_n}^{k_n}$, we obtain an isomorphism $\hat{u}: f \mapsto u f u\inv$ from $\Lin(\ch)$ to $\Lin(\bigoplus_{\veck \in \tilde{K}} \bigotimes_n \ch_{A_n}^{k_n})$ for which it is easy to compute that 
\be \label{eq: form of piveck A_n} \forall \veck \in \Tilde{K}, \forall n, \quad \hat{u}\left(\pi^\veck \ca_n\right) = \left( \Lin(\ch_{A_n}^{k_n}) \otimes \left( \bigotimes_{m \neq n} \id_{A_m^{k_m}} \right) \right) \oplus \left( \bigoplus_{\vecq \neq \veck} 0_{\ch^\vecq} \right) \, . \ee

\correction{
From now on we suppose that such an isomorphism has been applied, so that $\ch = \bigoplus_{\veck \in \tilde{K}} \bigotimes_n \ch_{A_n}^{k_n}$ and the $\pi^\veck \ca_n$'s are of the form of (\ref{eq: form of piveck A_n})'s RHS. This means that for any $\veck \in \tK$ and $n$, there exists an isomorphism $\mu_n^\veck : \Lin(\ch_{A_n}^{k_n}) \to \pi^\veck \ca_n$ consisting of the embedding in (\ref{eq: form of piveck A_n})'s RHS.

Furthermore, we also define $\ch^\ext := \bigotimes_n \ch_{A_n}$, where for every $n$, $\ch_{A_n}:= \bigoplus_{k_n \in K_n} \ch_{A_n}^{k_n}$. Its sectors can be rearranged as
\be
     \ch^\ext = \underbrace{\bigoplus_{\veck \in \tilde{K}} \bigotimes_{n} \ch_{A_n}^{k_n}}_{= \ch} \quad \oplus \quad \underbrace{\bigoplus_{\veck \in K \setminus \tilde{K}} \bigotimes_{n} \ch_{A_n}^{k_n}}_{=: \ch^\nul} \, ,
\ee
so we can see $\OM$ as living within $\Lin(\ch)$, itself living within $\Lin(\ch^\ext)$ as the set of the $f_\ch \oplus 0_{\ch^\nul}$'s.

We now fix an $n$ and a $k_n \in K_n$. For a $\vecq \in \tilde{K}$ such that $q_n = k_n$, we saw that the mapping $ \hatpi^\vecq|_{\pi^{k_n}_n \ca_n}: f \mapsto \pi^\vecq f$ from $\pi^{k_n}_n \ca_n$ to $\pi^\vecq \ca_n$ is an isomorphism by definition of $\pi^\vecq \ca_n$. Thus for any $\vecq, \vec{q'}$ with $q_n = q'_n = k_n$, $(\mu^{\vec{q'}}_n)\inv \circ \hatpi^{\vec{q'}}|_{\pi^{k_n}_n \ca_n} \circ (\hatpi^\vecq|_{\pi^{k_n}_n \ca_n})\inv \circ \mu^\vecq_n$ is a C* algebra automorphism of $\Lin(\ch_{A_n}^{k_n})$, which by the Skolem-Noether theorem can be written as a conjugation $f \mapsto v_n^{\vec{q'}, \vecq} f (v_n^{\vec{q'}, \vecq})^\dag$ for a certain unitary operator $v^{\vec{q'}, \vecq}_n \in \Lin (\ch_{A_n}^{k_n} )$.

Thus, fixing an arbitrary $\vecq$ with $q_n = k_n$, for an arbitrary $f \in \pi_n^{k_n} \ca_n$ we have (by (\ref{eq: form of piveck A_n}))
\be \pi^\vecq f = \left( \tilde{f} \otimes \left( \bigotimes_{m \neq n} \id_{A_m^{q_m}} \right) \right) \oplus \left( \bigoplus_{\vec{q''} \neq \vecq} 0_{\ch^{\vec{q''}}} \right) \, \ee
for a certain $\tilde{f} \in \Lin(\ch_{A_n}^{k_n})$, and the other $\pi^{\vec{q'}} f$'s are given by
\be \forall \vec{q'}, q_n'=k_n, \quad  \pi^{\vec{q'}} f = \left( v_n^{\vec{q'}, \vecq} \tilde{f} (v_n^{\vec{q'}, \vecq})^\dag \otimes \left( \bigotimes_{m \neq n} \id_{A_m^{q'_m}} \right) \right) \oplus \left( \bigoplus_{\vec{q''} \neq \vec{q'}} 0_{\ch^{\vec{q''}}} \right) \, . \ee
Defining the unitary
\be \label{eq: v_n} v_{n, k_n}:= \left( \bigoplus_{\vec{q'} \in \tilde{K}, q'_n=k_n} v^{\vec{q'}, \vec{q}}_n \otimes \left(\bigotimes_{m \neq n} \id_{A_m^{q'_m}} \right) \right) \oplus \left(\bigoplus_{\vec{q'} \in \tilde{K}, q'_n \neq k_n} \left(\bigotimes_{n} \id_{A_n^{q'_n}} \right) \right) \, , \ee
we then get
\be \forall \vec{q'}, q_n' = k_n, \quad (\hat{v}_{n, k_n})\inv (\pi^{\vec{q'}} f) = \left( \tilde{f} \otimes \left( \bigotimes_{m \neq n} \id_{A_m^{q'_m}} \right) \right) \oplus \left( \bigoplus_{\vec{q''} \neq \vec{q'}} 0_{\ch^{\vec{q''}}} \right) \, , \ee
so that 
\be \begin{split}
    (\hat{v}_{n, k_n})\inv(f) &= (\hat{v}_{n, k_n})\inv \left( \left( \sum_{\vec{q'} \in \tK} \pi^{\vec{q'}} \right) \pi_n^{k_n} f\right) \\
    &= (\hat{v}_{n, k_n})\inv \left( \left( \sum_{\substack{\vec{q'} \in \tilde{K}\\q_n'  = k_n}} \pi^{\vec{q'}} \right) \pi_n^{k_n} f\right) \\
    &=  \sum_{\substack{\vec{q'} \in \tilde{K}\\q_n'  = k_n}} (\hat{v}_{n, k_n})\inv \left(\pi^{\vec{q'}} f\right) \\
    &= \sum_{\substack{\vec{q'} \in \tilde{K}\\q_n'  = k_n}} \left(\left( \tilde{f}_{A_n} \otimes \left( \bigotimes_{m \neq n} \id_{A_m^{q'_m}} \right) \right) \oplus \left( \bigoplus_{\vec{q''} \neq \vec{q'}} 0_{\ch^{\vec{q''}}} \right) \right)\, . \end{split}\ee

    On the other hand, defining $\tilde{\bbpi}:= \sum_{\vec{q'} \in \tK} \pi^{\vec{q'}}$ (the projector of $\ch^\ext$ on $\ch$), we can compute
    \be \begin{split}
        \tilde{\bbpi} \left( \tilde{f}_{A_n} \otimes \bigotimes_{m \neq n} \id_{A_m}\right) &= \sum_{\vec{q'} \in \tK} \pi^{\vec{q'}} \pi_n^{k_n} \left( \tilde{f}_{A_n} \otimes \bigotimes_{m \neq n} \id_{A_m}\right) \\
        &= \sum_{\substack{\vec{q'} \in \tilde{K}\\q_n'  = k_n}} \pi^{\vec{q'}} \pi_n^{k_n} \left( \tilde{f}_{A_n} \otimes \bigotimes_{m \neq n} \id_{A_m}\right) \\
        &= \sum_{\substack{\vec{q'} \in \tilde{K}\\q_n'  = k_n}} \pi^{\vec{q'}} \left( \tilde{f}_{A_n} \otimes \bigotimes_{m \neq n} \id_{A_m}\right)  \\
    &= \sum_{\substack{\vec{q'} \in \tilde{K}\\q_n'  = k_n}} \left(\left( \tilde{f}_{A_n} \otimes \left( \bigotimes_{m \neq n} \id_{A_m^{q'_m}} \right) \right) \oplus \left( \bigoplus_{\vec{q''} \neq \vec{q'}} 0_{\ch^{\vec{q''}}} \right) \right)\, ,
    \end{split} \ee
    so we find
    \be (\hat{v}_{n, k_n})\inv(f) = \tilde{\bbpi} \left( \tilde{f}_{A_n} \otimes \bigotimes_{m \neq n} \id_{A_m}\right) \ee 
    and at the level of sets,
    \be \label{eq: form of pi An} (\hat{v}_{n, k_n})\inv \left(\pi_n^{k_n} \ca_n \right) = \tilde{\bbpi} \left( \Lin(\ch_{A_n}^{k_n}) \otimes \bigotimes_{m \neq n} \id_{A_m}\right) \, .\ee 

    Thus, for fixed $n$ and $k_n$, applying the isomorphism $(\hat{v}_{n, k_n})\inv$ maps us to a representation in which $\pi_n^{k_n} \ca_n$ has the form of (\ref{eq: form of pi An})'s RHS. We can repeat this procedure for other values of $n$ and $k_n$ without spoiling the form obtained for previous values, because the $v_{n,k_n}$ commute pairwise: one can see from (\ref{eq: v_n}) that they act on different factors if their $n$-value is different, and on different sectors if their $k_n$-value is different.
    
    We now assume that these isomorphisms were applied, so all $\pi_n^{k_n} \ca_n$'s have the form of (\ref{eq: form of pi An})'s RHS. Then, writing $\biguplus$ for linear spans, we have
    \be \begin{split}
        \forall n, \quad \ca_n &= \biguplus_{k_n} \pi_n^{k_n} \ca_n \\
        &= \biguplus_{k_n} \tilde{\bbpi} \left( \Lin(\ch_{A_n}^{k_n}) \otimes \bigotimes_{m \neq n} \id_{A_m}\right) \\
        &= \tilde{\bbpi} \left( \left( \biguplus_{k_n} \Lin(\ch_{A_n}^{k_n}) \right) \otimes \bigotimes_{m \neq n} \id_{A_m}\right) \\
        &= \tilde{\bbpi} \left( \Lin_\delta(\ch_{A_n}) \otimes \bigotimes_{m \neq n} \id_{A_m}\right) \\
     &= \left\{ \left(f \otimes \left(\bigotimes_{m \neq n} \id_{A_m}\right) \right) \, \tilde{\bbpi} \,\,\, \middle| \,\,\,  f \in \Lin_{\delta}\left(\ch_{A_n}\right) \right\} \, ,
    \end{split} \ee
    which is the form (\ref{eq: rep individual}).
}

Let us now determine the form of the $\ca_S$'s for $S \subseteq X$. We fix $S$. First, we define (denoting $\bar{S} := X \setminus S$)
\be \forall (k_n)_{n \in S}, \quad C^{(k_n)_{n \in S}} := \left\{(k_n)_{n \in \bar{S}} \, \middle| \, (k_n)_{n \in X} \in \tilde{K} \right\} \, , \ee
and aim to prove that if $(k_n)_{n \in S} \sim_{S} (l_n)_{n \in S}$, then $C^{(k_n)_{n \in S}} = C^{(l_n)_{n \in S}}$. Indeed, let us suppose that there exists a $(k_n)_{n \in \bar{S}}$ that is an element of $C^{(k_n)_{n \in S}}$ but not of $C^{(l_n)_{n \in S}}$. By Lemma \ref{lem: eta as cross blocks}, there exists an $f \in \ca_S$ such that 
\be \label{eq: comp C^kn} \left(\prod_{n \in S} \pi_n^{l_n} \right) f \left(\prod_{n \in S} \pi_n^{k_n} \right) \neq 0 \, . \ee
Furthermore, $(k_n)_{n \in S} \sim_{S} (l_n)_{n \in S}$ means that there exists a $q_S \in Q_S$ such that $(k_n)_{n \in S}$, $(l_n)_{n \in S} \in E^{q_S}$ and thus $\prod_{n \in S} \pi_n^{k_n}, \prod_{n \in S} \pi_n^{l_n} \leq \pi_S^{q_S}$, so that the LHS of (\ref{eq: comp C^kn}) is in the factor $\pi_S^{q_S} \ca_S$. Since $\prod_{n \in \bar{S}} \pi_n^{k_n} \in \ca_{\bar{S}} \subseteq \ca_{S}'$, multiplication by it is a homomorphism on $\pi_S^{q_S} \ca_S$. \correction{This homomorphism maps $\prod_{n \in S} \pi_n^{k_n}$ (which is an element of $\pi_S^{q_S} \ca_S$) to $\prod_{n \in X} \pi_n^{k_n}$, which is non-null since $(k_n)_{n \in \bar{S}} \in C^{(k_n)_{n \in S}}$. Thus this homomorphism is non-null, and therefore} injective by Lemma \ref{lem: homos on factors}, so \correction{applying it to the LHS of (\ref{eq: comp C^kn})} we find
\be
    0 \neq \left(\prod_{n \in \bar{S}} \pi_n^{k_n}\right) \left(\prod_{n \in S} \pi_n^{l_n} \right) f \left(\prod_{n \in S} \pi_n^{k_n} \right) = 0 \, ,
\ee
where \correction{the first inequality comes from injectivity, and the second equality from the fact that $(k_n)_{n \in \bar{S}} \not\in C^{(l_n)_{n \in S}}$ and therefore  $\left(\prod_{n \in \bar{S}} \pi_n^{k_n}\right) \left(\prod_{n \in S} \pi_n^{l_n} \right) = 0$.} This yields a contradiction.

\correction{Thus for every $(k_n)_{n\in S}$ we can label $C^{(k_n)_{n \in S}} = C^{q_S}$, where $q_S \in Q_S$ is the label for $(k_n)_n$'s equivalence class under $\sim_S$. (Note that $C^{q_S}$'s for different values of $q_S$ may be equal.)} We then decompose $\ch$ as
\be \ch = \bigoplus_{q_S \in Q_S} \ch_{A_S}^{q_S} \otimes \ch_{A_{\bar{S}}}^{q_S} \, , \ee
where \begin{subequations}
    \be\ch_{A_S}^{q_S} = \bigoplus_{(k_n)_{n \in S} \in E^{q_S}} \bigotimes_{n \in S} \ch_{A_n}^{k_n} \, , \ee
    \be \ch_{A_{\bar{S}}}^{q_S} = \bigoplus_{(k_n)_{n \in \bar{S}} \in C^{q_S}} \bigotimes_{n \in \bar{S}} \ch_{A_n}^{k_n} \, . \ee
\end{subequations} 

We fix $q_S \in Q_S$ and $(k_n)_{n \in \bar{S}} \in C^{q_S}$, and denote $\nu:= \pi_S^{q_S} \left( \prod_{n \in \bar{S}} \pi_n^{k_n} \right)$. $\hat{\nu} \OM$ is then the (factor) subalgebra of operators acting only on $\ch_{A_S}^{q_S} \otimes \bigotimes_{n \in \bar{S}} \ch_{A_n}^{k_n}$. Since $\ca_S \subseteq \ca_{\bar{S}}'$, we have in particular, for every $n \in \bar{S}$, $\ca_S \subseteq \ca_n'$, so $\hat{\nu} \ca_S \subseteq \hat{\nu} \ca_n'$. Since each $\hat{\nu} \ca_n$ is the subalgebra of $\hat{\nu} \OM$ of operators of the form $\id_{A_S^{q_S}} \otimes \bigotimes_{m \in \bar{S} \setminus \{n\}} \id_{A_m}^{k_m} \otimes f_{A_n^{k_n}}$, this entails that $\hat{\nu} \ca_S$ only contains operators of the form $f_{A_S^{q_S}} \otimes \bigotimes_{n \in \bar{S}} \id_{A_n^{k_n}}$.

Let us prove that it is in fact exactly the factor algebra $\cf$ of operators of that form. Each of the $\hat{\nu} \ca_n$'s, for $n \in S$, is included in $\hat{\nu} \ca_S$, and is the set of operators of the form $\left( \bigoplus_{(k_n)_{n \in S} \in E^{q_S}} f^{k_n}_{A_n^{k_n}} \otimes \bigotimes_{m \in S \setminus \{n\}} \id_{A_m^{k_m}}\right) \otimes \bigotimes_{m \in \bar{S}} \id_{A_m^{k_m}}$. Therefore $\hat{\nu} \ca_S$ contains their algebraic span, which is the algebra $\cg$ of operators of the form $\left( \bigoplus_{(k_n)_{n \in S} \in E^{q_S}}  f^{(k_n)_{n \in S}}_{\bigotimes_{n \in S} A_n^{k_n}}\right) \otimes \bigotimes_{m \in \bar{S}} \id_{A_m^{k_m}}$. Noticing that $\cg$ is a maximal subalgebra of $\hat{\nu} \ca_S$, in the sense that $\cg' \cap \hat{\nu} \ca_S \subseteq \cg$, we can conclude that $\cf = \hat{\nu} \ca_S$ by using the following lemma.

\begin{lemma}
    Let $\cg \subseteq \ct \subseteq \OM$ be C*-algebras, with $\ct$ and $\OM$ factors, and $\cg$ maximal, meaning that $\cg' \subseteq \cg$. Then $\ct = \OM$.
\end{lemma}

\begin{proof}
    Using the fact that $\ca \subseteq \cb \implies \cb' \subseteq \ca'$, we have $\ct' \subseteq \cg' \subseteq \cg \subseteq \ct$; therefore $\ct' \subseteq \ct' \cap \ct = \cz(\ct)$, which is the trivial algebra since $\ct$ is a factor. Therefore, by the double commutant theorem, $\ct = \ct'' = \OM$.
\end{proof}

\correction{This lemma can be used because $\hat{\nu}\ca_S$ is a factor: indeed by definition of $\nu$ it is equal to $\left( \prod_{n \in \bar{S}} \pi_n^{k_n} \right) \pi_S^{q_S} \ca_S$, where $\pi_S^{q_S} \ca_S$ is a factor by definition and multiplication by $\left( \prod_{n \in \bar{S}} \pi_n^{k_n} \right)$ is a non-null homomorphism on it (as these projectors commute with $\ca_S$) and thus injective by Lemma \ref{lem: homos on factors}, so it maps a factor to another factor.}

We thus get 
\be
    \hat{\nu} \ca_S = \left\{ f \otimes \bigotimes_{n \in \bar{S}} \id_{A_n^{k_n}} \, \middle| \, f \in \Lin(\ch_{A_S}^{q_S}) \right\} \, ,\ee
and varying over the possible $(k_n)_{n \in \bar{S}} \in C^{q_S}$, this yields
\be \label{eq: decomposing pi A_S} \begin{split}
    \pi_S^{q_S} \ca_S &\subseteq \bigvee_{(k_n)_{n \in \bar{S}} \in C^{q_S}} \hat{\nu} \ca_S \\
    &= \bigvee_{(k_n)_{n \in \bar{S}} \in C^{q_S}} \left\{ f \otimes \bigotimes_{n \in \bar{S}} \id_{A_n^{k_n}} \, \middle| \, f \in \Lin(\ch_{A_S}^{q_S}) \right\} \\
    &= \left\{ \bigoplus_{(k_n)_{n \in \bar{S}} \in C^{q_S}} f^{(k_n)_{n \in \bar{S}}}\otimes \bigotimes_{n \in \bar{S}} \id_{A_n^{k_n}} \, \middle| \, \forall (k_n)_{n \in \bar{S}} \in C^{q_S}, f^{(k_n)_{n \in \bar{S}}} \in \Lin(\ch_{A_S}^{q_S}) \right\} \, ,
\end{split} \ee
\correction{where the first inclusion comes from the fact that any $\pi_S^{q_S} f \in \pi_S^{q_S} \ca_S$ satisfies $\pi_S^{q_S} f = (\sum_{(k_n)_{n \in \bar{S}} \in K_{\bar{S}}} \prod_n \pi_n^{k_n}) \pi_S^{q_S} f = \sum_{(k_n)_{n \in \bar{S}} \in C^{q_S}} \prod_n \pi_n^{k_n} \pi_S^{q_S} f = \sum_{(k_n)_{n \in \bar{S}} \in C^{q_S}} \nu f$ (the change of sum comes from the fact that other terms are null by definition of $C$).}

Furthermore, for any $(k_n)_{n \in \bar{S}} \in C^{q_S}$, multiplication by $\pi^\veck := \prod_{n \in \bar{S}} \pi_n^{k_n}$ is (by commutation) a homomorphism \correction{$\hatpi^\veck|_{\pi_S^{q_S} \ca_S}$} on $\pi_S^{q_S} \ca_S$. This homomorphism is non-null by definition of $C^{q_S}$, \correction{and $\pi_S^{q_S} \ca_S$ is a factor since $\pi_S^{q_S} \in \Atproj(\cz(\ca_S))$,} so by Lemma \ref{lem: homos on factors}, it is injective. Picking an arbitrary reference $(l_n)_{n \in \bar{S}} \in C^{q_S}$ and defining $\kappa^{(k_n)_{n \in \bar{S}}}:= \hatpi^\veck|_{\pi_S^{q_S} \ca_S} \circ (\hatpi^{\vec{l}}|_{\pi_S^{q_S} \ca_S})\inv$, \correction{as a composition of isomorphisms it is an isomorphism, so}
\be \label{eq: form of picaS} \begin{split}
    \pi_S^{q_S} \ca_S &\subseteq \left\{ \bigoplus_{(k_n)_{n \in \bar{S}} \in C^{q_S}} \kappa^{(k_n)_{n \in \bar{S}}}\left(f\right) \otimes \bigotimes_{n \in \bar{S}} \id_{A_n^{k_n}} \, \middle| \, f \in \Lin(\ch_{A_S}^{q_S}) \right\} \\
    &= \left\{ \bigoplus_{(k_n)_{n \in \bar{S}} \in C^{q_S}} \left( \varphi^{(k_n)_{n \in \bar{S}}} f (\varphi^{(k_n)_{n \in \bar{S}}})^\dag \right) \otimes \bigotimes_{n \in \bar{S}} \id_{A_n^{k_n}} \, \middle| \, f \in \Lin(\ch_{A_S}^{q_S}) \right\}
\end{split} \ee
where \correction{the first inclusion comes from (\ref{eq: decomposing pi A_S}),} and we \correction{used Skolem-Noether to say} that $\kappa^{(k_n)_{n \in \bar{S}}}$'s action can be seen as conjugation by a certain unitary $\varphi^{(k_n)_{n \in \bar{S}}} \in \Lin(\ch_{A_S}^{q_S})$. 

\correction{Moreover, the map from an $f \in \Lin(\ch_{A_S}^{q_S})$ to the form given in (\ref{eq: form of picaS})'s last line is injective since the $\varphi$'s are unitary; thus the C* algebra described in this last line is isomorphic to $\Lin(\ch_{A_S}^{q_S})$. Yet we saw that for any $\veck \in C^{q_S}$, $\hatpi^\veck|_{\pi_S^{q_S} \ca_S}$ is an isomorphism from $\pi_S^{q_S} \ca_S$ to $\Lin(\ch_{A_S}^{q_S})$ (padded with some identities). Therefore, in (\ref{eq: form of picaS})'s inclusion, the subset is isomorphic (as a C* algebra) to the superset, so (\ref{eq: form of picaS}) is actually an equality.}

\correction{Let us pick $q_S \in Q_S$, $m \in S$, and $k_m \in K_m$ such that $\pi_m^{k_m} = \pi_m^{k_m} \pi_{S}^{q_S}$ . Since $\pi_m^{k_m} = \pi_m^{k_m} \pi_{S}^{q_S}$ and $\ca_m \in \ca_S$, we have $\pi_m^{k_m} \ca_m \subseteq \pi_S^{q_S} \ca_S$, so that its elements must be both of the forms (\ref{eq: form of pi An}) and (\ref{eq: form of picaS}). This shows that for any $(k_n)_{n \in \bar{S}} \in C^{q_S}$, $\hat{\varphi}^{(k_n)_{n \in \bar{S}}}$ restricts to the identity on $\Lin(\ch_{A_m}^{k_m}) \otimes \bigotimes_{n \in S \setminus \{m\}} \id_{A_n} = \pi_m^{k_m} \ca_m$, which is equivalent to $\varphi^{(k_n)_{n \in \bar{S}}}$ commuting with $\pi_m^{k_m} \ca_m$.

Thus more generally, for any choice of $(k_n)_{n \in \bar{S}} \in \tK_{\bar{S}}$, and any $q_S$ such that $(k_n)_{n \in \bar{S}} \in C^{q_S}$, $\varphi^{(k_n)_{n \in \bar{S}}}$ commutes with $\bigvee_{(k_n)_n \in E^{q_S}} \bigvee_{n \in S} \pi_n^{k_n} \ca_n$, and thus also with $ \bigvee_{(k_n)_n \in \tK_S} \bigvee_{n \in S} \pi_n^{k_n} \ca_n$ $= \bigvee_{n \in S} \ca_n$ (the additional terms in that latter span, corresponding to $q_S$'s such that $(k_n)_{n \in \bar{S}} \not\in C^{q_S}$, have null products with $\varphi^{(k_n)_{n \in \bar{S}}}$). Therefore, we find 
\be \varphi^{(k_n)_{n \in \bar{S}}} \in \ca_S \cap \left(\bigvee_{n \in S} \ca_n \right)' = \bigvee_{n \in S} \cz(\ca_n) \, , \ee 
where the second equality is obtained by applying Lemma \ref{lem: commuting with all parts} with $\OM = \ca_S$.

Since the atomic projectors of the commutative algebra $\bigvee_{n \in S} \cz(\ca_n)$ clearly are the $\prod_{n \in S} \pi_n^{k_n}$'s for $(k_n)_{n \in S} \in \tK_S$, $\varphi^{(k_n)_{n \in \bar{S}}}$ is a linear combination of those, and thus by unitarity is of the form
\be \varphi^{(k_n)_{n \in \bar{S}}} = \sum_{(k_n)_{n \in S} \in \tK_S} e^{i \al(\veck)} \prod_{n \in S} \pi_n^{k_n} \ee
for some phases $\al$ depending on $\veck = (k_n)_{n \in X}$.

We then define
\be \label{eq: def phi} \begin{split}
    \phi &= \sum_{(k_n)_{n \in \bar{S}} \in \tK_{\bar{S}}} \varphi^{(k_n)_{n \in \bar{S}}} \prod_{n \in \bar{S}} \pi_n^{k_n} \\
    &= \sum_{\substack{(k_n)_{n \in S} \in \tK_S \\ (k_n)_{n \in \bar{S}} \in \tK_{\bar{S}}}} e^{i \al(\veck)} \prod_{n \in X} \pi_n^{k_n} \\
    &= \sum_{\veck \in \tilde{K}} e^{i \al(\veck)} \prod_{n \in X} \pi_n^{k_n} \, ,
\end{split}\ee
where the last line's restriction of the sum to one over $\veck \in \tilde{K}$ comes from deleting null terms. This allows us to rewrite (\ref{eq: form of picaS}) as
\be \begin{split}
    \forall q_S \in Q_S, \quad \pi_S^{q_S} \ca_S &=  \hat{\phi} \left( \left\{  f \otimes \left(\bigoplus_{(k_n)_{n \in \bar{S}} \in C^{q_S}} \bigotimes_{n \in \bar{S}} \id_{A_n^{k_n}} \right) \, \middle| \, f \in \Lin(\ch_{A_S}^{q_S}) \right\} \right) \\
    &=  \hat{\phi} \left( \left\{  \left( f \otimes  \bigotimes_{n \in \bar{S}} \id_{A_n} \right) \tilde{\bbpi} \, \middle| \, f \in \Lin(\ch_{A_S}^{q_S}) \right\} \right)
\end{split} \ee
so 
\be \begin{split}
    \ca_S &= \biguplus_{q_S \in Q_S} \pi_S^{q_S} \ca_S \\
    &=  \hat{\phi} \left( \left\{  \left( \sum_{q_S \in Q_S} f^{q_S} \otimes  \bigotimes_{n \in \bar{S}} \id_{A_n} \right) \tilde{\bbpi} \, \middle| \, \forall q_S \in Q_S, f^{q_S} \in \Lin(\ch_{A_S}^{q_S}) \right\} \right) \\
    &=  \hat{\phi} \left( \left\{  \left( f \otimes  \bigotimes_{n \in \bar{S}} \id_{A_n} \right) \tilde{\bbpi} \, \middle| \,  f \in \Lin_{\sim_S}(\ch_{A_S}) \right\} \right) \\
\end{split} \ee
which is the form (\ref{eq: rep composite}). Additionally, (\ref{eq: def phi}) shows that $\phi$ is a unitary element of $\bigvee_{n \in X} \cz_n$, since it is a linear combination of its atomic projectors.} Note that $\phi$ may depend on the set $S$ that we fixed at the start.

\section{Proofs for Section \ref{sec: fermions}} \label{app: fermions}

For all $i$ in $\{1,...,N\}$, let $n_i = a^{\dagger}_i a_i $. Because two creation or annihilation operators on different modes anti-commute and because $a_ia_i = a^{\dagger}_ia^{\dagger}_i = 0$ and $a_ia^{\dagger}_i = \id - a^{\dagger}_ia_i = \id - n_i$ $\forall i$, elements $y$ of $\OM$ can be written in a unique way as 
\begin{equation}
\label{eq: canonical decomposition fermions}
    y = \sum_{\Vec{k} \in \{0,...,3\}^N} \alpha_{\vec{k}} \prod_i x_i(k_i) \, ,
\end{equation} where $\forall i \in X$, $x_i(0) = \id$, $x_i(1) = a^{\dagger}_i$, $x_i(2) = a_i$, $x_i(3) = n_i$ \correction{and the $\alpha_{\vec{k}}$ are complex numbers}. We call this decomposition the canonical decomposition of $y$. 

\subsection{Proof of Proposition \ref{prop: atomic projectors fermions}} \label{app: proof atomic projectors fermions}

\correction{We recall that $\OM = \ca_X = \textrm{Span}(\mathcal{M}_X)$ is the algebra of physical fermionic operators on all the modes of $X$}. We start by computing $\mathcal{Z}(\OM)$ and its atomic projectors. Let $z \in \mathcal{Z}(\OM)$; this is equivalent to asking that $z$ commutes with every $a_ia_j$, $a^{\dagger}_ia^{\dagger}_j$, $a_ia^{\dagger}_j$, $a^{\dagger}_ia_j$ for $i,j \in \{1,...,N\}$.

\correction{Let us fix $i \in \{1,...,N\}$ and see what it means for $z$ to commute with $n_i$.} By definition of the algebra $\OM$, $z$ can be decomposed in a unique way as $z = z_0 + a_iz_1 + a^{\dagger}_iz_2 + n_iz_3$ where the $z_j$'s contain no appearance of $a_i$ or $a^{\dagger}_i$. We can then compute 
\begin{equation}
\begin{split}
    0 = [z, n_i] & = [z_0,n_i] + [a_iz_1,n_i] + [a^{\dagger}_iz_2,n_i] + [n_iz_3,n_i] \\ 
    & = 0 + [a_i,n_i]z_1 + [a^{\dagger}_i,n_i]z_2 + 0 \\
    & = a_iz_1 - a^{\dagger}_iz_2 \, ,
\end{split}
\end{equation}
and thus $z_1 = z_2 =0$. \correction{This implies that the only operator acting on the $i^{\text{th}}$ mode appearing in $z$ is $n_i$. As $z$ must commute with $n_j$ for all $j \in \{1,...,N\}$, we get that for any $j \in \{1,...,N\}$ the only operator acting on the $j^{\text{th}}$ mode appearing in $z$ is $n_j$.}

Next, let us see what it means for $z$ to commute with $a^{\dagger}_i a^{\dagger}_j$, $i,j \in \{1,...,N\}$. By definition of the algebra $\OM$ and the previous commutation result, $z$ can be decomposed in a unique way as $z = z_0 + n_iz_1 + n_jz_2 + n_in_jz_3$ where the $z_k$ contain no appearance of $a_i$, $a^{\dagger}_i$, $a_j$, $a^{\dagger}_j$. We can then compute 
\begin{equation}
\begin{split}
    0 = [z,a^{\dagger}_i a^{\dagger}_j] & = [z_0,a^{\dagger}_i a^{\dagger}_j] + [n_i,a^{\dagger}_i a^{\dagger}_j]z_1 + [n_j,a^{\dagger}_i a^{\dagger}_j]z_2 + [n_in_j,a^{\dagger}_i a^{\dagger}_j]z_3 \\
    & = 0 + a^{\dagger}_i a^{\dagger}_jz_1 + a^{\dagger}_i a^{\dagger}_jz_2 + a^{\dagger}_i a^{\dagger}_jz_3\\
    & = a^{\dagger}_i a^{\dagger}_j(z_1 + z_2 + z_3) \, .
\end{split}
\end{equation}
Thus the fact that $z$ commutes with $a^{\dagger}_i a^{\dagger}_j$ is equivalent to the fact that $z_1 + z_2 + z_3 = 0$. The commutation of $z$ with $a_ia_j$ gives the exact same relation.

Finally let us see what it means for $z$ to commute with $a^{\dagger}_ia_j$, $i,j \in \{1,...,N\}$. We decompose $z$ as $z = z_0 + n_iz_1 + n_jz_2 + n_in_jz_3$ where the $z_k$ contain no appearance of $a_i$, $a^{\dagger}_i$, $a_j$, $a^{\dagger}_j$. We get that 
\begin{equation}
\begin{split}
    [z,a^{\dagger}_ia_j] & = [z_0, a^{\dagger}_ia_j] + [n_i,a^{\dagger}_ia_j]z_1 + [n_j,a^{\dagger}_ia_j]z_2 + [n_in_j,a^{\dagger}_ia_j]z_3 \\
    & = 0 + a^{\dagger}_ia_jz_1 - a^{\dagger}_ia_jz_2 + 0  \\
    & = a^{\dagger}_ia_j (z_1 - z_2)
\end{split}
\end{equation}
so that $z_1 = z_2$. The commutation with $a_ia^{\dagger}_j$ gives the exact same relation.

Combining these relations, we get that $z$ is in $\mathcal{Z}(\OM)$ if and only if, $\forall i \neq j \in X$, it can be decomposed as 
\begin{equation}
     z = z_0^{ij} + n_iz_1^{ij} + n_jz_1^{ij} -2n_i\correction{n_j}z_1^{ij}
\end{equation}
where the $z^{ij}_k$'s contain no appearance of $a_i$, $a^{\dagger}_i$, $a_j$, $a^{\dagger}_j$.
Let $z$ be such an element, by adding a multiple of the identity we can suppose that the coefficient of $\id$ in the canonical decomposition \correction{(defined in Equation \ref{eq: canonical decomposition fermions})} of $z$ is $0$. Now up to multiplication by a scalar we can distinguish two possibilities: the coefficient of $n_1$ in the canonical decomposition of $z$ is either $1$ or $0$.
Suppose that it is $1$. Let $j \neq 1$; we can write the decomposition 
\begin{equation}
\label{eq: decomposition center 1j}
    z = z_0^{1j} + n_1z_1^{1j} + n_jz_1^{1j} -2n_1\correction{n_j}z_1^{1j} \, ,
\end{equation}
where, because the coefficient of $n_1$ in the canonical decomposition of $z$ is $1$, $\id$ has coefficient $1$ in the canonical decomposition of $z_1^{1j}$. It follows that $n_j$ has coefficient $1$ and $n_1n_j$ has coefficient $-2$ in the canonical decomposition of $z$. By applying this reasoning for the different $i,j$ we get that $\forall i < j \in \{1,...,N\}$, the coefficient of $n_i$ in the canonical decomposition of $z$ is $1$ and the coefficient of $n_in_j$ is $-2$. Let now $1 \neq j < k$; \correction{by considering the decomposition given in Equation \ref{eq: decomposition center 1j}}, we know because of the previous step that $n_k$ has coefficient $-2$ in the canonical decomposition of \correction{$z_1^{1j}$} and thus that $n_1n_jn_k$ has coefficient $+4$ in the canonical decomposition of $z$. We apply the same reasoning for every $i < j < k$ and get that $+4$ is the coefficient of $n_in_jn_k$ in the canonical decomposition of $z$ for all such triplets.
Iterating this reasoning up to the N-tuple gives us that 
\begin{equation}
    z = \sum_{\emptyset \neq Y \subseteq X} (-2)^{|Y|-1} \prod_{i \in Y} n_i \, .
\end{equation} 
% For example, on a $3$-modes fermionic system we get that 
% \begin{equation}
% \pi = n_1 + n_2 + n_3 - 2(n_1n_2 + n_2n_3 + n_1n_3) + 4n_1n_2n_3 \, .
% \end{equation}
On the other hand if the coefficient of $n_1$ is $0$ in the canonical decomposition of $z$, we can apply the same reasoning except that all coefficients are null and thus $z = 0$. 

It follows that 
\begin{equation}
    \mathcal{Z}(\OM) = \textrm{Span}(\{ \id, \pi\})
\end{equation}
with
\begin{equation}
    \pi = \sum_{\emptyset \neq Y \subseteq X} (-2)^{|Y|-1} \prod_{i \in Y} n_i \, .
\end{equation}
\correction{Let us now present the following lemma, which we will use to prove that $\pi$ is a projector.
\begin{lemma}
\label{lem: computation projection}
    Let $U$ be a non-empty finite set, then 
    \be
    \left( \sum_{\substack{\emptyset \neq Y \subseteq U \\ V \subseteq Y}} (-2)^{|U|+|V|-2} \right) - (-2)^{|U|-2} = (-2)^{|U|-1} \, .
    \ee
\end{lemma}
\begin{proof}
    Let $U$ be such a set, we compute that 
    \begin{equation}
    \begin{split}
        & \left( \sum_{\substack{\emptyset \neq Y \subseteq U \\ V \subseteq Y}} (-2)^{|U|+|V|-2} \right) - (-2)^{|U|-2}\\
        = & \left( \sum_{k=1}^{|U|} \binom{|U|}{k} \left( \sum_{i = 0}^{k} \binom{k}{i} (-2)^{|U|+i-2} \right) \right) - (-2)^{|U|-2} \\
        = & (-2)^{|U|-2} \left( \left( \sum_{k=1}^{|U|} \binom{|U|}{k} \left( \sum_{i = 0}^{k} \binom{k}{i} (-2)^{i} \right) \right) -1 \right) \\
        = & (-2)^{|U|-2} \left( \left( \sum_{k=1}^{|U|} \binom{|U|}{k} (-1)^{k} \right) -1 \right) \\
        = & (-2)^{|U|-2} \left( \left( \sum_{k=0}^{|U|} \binom{|U|}{k} (-1)^{k} \right) -2 \right) \\
        = & (-2)^{|U|-2} \times (-2) \\
        = & (-2)^{|U|-1} \, .
    \end{split}    
    \end{equation}
\end{proof}}
Let us now prove that the atomic projectors of $\cz(\OM)$ are  $\pi$ and $\id - \pi$. Indeed, \correction{1)} $\pi$ and $\id-\pi$ are obviously non-null as we have made explicit the coefficient of $\pi$; \correction{2)} they are orthogonal projectors as \correction{
\begin{equation}
\begin{split}
    \pi^2 & = \left( \sum_{\emptyset \neq Y \subseteq X} (-2)^{|Y|-1} \prod_{i \in Y} n_i \right) \left( \sum_{\emptyset \neq Z \subseteq X} (-2)^{|Z|-1} \prod_{i \in Z} n_i \right) \\
    & = \sum_{\emptyset \neq U \subseteq X} \left( \sum_{\substack{\emptyset \neq Y,Z \subseteq X \\ Y \cup Z = U}} (-2)^{|Y|+|Z|-2} \right) \prod_{i \in U} n_i\\
    & = \sum_{\emptyset \neq U \subseteq X} \left( \left( \sum_{\substack{\emptyset \neq Y \subseteq U \\ V \subseteq Y}} (-2)^{|U|+|V|-2} \right) - (-2)^{|U|-2} \right) \prod_{i \in U} n_i\\
    & = \sum_{\emptyset \neq U \subseteq X} (-2)^{|U|-1} \prod_{i \in U} n_i \\
    & = \pi
\end{split}    
\end{equation}
where the third equality comes from summing over $V = Y \cap Z$ instead of $Z$ as well as remarking that as it is written this sum allows for the case where $Y = U$ and $Z = V = \emptyset$ whose term must then be subtracted. The fourth equality comes from applying Lemma \ref{lem: computation projection}. We can finally check that}
\begin{equation}
    (\id - \pi)^2  = \id -2\pi + \pi^2 = \id - \pi
\end{equation}
and that $\pi^{\dagger} = \pi$ as the $n_i$ are Hermitian and \correction{pairwise} commute; \correction{3)} they are orthogonal to each other as 
\begin{equation}
    \pi (\id - \pi ) = \pi - \pi^2 = 0 = \pi^2 - \pi = (\id - \pi) \pi \, ;
\end{equation}
and \correction{4)} they form a basis of $\mathcal{Z}(\OM)$ as
\begin{equation}
    \mathcal{Z}(\OM) = \textrm{Span}(\{\pi, \id - \pi \}) \, .
\end{equation}

To give intuition, it is interesting to remark that $\pi$ is `the projector on the subspace where we have an odd number of fermions in the system'. Indeed, consider a state $ \ket{m} = \ket{m_1...m_N}$ \correction{where $m_i \in \{0,1\}$ for all $i \in \{1,...,N\}$} and such that $m_1 + ... + m_N = k$. We first remark that for $Y \subseteq X$
\begin{equation}
(\prod_{i \in Y} n_i) \ket{m_1...m_N} = 
\begin{cases}
    \ket{m_1...m_N} & \mbox{ if } Y \subseteq \{j | m_j = 1 \} \\
    0 & \mbox{ otherwise } \, .
\end{cases}
\end{equation}
It follows that 
\begin{equation}
\begin{split}
    \pi \ket{m} & = (\sum_{\emptyset \neq Y \subseteq X} (-2)^{|Y|-1} \prod_{i \in Y} n_i)\ket{m_1...m_N} \\
     & = \sum_{1 \leq j \leq N} \binom{k}{j} (-2)^{j-1} \ket{m_1...m_N} \\
     & = - \frac{1}{2} (\sum_{1 \leq j \leq k} \binom{k}{j} (-2)^{j}) \ket{m_1...m_N} \\
     & = \frac{1-(-1)^k}{2} \ket{m_1...m_N} \\
     & = 
     \begin{cases}
    \ket{m_1...m_N} & \mbox{ if } k \mbox{ is odd} \\
    0 & \mbox{ if } k \mbox{ is even} \, .
    \end{cases}
\end{split}
\end{equation}
Conversely, $\id- \pi$ is `the projector on the subspace where we have an even number of fermions in the system'.

Because the difference with the special case of $\OM$ is just the choice of the modes, for any $\emptyset \neq S \subseteq X$ the atomic projectors of $\cz(\mathcal{A}_S)$ can be computed in the same way: they are $\{\pi_S, \id - \pi_S\}$ where 
\begin{equation}
    \pi_S = \sum_{\emptyset \neq Y \subseteq S} (-2)^{|Y|-1} \prod_{i \in Y} n_i \,.
\end{equation}

\subsection{Proof of Proposition \ref{prop: conjunctions of centres fermions}}
Let $S$ and $T$ be two disjoint subsets of $X$; then 
\begin{equation}
\begin{split}
    \pi_{S \sqcup T} & = \sum_{\emptyset \neq Y \subseteq S \sqcup T} (-2)^{|Y|-1} \prod_{i \in Y} n_i \\
    & = \sum_{\emptyset \neq Y \subseteq S} (-2)^{|Y|-1} \prod_{i \in Y} n_i + \sum_{\emptyset \neq Y \subseteq T} (-2)^{|Y|-1} \prod_{i \in Y} n_i + \sum_{\substack{\emptyset \neq Y \subseteq S \sqcup T \\ Y \nsubseteq S \\ Y \nsubseteq T }} (-2)^{|Y|-1} \prod_{i \in Y} n_i \\
    & = \pi_S + \pi_T + \sum_{\emptyset \neq Y_1 \subseteq S} \sum_{\emptyset \neq Y_2 \subseteq T} (-2)^{|Y_1|+|Y_2|-1} \prod_{i \in Y_1} n_i \prod_{j \in Y_2} n_j \\
    & = \pi_S + \pi_T - 2 \pi_S \pi_T \, .
\end{split}
\end{equation}
    
\subsection{Proof of Proposition \ref{prop: partition fermions}}

We now prove that $(\mathcal{A}_S)_{S \subseteq X} \vdash \OM$. 

Let $S$ and $T$ be two disjoint subsets of $X$; $\mathcal{M}_S \subseteq \mathcal{M}_{S \sqcup T}$ and $\mathcal{M}_T \subseteq \mathcal{M}_{S \sqcup T}$, thus because $\mathcal{A}_U = \textrm{Span}(\mathcal{M}_U)$ $\forall U \subseteq X$, $\mathcal{A}_S$ and $\mathcal{A}_T$ are sub-$\mathbb{C}^*$ algebras of $\mathcal{A}_{S \sqcup T}$. 

The condition $\mathcal{Z}(\mathcal{A}_{S \sqcup T}) \subseteq \mathcal{Z}(\mathcal{A}_S) \vee \mathcal{Z}(\mathcal{A}_T)$ is equivalent to showing that $\pi_{S \sqcup T} \in \mathcal{Z}(\mathcal{A}_S) \vee \mathcal{Z}(\mathcal{A}_T)$ as $\mathcal{Z}(\mathcal{A}_{S \sqcup T})$ is spanned by $\pi_{S \sqcup T}$ and $\id$ which we know is in $\mathcal{Z}(\mathcal{A}_S) \vee \mathcal{Z}(\mathcal{A}_T)$. This is a direct consequence of Proposition \ref{prop: conjunctions of centres fermions}.

Finally it remains to show that 
\begin{subequations}
\begin{align}
\pi_{S \sqcup T} \mathcal{A}_S' & = \pi_{S \sqcup T} \mathcal{A}_T  \, ,\\
 \label{eq: second equality}(\id - \pi_{S \sqcup T}) \mathcal{A}_S' & = (\id -\pi_{S \sqcup T}) \mathcal{A}_T \, .
\end{align}
\end{subequations}
First we can remark that because of the anticommutation relations and the fact that elements of $\mathcal{A}_S$ and $\mathcal{A}_T$ are generated by even numbers of creation and annihilation operators, $\mathcal{A}_T \subseteq \mathcal{A}_S'$ and thus $\pi_{S \sqcup T} \mathcal{A}_T \subseteq \pi_{S \sqcup T} \mathcal{A}_S'$.

Now let $z \in \mathcal{A}_S'$; by the same reasoning that we used to deduce some information on the elements of the centre \correction{in the proof of Proposition \ref{prop: atomic projectors fermions}}, this is equivalent to the fact that $\forall i \neq j \in S$, $z$ can be decomposed as 
\begin{equation}
     z = z_0^{ij} + n_iz_1^{ij} + n_jz_1^{ij} -2n_in_jz_1^{ij}
\end{equation}
where the $z^{ij}_k$ contain no appearance of $a_i$, $a^{\dagger}_i$, $a_j$, $a^{\dagger}_j$. \correction{As we only know that this decomposition exists for $i,j \in S$, we will not compare it to the canonical decomposition but instead to the following unique decomposition of elements of $\mathcal{A}_{S \sqcup T}$, as}
\begin{equation}
    z = \sum_{\Vec{k} \in \{0,...,3\}^{|S|}} z_{\vec{k}} \,  \prod_{i \in S} x_i(k_i)
\end{equation}
where $z_{\correction{\vec{k}}} \in \mathcal{A}_T$ and $\forall i \in \correction{S}$, $x_i(0) = \id$, $x_i(1) = a^{\dagger}_i$, $x_i(2) = a_i$ and $x_i(3) = n_i$.

We can now use the same reasoning as in Section \ref{app: proof atomic projectors fermions}, with the only difference being that coefficients are not complex numbers but elements of $\mathcal{A}_T$. Going through all the steps of the computation gives us that 
\begin{equation}
    z = z_0 + z_T \pi_S
\end{equation}
where $z_0,z_T \in \mathcal{A}_T$. Then 
\begin{equation}
\begin{split}
    \pi_{S \sqcup T}z & = \pi_{S \sqcup T}(z_0 + z_T \pi_S) \\
    & = \pi_{S \sqcup T}z_0 + (\pi_S + \pi_T - 2 \pi_S\pi_T)(z_T \pi_S) \\
    & = \pi_{S \sqcup T}z_0 + (\pi_Sz_T - \pi_S\pi_Tz_T) \, .
\end{split}
\end{equation}
However, since
\begin{equation}
\begin{split}
    \pi_{S \sqcup T}(z_T - \pi_T\correction{z_T}) & = (\pi_S + \pi_T - 2 \pi_S\pi_T)\correction{(z_T - \pi_Tz_T)}\\
    & = \pi_Sz_T + \pi_Tz_T - 2\pi_S\pi_Tz_T - \pi_S\pi_Tz_T - \pi_Tz_T + 2\pi_S\pi_Tz_T\\
    & = \pi_Sz_T - \pi_S\pi_Tz_T \,,
\end{split}
\end{equation}
we have that 
\begin{equation}
\begin{split}
    \pi_{S \sqcup T}z & = \pi_{S \sqcup T}z_0 + \pi_{S \sqcup T}(z_T - \pi_T\correction{z_T})\\
    & = \pi_{S \sqcup T}(z_0 + z_T - \pi_T\correction{z_T})
\end{split}
\end{equation}
where $z_0 + z_T - \pi_Tz_T \in \mathcal{A}_T$. This proves that $\pi_{S \sqcup T} \mathcal{A}_S' \subseteq \pi_{S \sqcup T} \mathcal{A}_T$ and thus that $\pi_{S \sqcup T} \mathcal{A}_S' = \pi_{S \sqcup T} \mathcal{A}_T$.

The proof of (\ref{eq: second equality}) is very similar. We have that $(\id - \pi_{S \sqcup T}) \mathcal{A}_T \subseteq (\id -\pi_{S \sqcup T}) \mathcal{A}_S'$ from the anti-commutation relations. And letting $z = z_0 + z_T \pi_S \in \mathcal{A}_S'$,
\begin{equation}
\begin{split}
    (\id - \pi_{S \sqcup T})z & =  (\id - \pi_{S \sqcup T})z_0 + \pi_Sz_T - \pi_{S \sqcup T}\pi_Sz_T \\
    & = (\id - \pi_{S \sqcup T})z_0 + \pi_S\pi_Tz_T \\
    & = (\id - \pi_{S \sqcup T})z_0 + (\id - \pi_{S \sqcup T})\pi_Tz_T\\
    & = (\id - \pi_{S \sqcup T})(z_0 + \pi_Tz_T) \,.
\end{split}
\end{equation}
It follows that $(\id - \pi_{S \sqcup T}) \mathcal{A}_S' = (\id -\pi_{S \sqcup T}) \mathcal{A}_T$, which concludes the proof. 

\subsection{Proof of Proposition \ref{prop: non-representable fermions}}
To show this result we adapt to our formalism a proof of an analogous result given in \cite{guaita2024locality}.
Let $N \geq 3$ be an integer and $(\mathcal{A}_S)_{S \subseteq X}$ be the partitioned algebra of physical operators on an $N$-mode fermionic system. For $i,j \in X$ with $i \neq j$, we define 
\begin{equation}
    b_{ij} = a_ia_j + a_ia^{\dagger}_j + a^{\dagger}_ia_j + a^{\dagger}_ia^{\dagger}_j = (a_i+a^{\dagger}_i)(a_j+a^{\dagger}_j) \,.
\end{equation}
We remark that for $i,j,k,l \in X$ with $i \neq j$ and $k \neq l$,
\begin{equation}
    b_{ij}b_{kl} = (-1)^{\delta_{ik} + \delta_{il} + \delta_{jk} + \delta_{jl}} b_{kl}b_{ij} \,;
\end{equation}
in other words, $b_{ij}$ and $b_{kl}$ anti-commute if they have exactly one index in common. We also remark that 
\begin{equation}
\begin{split}
    b_{ij}b_{jk}b_{ki} & = (a_i+a^{\dagger}_i)(a_j+a^{\dagger}_j)(a_j+a^{\dagger}_j)(a_k+a^{\dagger}_k)(a_k+a^{\dagger}_k)(a_i+a^{\dagger}_i) \\
    & = (a_ia^{\dagger}_i+a^{\dagger}_ia_i)(a_ja^{\dagger}_j+a^{\dagger}_ja_j)(a_ka^{\dagger}_k+a^{\dagger}_ka_k) \\
    & = \id \,.
\end{split}
\end{equation}

Now, suppose the partition is fully representable; we can then fix a representation $\iota$ of it such that
\be \forall S \subseteq X, \quad \iota(\ca_S) = \left\{ \Big(f_{A_n, \, n \in S} \otimes \id_{A_n, \, n \not\in S}\Big) \, \tilde{\bbpi} \,\,\, \middle| \,\,\,  f \in \Lin_{\sim_S}\left(\bigotimes_{n \in S} \ch_{A_n}\right) \right\} \, . \ee
In particular, as $b_{ij} \in \mathcal{A}_{\{i,j\}}$ we can define 
\be
B_{ij} = \iota(b_{ij}) = (\tilde{B}_{ij} \otimes \id_{X\setminus\{i,j\}}) \tilde{\bbpi} 
\ee
where $\tilde{B}_{ij} \in \Lin_{\sim_{\{i,j\}}}(\ch_{A_i} \otimes \ch_{A_j})  \subseteq \Lin(\ch_{A_i} \otimes \ch_{A_j})$. By orthogonal tensor decomposition we can write these operators as
\begin{equation}
    \tilde{B}_{ij} = \sum_{k = 1}^r \alpha_k^{ij} B^{(ij;i)}_k \otimes B^{(ij;j)}_k
\end{equation}
where the $B^{(ij;i)}_k$'s are pairwise orthogonal operators acting on $\ch_{A_i}$, and the $B^{(ij;j)}_k$'s are pairwise orthogonal operators acting on $\ch_{A_j}$; which means that 
\begin{equation}
\label{eq: orthonormality of the Bs}
    \Tr((B^{(ij;i)}_k)^{\dagger}B^{(ij;i)}_{k'}) =  \Tr((B^{(ij;j
    )}_k)^{\dagger}B^{(ij;j)}_{k'}) = \delta_{kk'} \,.
\end{equation}

Let $i,j,k$ be three different indices; since $b_{ij}$ and $b_{jk}$ anti-commute in $\mathcal{A}_{\{i,j,k\}}$ we have that
 \begin{equation}
    0 = \{\tilde{B}_{ij},\tilde{B}_{jk}\} = \sum_{l,l'} \alpha_l^{ij} \alpha_{l'}^{jk} B^{(ij;i)}_l \otimes \{B^{(ij;j)}_l , B^{(jk;j)}_{l'} \} \otimes B^{(jk;k)}_{l'} \otimes \id_{X \setminus \{i,j,k\}} \,.
\end{equation}
\correction{ For any indices $m,m' \in \{1,...,r\}$, multiplying this expression by $(B^{(ij;i)}_{m})^{\dagger} \otimes \id_{X \setminus \{i\}}$ and $(B^{(jk;k)}_{m'})^{\dagger} \otimes \id_{X \setminus \{k\}}$, and then tracing on $\mathcal{H}_{A_i}$ and $\mathcal{H}_{A_k}$, we obtain that 
\begin{equation}   
    %& 0 = \sum_{l,l'} \alpha_l^{ij} \alpha_{l'}^{jk} B^{(ij;i)}_l \otimes \{B^{(ij;j)}_l , B^{(jk;j)}_{l'} \} \otimes B^{(jk;k)}_{l'} \otimes \id_{X \setminus \{i,j,k\}} \\
    %\Rightarrow & 0 = \sum_{l,l'} \alpha_l^{ij} \alpha_{l'}^{jk} (B^{(ij;i)}_{l''})^{\dagger} B^{(ij;i)}_l \otimes \{B^{(ij;j)}_l , B^{(jk;j)}_{l'} \} \otimes B^{(jk;k)}_{l'}(B^{(jk;k)}_{l'''})^{\dagger} \\
     \sum_{l,l'} \alpha_l^{ij} \alpha_{l'}^{jk} \Tr((B^{(ij;i)}_{m})^{\dagger} B^{(ij;i)}_l) \otimes \{B^{(ij;j)}_l , B^{(jk;j)}_{l'} \} \otimes \Tr(B^{(jk;k)}_{l'}(B^{(jk;k)}_{m'})^{\dagger}) = 0 \, .
\end{equation}
It follows, by applying Equation \ref{eq: orthonormality of the Bs}, that
\begin{equation}
    \{B^{(ij;j)}_{m} , B^{(jk;j)}_{m'} \} = 0 \, ,
\end{equation}
and we can conclude that for any $i,j,k \in X$ and any appropriate indices $l,l'$
\begin{equation}
    \Tr(B^{(ij;j)}_{l} B^{(jk;j)}_{l'}) = \frac{1}{2} \Tr(\{B^{(ij;j)}_{l} , B^{(jk;j)}_{l'} \}) = 0 \,.
\end{equation}}

Finally, let $i,j,k$ be three different elements of $X$. As $b_{ij}b_{jk}b_{ki} = \id$ we have that 
\begin{equation}
    B_{ij}B_{jk}B_{ki} = \id_X \,.
\end{equation}
It follows that 
\begin{equation}
    \correction{(\tilde B_{ij} \otimes \id_{k}) (\tilde B_{jk} \otimes \id_{i}) (\tilde B_{ki} \otimes \id_{j}) = \id_{\{i,j,k\}}}
\end{equation}
which can be developed as
\begin{equation}
    \sum_{l,l',l''} \alpha_l^{ij} \alpha_{l'}^{jk} \alpha_{l''}^{ki} (B^{(ij;i)}_l B^{(ki;i)}_{l''}) \otimes  (B^{(ij;j)}_{l} B^{(jk;j)}_{l'}) \otimes  (B^{(jk;k)}_{l'} B^{(ki;k)}_{l''}) = \id_{\{i,j,k\}} \,.
\end{equation}
However this implies that 
\begin{equation}
\begin{split}
    &\Tr( \tilde B_{ij} \tilde B_{jk} \tilde B_{ki}) \\
    & = \sum_{l,l',l''} \alpha_l^{ij} \alpha_{l'}^{jk} \alpha_{l''}^{ki} \Tr(B^{(ij;i)}_l B^{(ki;i)}_{l''}) \Tr(B^{(ij;j)}_{l} B^{(jk;j)}_{l'}) \Tr(B^{(jk;k)}_{l'} B^{(ki;k)}_{l''}) \\
    & = \Tr(\id_{\{i,j,k\}}) \neq 0
\end{split}    
\end{equation}
which is impossible: we saw that all the $\Tr(B^{(ij;i)}_l B^{(ki;i)}_{l''})$'s are null and thus $\Tr( \tilde B_{ij} \tilde B_{jk} \tilde B_{ki})=0$. We have reached a contradiction, which concludes the proof.

\end{document}